\documentclass[usenatbib,twocolumn]{texstuff/aastex631}
\usepackage{graphicx}
\usepackage{xcolor}
\usepackage{amssymb,amsmath}
\usepackage{natbib}
\usepackage[T1]{fontenc}
\usepackage{verbatim}

\newcommand{\ha}{\mbox{H$\alpha$}}
\newcommand{\hii}{H\,{\textsc{II}~}}
\newcommand{\hub}{{$H_0$}}
\newcommand{\galnum}{1,872,654} 
\newcommand{\degsq}{deg$^2$}
\newcommand{\phistar}{$\phi^{*}$}
\newcommand{\Lstar}{$L^{*}$}

\begin{document}

\title{Completeness of the NASA/IPAC Extragalactic Database (NED) - Local Volume Sample}

\author[0000-0002-6877-7655]{D.O. Cook}
\affiliation{Caltech/IPAC, 1200 E. California Boulevard, Pasadena, CA 91125, USA}
\author{J.M. Mazzarella}
\affiliation{Caltech/IPAC, 1200 E. California Boulevard, Pasadena, CA 91125, USA}
\author{G. Helou}
\affiliation{Caltech/IPAC, 1200 E. California Boulevard, Pasadena, CA 91125, USA}
\author{A. Alcala}
\affiliation{Caltech/IPAC, 1200 E. California Boulevard, Pasadena, CA 91125, USA}
\author{T.X. Chen}
\affiliation{Caltech/IPAC, 1200 E. California Boulevard, Pasadena, CA 91125, USA}
\author{R. Ebert}
\affiliation{Caltech/IPAC, 1200 E. California Boulevard, Pasadena, CA 91125, USA}
\author{C. Frayer}
\affiliation{Caltech/IPAC, 1200 E. California Boulevard, Pasadena, CA 91125, USA}
\author{J. Kim}
\affiliation{Caltech/IPAC, 1200 E. California Boulevard, Pasadena, CA 91125, USA}
\author{T. Lo}
\affiliation{Caltech/IPAC, 1200 E. California Boulevard, Pasadena, CA 91125, USA}
\author{B.F. Madore}
\affiliation{Department of Astronomy \& Astrophysics, University of Chicago, 5640 South Ellis Avenue, Chicago, IL 60637, USA}
\affiliation{Observatories of the Carnegie Institution for Science, Pasadena, CA}
\author{P.M. Ogle}
\affiliation{Space Telescope Science Institute, Baltimore, MD}
\author{M. Schmitz}
\affiliation{Caltech/IPAC, 1200 E. California Boulevard, Pasadena, CA 91125, USA}
\author{L.P. Singer}
\affiliation{NASA Goddard Space Flight Center, Code 661, Greenbelt, MD 20771, USA}
\author{S. Terek}
\affiliation{Caltech/IPAC, 1200 E. California Boulevard, Pasadena, CA 91125, USA}
\author{J. Valladon}
\affiliation{Department of Astronomy, San Diego State University, 5500 Campanile Drive, San Diego, CA, 92812-1221, USA}
\affiliation{Dept. of Physics \& Astronomy, San Jose State University, San Jose, CA}
\author{X. Wu}
\affiliation{Caltech/IPAC, 1200 E. California Boulevard, Pasadena, CA 91125, USA}



\begin{abstract}

We introduce the NASA/IPAC Extragalactic Database (NED) Local Volume Sample (NED-LVS), a subset of $\sim$1.9 million objects with distances out to 1000~Mpc. We use UV and IR fluxes available in NED from all-sky surveys to derive physical properties, and estimate the completeness relative to the expected local luminosity density. The completeness relative to NIR luminosities (which traces a galaxy's stellar mass) is roughly 100\% at $D<$30~Mpc and remains moderate (70\%) out to 300~Mpc. For brighter galaxies ($\gtrsim L_{*}$), NED-LVS is $\sim$100\% complete out to $\sim$400~Mpc. When compared to other local Universe samples (GLADE and HECATE), all three are $\sim$100\% complete below 30~Mpc. At distances beyond $\sim$80~Mpc, NED-LVS is more complete than both GLADE and HECATE by $\sim$10-20\%. NED-LVS is the underlying sample for the NED gravitational wave follow-up (NED-GWF) service, which provides prioritized lists of host candidates for GW events within minutes of alerts issued by the LIGO-Virgo-KAGRA collaboration. We test the prioritization of galaxies in the volume of GW170817 by 3 physical properties, where we find that both stellar mass and inverse specific star formation rate place the correct host galaxy in the top ten. In addition, NED-LVS can be used for a wide variety of other astrophysical studies: galaxy evolution, star formation, large-scale structure, galaxy environments, and more. The data in NED are updated regularly, and NED-LVS will be updated concurrently. Consequently, NED-LVS will continue to provide an increasingly complete sample of galaxies for a multitude of astrophysical research areas for years to come.

\end{abstract}


\section{INTRODUCTION}
Large, modern surveys are fundamental for understanding the demographics of galaxies, their distribution in space, and their evolution over cosmic time. Imaging the night sky is an essential first step to acquire coordinates, fluxes, and morphologies needed to identify the galaxies, but mapping the Universe in three dimensions requires distances either by redshift surveys \citep[e.g.,][]{Huchra83} or redshift-independent distances \citep[e.g.,][]{cosmicflows4}. As advances in detectors and computing power have led to an explosion of photometric imaging in different spectral regions and distance measurements, the NASA/IPAC Extragalactic Database (NED)\footnote{https://ned.ipac.caltech.edu/} has maintained a comprehensive census of objects beyond the Milky Way and a fusion of their most fundamental measurements across the electromagnetic spectrum. While the completeness of the NED census of distinct objects in a given volume is dependent on the observational selection effects of the largest sky surveys, the relative scarcity of redshifts is the most limiting factor in constructing a thorough 3D galaxy census map.


Understanding the completeness of redshift measurements in NED is important for a wide range of applications. Studies of the variation in space density, star formation rate density, and luminosity functions of galaxies as a function of look-back time, position on the sky, or differing environments in the cosmic web can be biased by incompleteness in galaxy detections or redshift measurements \citep[e.g.,][]{2014ApJ...795L..13H,Kulkarni18}. Failure to apply completeness corrections can lead to incorrect interpretations of astrophysical phenomena \citep[e.g.,][]{2021ApJS..257...60Z,2022MNRAS.509.5836L}. Similarly, searches for galaxy group/cluster members and satellites can be contaminated by foreground or background objects in regions with poor redshift coverage. Time-critical searches for the host galaxies to gamma ray burst (GRB) events \citep[][]{Singer15,Andreoni21,Ahumada22}, gravitational wave (GW) events \citep{Evans16,Coughlin19a}, high-energy neutrino events \citep{IceCube170922Ahost}, and other transients \citep{De20} are enhanced by the availability of redshifts to narrow down the list of plausible candidates and confirm distance or galaxy environment. In these, and many other cases, it is important to understand how the completeness of galaxy redshifts varies with distance and across the sky. The term ``completeness" hereafter refers to the fraction of cataloged galaxies with available redshifts compared to how many we expect in a given volume \citep{kopparapu08,gehrels16,Kulkarni18}.



One of the main science drivers for constructing the NED Local Volume Sample (NED-LVS) is its utilization in the search for electromagnetic counterparts to gravitational wave (GW) events. The detection of GWs \citep{GW150914} by the LIGO\footnote{Laser Interferometer Gravitational-Wave Observatory; https://www.ligo.caltech.edu/}-Virgo collaboration (and later by the LIGO-Virgo-KAGRA collaboration; hereafter LVK) has provided exciting new tests of basic physics and general relativity \citep{gw170817grtests}, but many of the most exciting discoveries are only possible when the electromagnetic (EM) counterpart is detected and studied concurrently. The EM counterpart to only one such event has been found to date (GW170817), yet has yielded remarkable results: constraints on the speed of GWs \citep{gw170817grtests}, the origin of most metals heavier than iron \citep{kasliwal17}, and an independent measure of the Hubble constant \citep{GW170817H0}. Now the race is on to find a population of these events. However, due to the poor sky localizations of GW events, these searches will be difficult for the next several years where they will be limited by the availability and quality of galaxy catalogs. The redshift completeness inside the sensitivity horizons of GW detectors will impact future searches for EM signatures to GW events.

The L-shaped interferometer detectors employed by LVK are sensitive to GWs emitted during the merger of two compact, massive objects: binary neutron stars (BNS), binary blackholes (BBH), or a mix of the two (NSBH). BNS and NSBH events are expected to produce EM counterparts via r-process radioactive decay of neutron-rich material expelled during the merger \citep[][]{Li98,Metzger10,Kasen15} referred to as a kilonova, whereas no EM counterparts are expected for BBH events \citep[but see][]{Graham20}. 

To date, there has been three observing runs aimed at detecting GWs: O1 (2015), O2 (2017), and O3 (2019), and two more are scheduled: O4 in early 2023 and O5 in 2025. However, no EM counterparts were discovered during the O3 run. A major factor in the absence of an O3 detection was poor sky localizations \citep{Petrov21}, where the median localization area during O3 was 4,500~deg$^2$. While there exist optical survey telescopes with large fields-of-view that can image significant sections of these areas, such as the Zwicky Transient Facility \citep[ZTF;][]{ztf1,ztf2} or Vera Rubin Observatory \citep[][]{rubin}, these surveys encounter a ``needle in a haystack" problem where transient candidates in sky localization areas are dominated by Galactic flare stars and distant supernovae \citep{kasliwal14}. However, the amplitude of a GW signal provides an independent estimate of the distance to the source, which is reported in an event's public notification. Thus, some groups chose to only target known galaxies in the 3-dimensional localization (including distance) which reduces the number of candidate sources by a factor of $\sim$100 \citep{nissanke13,gehrels16,ducoin20}. 


However, in the largest event localizations, even a 3D galaxy-targeted strategy will have limitations due to large numbers of host candidates in the event's volume \citep[e.g.,][]{GCN_S200213t}. This problem can be mitigated by prioritizing galaxies via their properties. The host of the only EM counterpart found to date \citep[NGC 4993 associated with GW170817;][]{coulter17} is an early-type galaxy with high mass and low SFR with an older stellar population \citep[e.g.,][]{Levan17,Pan17}. Theoretical studies have suggested that BNS events are more likely to occur in more massive galaxies given the long merger delay times of neutron-star pairs \cite[$>1$~Gyr;][]{Dominik12,Belczynski18}. However, the hosts of SNIa events, which are also thought to come from older populations, depend on both stellar mass and SFR \citep{scannapieco05,Sullivan06,Pan14,Wiseman21}. Given a BNS-host sample size of one, the relationship between BNS events and host galaxy properties is still hotly debated and not fully understood \citep{mapelli18,Artale19,Toffano19,Adhikari20,ducoin20,Fong22,Nugent22,Zevin22}. Thus, a galaxy catalog with several derived properties related to various population tracers would provide a versatile tool in the search for GW counterparts. 


An additional complicating factor is that the LIGO \citep[][]{aligo}, Virgo \citep[][]{virgo}, and KAGRA \citep{kagra} detectors for O4 will be sensitive to even larger distances than in previous runs, where the BNS and NSBH sensitivity distances will extend to 200 and 350~Mpc, respectively. However, the majority of BNS and NSBH events during the O3 run had distances greater than their predicted sensitivity limits \citep{Petrov21}, which suggests that some fraction of the upcoming O4 events may also have greater distances than expected. As a result, the searches for EM counterparts to GW events will be required to rely on galaxy catalogs that are less complete at these greater distances which may hinder these searches. Constructing a galaxy catalog that is highly complete out to the sensitivity distances of \emph{both} BNS and NSBH events would provide a higher probability of finding more EM signatures to GW events.

\textit{How well do we know the local Universe?} Considerable effort has been made to construct homogenized galaxy catalogs for the purposes of aiding the searches for kilonova signatures from GW events: Compact Binary Coalescence Galaxy catalog \citep[CBCG;][]{kopparapu08}, Gravitational Wave Galaxy Catalog \citep[GWGC;][]{white11}, Galaxy List for the Advanced Detector Era \citep[GLADE;][]{glade}, Census of the Local Universe \citep[CLU;][]{cook19b}, and the Heraklion Extragalactic Catalogue \citep[HECATE;][]{kovlakas21}. While each of these catalogs are well constructed and provide unique features, the distances for CBCG and GWGC only extend to 100~Mpc while CLU and HECATE are limited to 200~Mpc and GLADE has no formal distance limitation. Thus, most of these catalogs will not be useful for many NSBH and even some BNS events if the majority of the event distances remain as high as observed in O3.

In addition to the distance limits of these catalogs, their estimated completeness can have a significant impact finding kilonova counterparts. Only two previous catalogs estimate completeness out to 200~Mpc. The HECATE catalog reports a completeness (by $B-$ and $K-$band) of less than 30-40\% at 200~Mpc, and GLADE reports 40\% (by $B-$band) out to 200~Mpc. However, more recent studies \citep{Kulkarni18,fremling20} using an independent method to estimate galaxy completeness by counting the number of volume-limited SN hosts with a cataloged distance suggest that completeness could be as high as 60-80\% out to 200~Mpc when using distance information from multiple sources including NED. The lower completeness estimates from current galaxy catalogs relative to SN hosts with available redshifts suggests that these compilations may be missing objects with known redshifts and will suffer from even lower completeness past 200~Mpc where future NSBH (and possibly BNS) events are more likely to occur. 


In this study, we construct a sample of galaxies in the local Universe out to 1000~Mpc from NED, which will benefit from the increasing completeness of extragalactic objects with redshift or distance measurements as NED is regularly updated. In addition, we provide panchromatic fluxes and derive physical galaxy properties (e.g., SFRs, stellar masses, etc.) to enable quick prioritization during follow-up observational campaigns. The sample construction and completeness results are detailed in Sections 2 and 3, respectively. In section 4, we provide a comparison to other local Universe samples, detail some science use cases of NED-LVS, and discuss limitations to the sample. 



\section{Sample Construction} \label{sec:catalog}

The galaxy sample utilized in this study is constructed from the NED \citep[][]{ned_doi} and is limited to a distance of 1000~Mpc (z$\sim$0.2). This distance was chosen to fully encompass the expected distance limits for detecting BNS and NSBH mergers in the LVK detectors during the observing runs out to O5: D$\sim$330 and D$\sim$590~Mpc for BNS and NSBH, respectively \citep{ligoo4}. Since the objects studied here are a sub-sample of NED limited to the local Universe, we refer to this catalog as the NED Local Volume Sample (hereafter NED-LVS). In total, the final list contains \galnum~extragalactic objects that were extracted in September of 2021. 

The observable quantities collected for NED-LVS are taken directly from the NED database. In this section, we describe the object selection, how the final distances are chosen when multiple measurements exist, the fluxes that are extracted, and the methods used to derive physical properties. We note that future updates to this catalog will include photometry from additional large-area surveys and additional objects with newly measured redshifts that are routinely ingested by NED (see \S\ref{sec:summary}).


\subsection{Sample Definition \& Object Selection}

As of September 2021, NED contains over 1.1 billion unique objects\footnote{https://ned.ipac.caltech.edu/CurrentHoldings} that are integrated from both the astrophysical literature and from large-area surveys where names, positions, redshifts, redshift independent distances, and photometry can be extracted. However, not all of these objects are extragalactic. In fact, the unique objects in NED are dominated by unclassified sources with no distance information (81\%) due to the ingestion of several all-sky surveys: 2MASS \citep{2mass}, AllWISE \citep{wise}, and GALEX \citep{galex}.\footnote{https://ned.ipac.caltech.edu/Documents/Holdings/Sets} In this section, we describe how extragalactic sources were selected from the database. 


We start the selection process by extracting all unique objects that have a redshift-based or a redshift-independent distance \cite[e.g., from Cepheid pulsation timing, SNIa light curve peak luminosities, etc.;][]{NEDd} out to 1000~Mpc. We also require that the objects are a single, localized system (e.g., galaxy or QSO) and not the center of a large collection of localized systems (i.e., galaxy clusters); though the individual components are retained. However, we do allow galaxy pairs and triples as the individual centers of these systems cannot always be cleanly disentangled and the distances to the individual components may not be measured. We have also excluded objects that are classified as parts of galaxies (e.g., \hii~regions, star clusters, etc.). However, we do not make any exclusions on galaxy or activity type. Thus, NED-LVS will contain the full range of galaxy-scale objects found in the local Universe.

Table~\ref{tab:objtype} presents the various object types in the final NED-LVS list, and the definitions of the object type codes can be found on the NED website.\footnote{https://ned.ipac.caltech.edu/Documents/Guides/Database} The makeup of the final list is heavily dominated by individual galaxies ($\sim$97\%) followed by unclassified visual and infrared sources (each with $\sim$1\%). These objects represent the final list after visual inspection was performed to identify and remove those with types that are not consistent with a galaxy-scale object or have unreliable redshifts.

\begin{table}

\centering
\begin{tabular}{llrr}
\hline
\hline
Object  & Type	        & N  & \% of \\ 
Type    & Description   &    & Total \\ 
\hline

G & Galaxy & 1,811,810 & 96.757 \\
VisS & Visual Src & 16,096 & 0.860 \\
IrS & Infrared Src & 15,718 & 0.839 \\
UvS & Ultraviolet Src & 11,061 & 0.591 \\
GTrpl & Galaxy Triple & 5,380 & 0.287 \\
QSO & Quasar & 4,489 & 0.240 \\
RadioS & Radio Src & 3,796 & 0.203 \\
GPair & Galaxy Pair & 2,782 & 0.149 \\
UvES & Ultraviolet Emission Src & 604 & 0.032 \\
XrayS & X-ray Src & 334 & 0.018 \\
GammaS & Gamma Ray Src & 164 & 0.009 \\
AbLS & Absorption-Line System & 153 & 0.008 \\
G\_Lens & Lensed Image of a G & 107 & 0.006 \\
EmLS & Emission-Line Src & 45 & 0.002 \\
Q\_Lens & Lensed Image of a QSO & 3 & $<$0.001 \\
EmObj & Emission Object & 1 & $<$0.001 \\
Other & Other Entities & 1 & $<$0.001 \\
\hline
\end{tabular}

\caption{A list of the object types in NED-LVS extracted from the NED database. The primary objects are individual galaxies. The secondary objects are composed of unclassified sources (visual, ultra-violet, and infrared), QSOs, and galaxy pairs and triples.}
\label{tab:objtype}

\end{table}

\begin{figure}
  \begin{center}
  \includegraphics[scale=0.51]{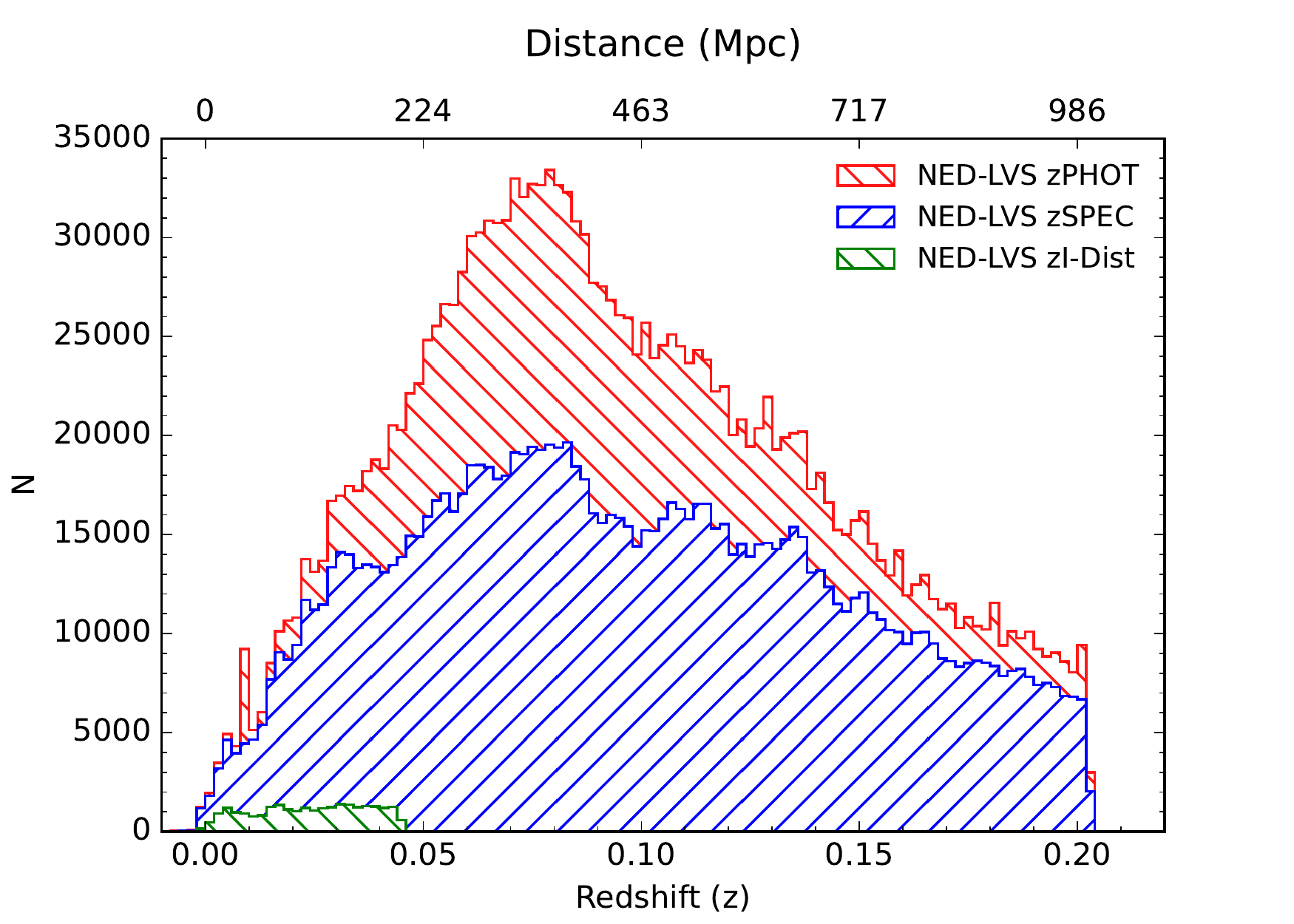}
  \caption{The distance distribution of each object in NED-LVS out to z$\sim$0.2, where the top axis shows the corresponding distance out to $\sim$1000~Mpc. The red, blue, and green-hashed histograms represent sources with final distances with photometric redshifts (zPHOT), spectroscopic redshifts (zSPEC), and redshift-independent distances (zI-Dist). These data are presented as a stacked histogram chart with zSPEC on top of zI-Dist and zPHOT on top of zSPEC in each bin. }
   \label{fig:zhist}
   \end{center}
\end{figure}  

\begin{figure*}
  \begin{center}
  \includegraphics[scale=0.82]{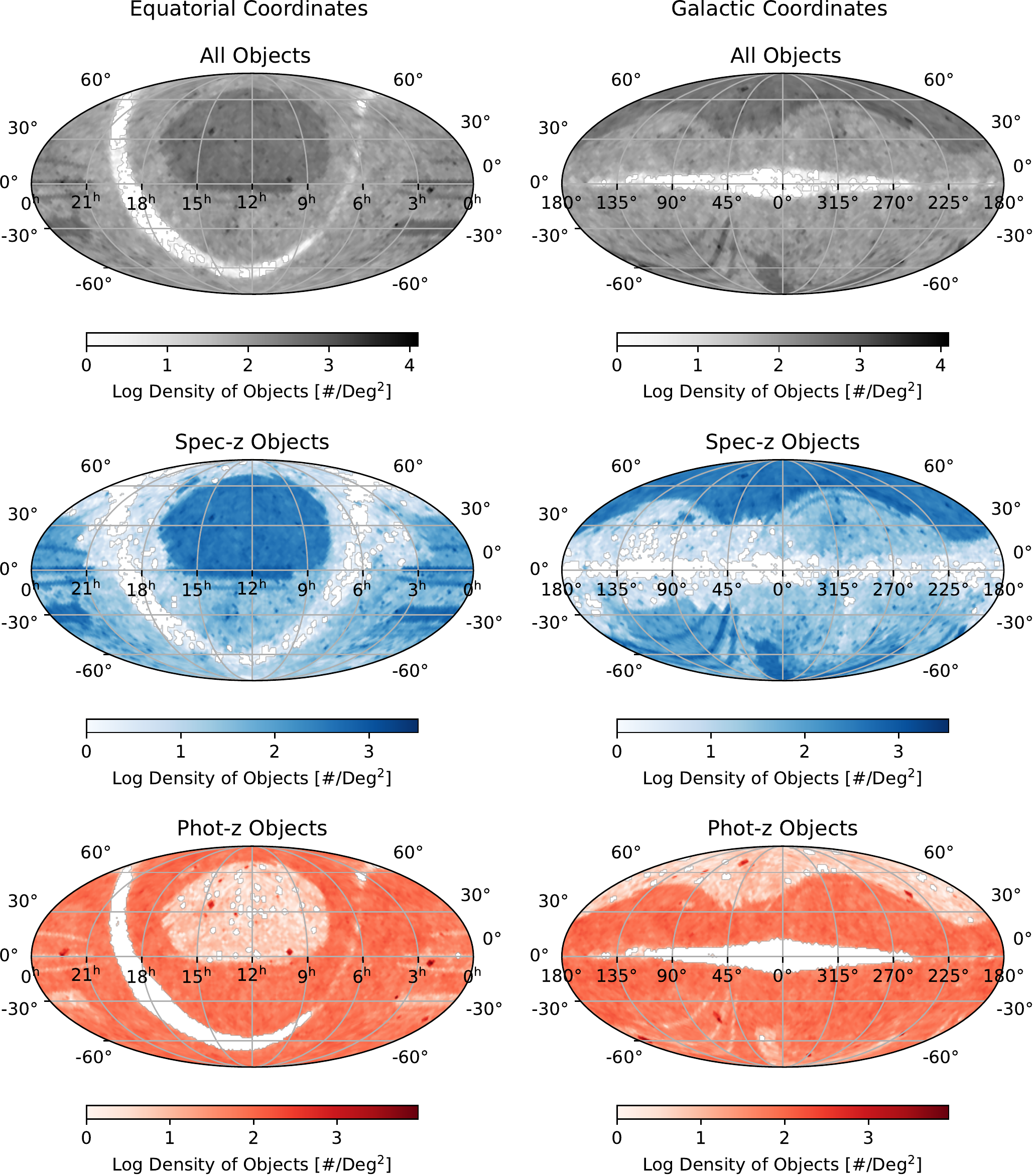}
  \caption{All-sky density plots of NED objects with distance measurements in equatorial (J2000) coordinates \textbf{(left column)} and Galactic coordinates \textbf{(right column)}. The top, middle, and bottom rows show all objects, those with spectroscopic redshifts, and photometric redshifts, respectively. Each pixel in the HEALPix maps is $\rm 2~deg^2$. }
   \label{fig:skyplot}
   \end{center}
\end{figure*}  

\begin{figure}
  \begin{center}
  \includegraphics[scale=0.5]{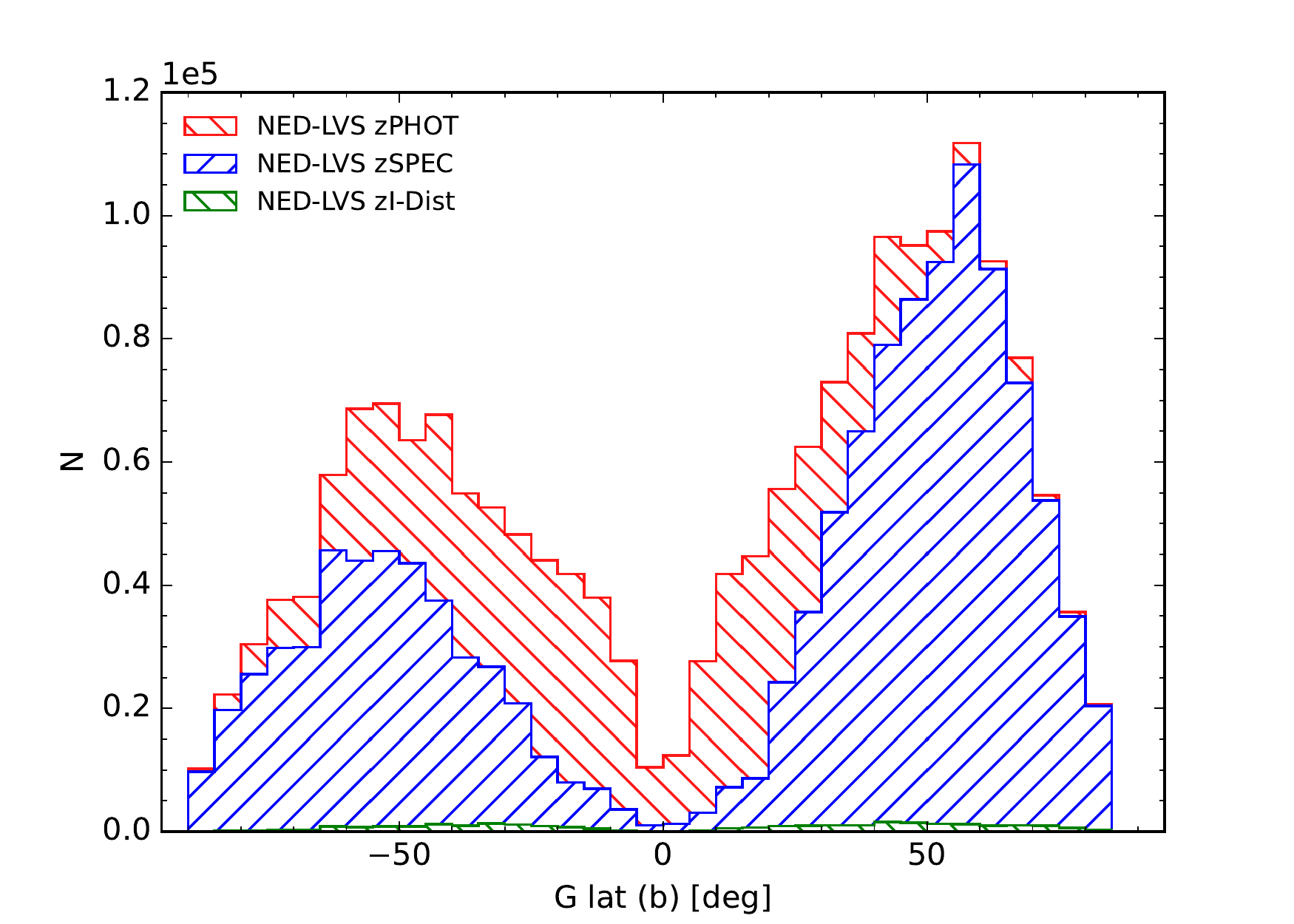}
  \caption{The distribution of NED-LVS across the Galactic Plane, where the color-coded histograms are the same as those in Figure~\ref{fig:zhist}.}
   \label{fig:galbhist}
   \end{center}
\end{figure}


We made significant efforts to clean NED-LVS of contamination (see Appendix \ref{sec:app_visclass} for the full details). Using a combination of visual inspection and automated assessment, we identified and cleaned over 200,000 objects, where we updated object types and/or set redshift quality flags. While a significant fraction of this cleaning effort came from automated assessments and updates, we did visually classify tens of thousands of galaxies primarily at $z<0.002$ or from individual publications whose data warranted additional investigation.

NED-LVS entries whose object types are not consistent with a galaxy-scale object were removed from the sample, and include: stars, nebulae in the Milky Way, parts of nearby galaxies (individual stars, star clusters, planetary nebulae, \hii\ regions, etc.), and image artifacts where spectra were taken of blank sky. The full list of objects with updated types is shown in Table~\ref{tab:objtypeup}. We also flagged individual redshift measurements with a redshift quality flag (`z\_qual') equal to `True' that were deemed unreliable for the following reasons: noisy spectra, unphysical recessional velocities, or unphysical patterns in the redshift distributions of objects from a given publication. The objects whose current fiducial redshifts that are flagged were not removed from the sample since more reliable measurements may be in NED or added in the future. These quality flags are currently being incorporated back into NED which will inform future versions of NED's fiducial redshift algorithm (see \S\ref{sec:fidz}) and the updated redshifts will be available in a future version of NED-LVS. The full list of objects whose redshift quality flags were set is available in Table~\ref{tab:zqualup}.

\subsection{Redshifts \& Distances}

We first obtain all of the redshifts and redshift-independent distances available for each object in NED-LVS, and determine a fiducial value for both redshifts and redshift-independent distances. Then, we determine a final distance for objects with both redshifts and redshift-independent distances. Figure~\ref{fig:zhist} shows the final redshift/distance histogram of NED-LVS where the values are color-coded based on the method used to derive the final distance: redshift-independent distances (1.4\%), spectroscopic redshifts (62.0\%), and photometric redshifts (31.5\%). We also provide visualizations of NED-LVS across the sky via Figures~\ref{fig:skyplot} and \ref{fig:galbhist} which show the distribution of NED-LVS objects across the sky and across Galactic latitude, respectively. The next three subsections describe how fiducial and final redshifts/distances are determined.

\subsubsection{Fiducial Redshifts} \label{sec:fidz}
There are $\sim$1.9 million objects with at least one redshift smaller than 0.2 in NED after removing contaminants (see Appendix~\ref{sec:app_visclass}). The measurements presented in this section and those discussed throughout the paper are heliocentric redshifts. To determine a fiducial redshift for each object, we utilize a new decision tree developed by the NED team specifically for the NED database that is implemented in an automated process. The details of this decision tree can be found in Cook et al. (in prep) and on the NED website\footnote{https://ned.ipac.caltech.edu/Documents/Guides/Database}, but we provide a brief overview below.

Of the total number of objects with redshifts in the local Universe, 1.06 million (52\%) have a single redshift measurement which is set as the fiducial value by default. Next, the fiducial values for the remaining multi-measurement objects are selected based on both the published uncertainties and the technique used to measure the redshift. The techniques (`\textit{z\_tech}' column) captured by NED are: spectroscopic (SPEC), photometric (PHOT), modeled (MOD), multiple methods (MULT), inferred (INFD), and unknown (UNKN); listed in order of highest to lowest priority. The majority of the fiducial redshifts in NED-LVS have \textit{z\_tech} flags of SPEC ($\sim$65\%) or PHOT ($\sim$32\%) while the other flags make up less than a few percent. 

The redshifts of the multi-measurement objects fall into two categories: 1) At least one redshift uncertainty exists and the measurement with the lowest uncertainty is chosen as the fiducial; 2) No redshift uncertainties exist and the fiducial value is chosen from the priority-ordered technique list. In the cases where an object has more than one lowest uncertainty value, the fiducial value is chosen from the priority-ordered technique listed. In the cases where there exists a tie between two or more priority-ordered technique values, the most recent value is chosen as the fiducial. The most recent value was chosen based on the assumption that a new measurement is more likely to be made with more sophisticated instruments and techniques resulting in a more accurate measurement. 

To assess the reliability of these redshift values, we first compare the new, algorithm-based ``fiducial'' redshifts to the previous ``preferred'' redshifts using the objects in the NED-LVS sample. The NED preferred redshifts were chosen with human intervention, which has become unsustainable in recent years thus motivating the new algorithm. For objects with N$>$1 redshifts, we find that $\sim$58\% of the objects show no change in redshift and that most (80\%) show less than a 10\% change. Furthermore, the median offset and median absolute deviation of the redshift difference is $1.4\times 10^{-9}$ and $2.0\times 10^{-3}$, respectively. We note here that the absolute median deviation was used due to the highly peaked $\Delta$z distribution with small tails of large differences (resembling a double exponential or Cauchy distribution) that skewed the standard deviation metric to unreasonably high values. In the rare cases where a photometric redshift is chosen over a spectroscopic value (0.6\%), the median redshift offset and scatter to the preferred values is small (an offset of 0.002 with a median absolute deviation of 0.2, respectively). 

We have also compared the fraction of objects with spectroscopic redshifts between the ``preferred'' and ``fiducial'' values. Using only those objects that have the older ``preferred'' values, we find that both have high fractions of spectroscopic measurements: 94\% and 98\% for the ``preferred'' and ``fiducial'', respectively. The selection of spectroscopic measurements by the ``fiducial'' algorithm is likely a consequence of the relative errors between spectroscopic and photometric redshifts, where we find that the median errors in the NED database are 0.015 and 0.00001 for photometric and spectroscopic entries, respectively. As a consequence, our selection method of choosing the lowest error measurement typically selects a spectroscopic value, and provides confidence that more secure measurements are selected preferentially.


Another means to test the reliability of the new fiducial redshifts is to compare these values to the median of the redshifts for individual objects, where it is reasonable to assume that the median is a good proxy for a statistically accurate redshift. The median offset and median absolute deviation of the redshift difference (fiducial minus median for each object) is $1.6\times 10^{-8}$ and $9.3\times 10^{-5}$, respectively. This is an order of magnitude lower scatter than that found for old, preferred redshifts. In addition, we quantify potential discrepant redshifts via the standard deviation of redshifts for each object. We find that only 1266 (0.1\%) objects show a $>$3$\sigma$ deviation from the median. The significantly reduced scatter and the small fraction of objects with significant deviations between the fiducial and median redshifts suggest that the new values are more reliable. 

To summarize, the new fiducial redshifts are chosen based on a decision tree which takes into account measurement technique and uncertainties. The new fiducial values agree with the old preferred ones as evidenced by the low scatter in the differences of the redshifts and the large fraction of objects with little-to-no change in their redshifts (58\% show no change and 80\% show $<$10\% change). Furthermore, the increased fraction of spectroscopic redshifts chosen by the fiducial algorithm (66\%) compared to the fraction in the preferred values (40\%) provides confidence that more reliable redshifts are being chosen. Finally, the reduced scatter found in the difference between fiducial and median redshifts (relative to the old preferred redshifts) suggests that the fiducial redshifts reflect a statistically more accurate selection.


\subsubsection{Fiducial Redshift-Independent Distances} \label{sec:zidist}

There are approximately 145,000 objects in NED with a redshift-independent distance at $D<1000$~Mpc. To determine a fiducial distance, we employ a simple decision tree that relies on the measurement method, but takes into account the uncertainties. First, the majority of NED-LVS objects with  distances have only a single measurement (90.2\%) which is set as the fiducial value. Next, the distances for multi-measurement objects are prioritized by the method. 

Our choice to select a fiducial redshift-independent distance based on method over uncertainty is driven by the large diversity of methods and large relative uncertainty between these methods. Most methods are grouped into two categories: primary and secondary indicators. Primary indicators are derived from standard candles or rulers which provide distance scales with an absolute calibration, and have a high internal accuracy with typical uncertainties of 10\% or less \citep[e.g., Cepheids and masers;][]{ferrarese00a,freedman10}. Secondary indicators in general provide more indirect measurements requiring a zero point tied to a primary indicator (e.g., fundamental plane and Tully-Fisher relationship) and typically have larger errors of $\sim$25\% \citep{cosmicflows3}. In addition, for a given object there can be multiple measurements based on methods within an indicator group that probe distances using different physics (Cepheid pulsations, helium flash, etc.) making the selection of one measurement difficult to justify. As a result, we choose to take the weighted average of distances (using the square inverse uncertainties) for an object within an indicator group, and favor weighted primary over secondary values. We note here that the vast majority (95\%) of redshift-independent measurements in NED have a published uncertainty. However, in the cases where all measurements within the indicator group have no uncertainties, we take a simple median as the distance considered for that group. 

For this work, we assign the following redshift-independent methods as primary: carbon stars, Cepheids, eclipsing binaries, gravitational waves, horizontal branch, masers, Mira variables, red clump, RR Lyrae, and tip of the red giant branch (TRGB). All other methods are considered secondary here \citep[for a full list of methods see][]{NEDd}. The most common secondary method ($\sim$89\% of all objects with a distance) is the fundamental plane (FP) technique for elliptical galaxies.


To assess the accuracy of the fiducial redshift-independent values derived here, we compare our distances to those from another large compilation of redshift-independent distances: COSMICFLOWS-3 \citep[hereafter CS3;][]{cosmicflows3} as tabulated by The Extragalactic Distance Database \citep[EDD;][]{EDD}.\footnote{https://edd.ifa.hawaii.edu/} The CS3 distances are a weighted average of multiple methods: Cepheids, TRGB, surface brightness fluctuation, SNIa, Tully-Fisher relation, fundamental plane, and miscellaneous high-quality RR Lyrae, horizontal branch, eclipsing binary, or maser measurements. This catalog contains distances for $\sim$18,000 nearby galaxies where the measurements from the different methods are selected from various literature sources and have been corrected to a common distance scale or zero point. Thus, the CS3 catalog provides an independent data set check NED-LVS for any biases inherent to compiling distances derived from different sources and methods.

\begin{figure}
  \includegraphics[scale=0.52]{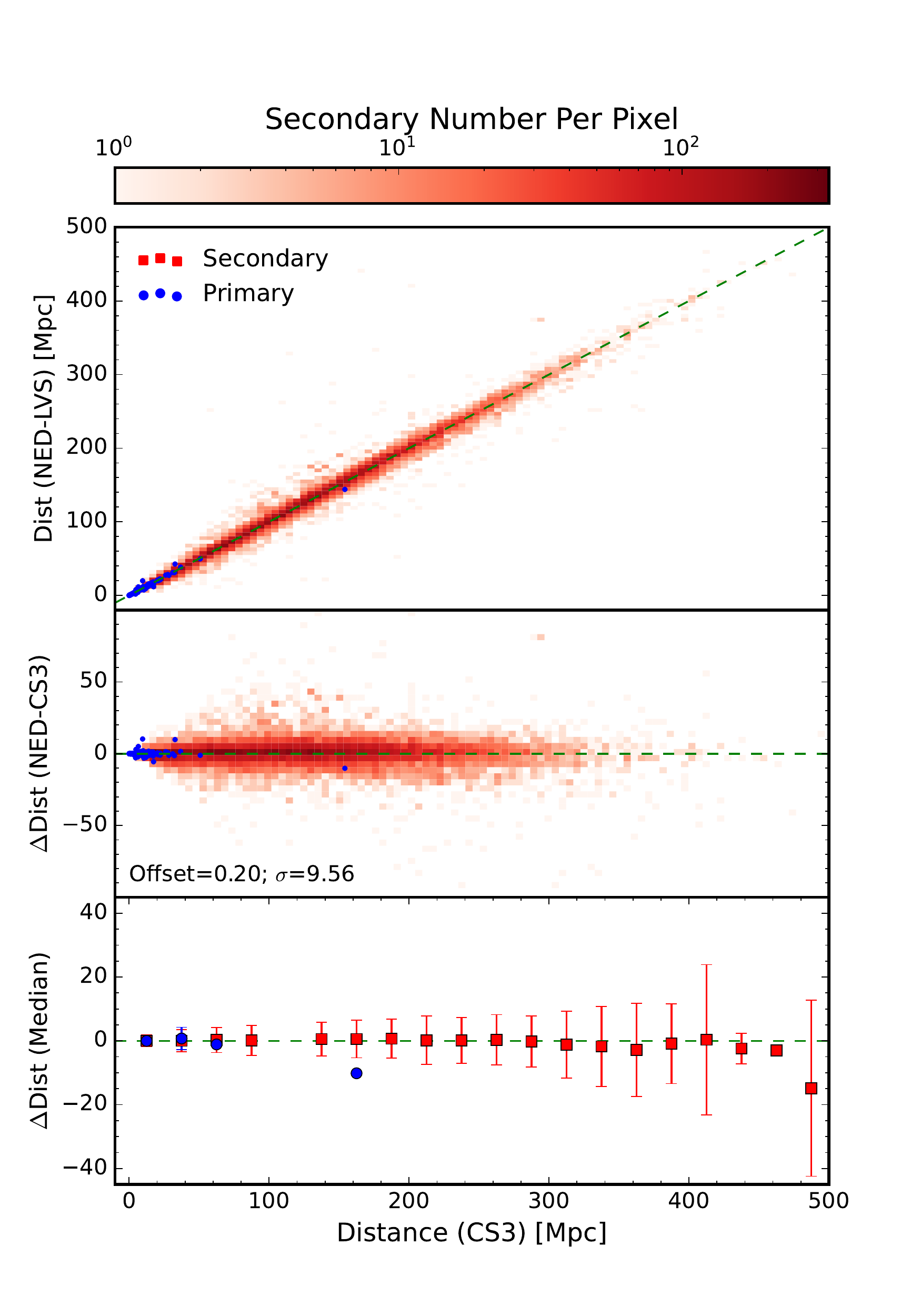}
  \caption{The comparison of redshift-independent distances between COSMICFLOWS-3 and NED-LVS, where the objects with secondary distances are shown as red, 2D histograms and those with primary distances are shown as blue circles. The \textbf{top and middle panels} show the 1-to-1 comparison and residual plots which show good agreement. The \textbf{bottom panel} shows the median distance offset between CS3 and NED-LVS, where the error bars represent the standard deviation of distance offsets.  We find good agreement with CS3 and no trends with distance to indicate a bias in the NED-LVS values. }
   \label{fig:eddcomp}
\end{figure}  


Figure~\ref{fig:eddcomp} shows the one-to-one comparison and the residual scatter plot for the NED-LVS and CS3 distances, where we have color-coded the method indicators. We find good agreement with a median offset of 0.2~Mpc and a scatter of $\sim10\%$. In addition, we find no obvious biases or trends with distance. We also find no major differences when breaking the sample into primary and secondary indicators as defined in NED-LVS; other than the primary indicator measurements tending to be at lower distances as one would expect. 

We do find roughly a dozen objects whose final redshift-independent distances in NED-LVS show large differences relative to CS3 ($>$100~Mpc). The majority of these have only a few measurements from different studies showing significant discrepancies; even within the same distance method (i.e., two fundamental plane or SNIa distances that disagree by more than 30\%). However, these small numbers of large outliers do not significantly affect the results of this study.

In summary, we split the NED redshift-independent distances into those derived from primary and secondary indicators, where we select primary distances when available and secondary distances otherwise. A comparison with another high-quality compilation of redshift-independent distances shows good agreement.



\subsubsection{Final Distances} \label{sec:finaldist} 

To provide a final distance for NED-LVS we next make a decision between redshifts and redshift-independent distances for objects with both types of measurements ($N=$142,424). While the vast majority of galaxies in NED-LVS have redshift measurements, a luminosity distance computed from the Hubble law may not be accurate for the closest galaxies. This is especially true for galaxies at distances less than 40~Mpc ($\sim$3000~km~s$^{-1}$) where peculiar velocities are typically a few 100s~km~s$^{-1}$  \citep{Watkins15,Anand19,Lilow21,cosmicflows4} but can be as high as 1000~km~s$^{-1}$ and a significant fraction of the observed velocity \citep[e.g.,][]{Karachentsev03,Anand18}. For this reason, we choose a final distance based on redshift-independent measurements for the nearest galaxies when available. 

\begin{figure}
  \includegraphics[scale=0.51]{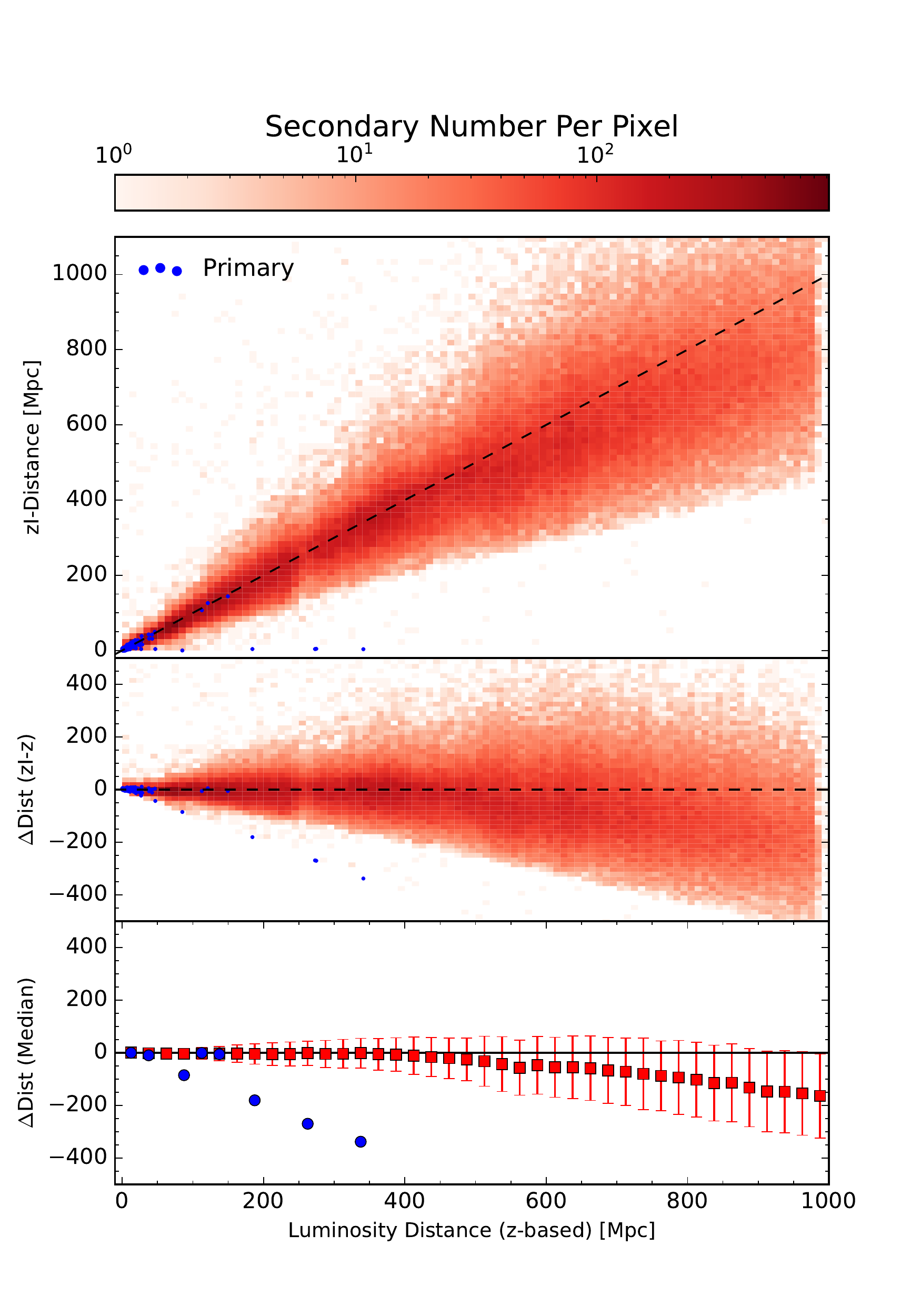}
  \caption{The comparison of redshift-independent distances and those derived from redshifts in NED-LVS. The \textbf{top and middle panels} show the 1-to-1 comparison and residual plots, where the objects with secondary distances are shown as red, 2D histograms and those with primary distances are shown as blue circles. The vertical structure at $\sim$250~Mpc in the top two panels is the result of an upper velocity limit in the fundamental plane distances of \cite{Springob14}. The \textbf{bottom panel} shows the median distance offset, where the error bars represent the standard deviation of distance offsets. Both primary and secondary redshift-independent distances show low scatter at small distances, but secondary measurements show an increase in scatter past 200~Mpc and a systematic offset from redshift-based distances past 400~Mpc. Due to these issues, we use redshift-based distances past 200~Mpc. }
   \label{fig:zzI}
\end{figure}

However, it is not clear if redshift-independent measurements will provide improved accuracy compared to redshifts at greater distances where the Hubble flow will dominate the observed velocity. It is also possible that the numerous different methods used to derive redshift-independent distances may result in increased scatter relative to redshift measurements at these distances. Figure~\ref{fig:zzI} shows a comparison of distances derived from redshifts and redshift-independent measurements, where the bottom panel shows the median offset and scatter. Below 200~Mpc we find that the primary and secondary redshift-independent measurements show small offsets and scatter compared to those at greater distances. We also find that the secondary redshift-independent distances show increasing absolute scatter past 200~Mpc (though the scatter relative to the distance is roughly 20\% at most distances), and show a systematic offset from redshift-based distances past 400~Mpc.

The objects with secondary distances past 400~Mpc in Figure~\ref{fig:zzI} are dominated by FP measurements from a single study: \cite{Saulder16}. Their Figure~19, using only data out to 400~Mpc, shows that this offset actually starts at $\sim$200~Mpc, and they suggest that the systematic offset between fundamental plane and redshifts-based distances is likely due to environmental effects on galaxy properties \citep[e.g., mass-to-light ratios are differnt for central and satellite galaxies;][]{Joachimi15}. Our analysis (Figure~\ref{fig:zzI}) demonstrates that this trend continues to greater distances where the discrepancy increases to a maximum of $\sim$15\% of the distance at 1000~Mpc. While environmental factors may have an effect, several studies on the deviations of fundamental plane distances have found that more than one effect can be present and that these effects can be difficult to disentangle \citep[selection effects, Malmquist bias corrections, cluster/field environments, passive evolution, etc.][]{Bernardi03,Fernandez11,Saulder13,vandeSande14,Joachimi15,Singh21}.

In light of the systematic offset found in fundamental plane distances that make up the majority of redshift-independent distances past 200~Mpc (98.3\%), we select redshift-based distances past 200~Mpc. We note that there are only $\sim$500 objects (0.03\%) with no redshifts available past 200 in which case a secondary distance indicator must be used to keep them in the sample (i.e., it is the only distance available).



In addition, there are five objects with primary indicator distances that are significantly discrepant from the redshift-based distance. Upon further inspection, the redshifts of these objects may be suspect, where one is a photometric redshift, three are older measurements of objects located behind the plane of the MW, and one comes from an HI measurement that was given a `fair' quality flag in the source study. We note that for these five objects our selection methods correctly identify the more accurate primary redshift-independent indicators.



For objects whose final distance estimate comes from a redshift, we calculate luminosity distances assuming a Hubble constant (\hub) value of 69.6 $\rm{km}~\rm{s}^{-1}~\rm{Mpc}^{-1}$ based on local estimates calibrated to TRGB measurements \citep{freedman20,freedman21} and a flat cosmology with $\Omega_m=0.3$. We also note here that we do not correct our redshifts for local flow deviations for the members of galaxy clusters \citep[e.g., Virgo Cluster;][]{kovlakas21} since this will only affect a small fraction of our sample. However, these corrections are left for future work and updates to the sample.



\subsection{Photometry} \label{sec:phot}

In this section we describe the fluxes extracted from three nearly all-sky surveys currently available in NED: GALEX, 2MASS, and AllWISE. Of the \galnum~objects in NED-LVS, 789,080 (42\%),  1,481,114 (79\%), 1,649,491 (88\%) have at least one flux from the GALEX, 2MASS, and AllWISE catalogs, respectively. We note here that the GALEX and AllWISE fluxes for highly extended galaxies are supplemented with custom, large-aperture fluxes that have been published for galaxies out to 50~Mpc (see \S\ref{sec:z0mgs}). 

\begin{figure*}
  \begin{center}
  \includegraphics[scale=0.47]{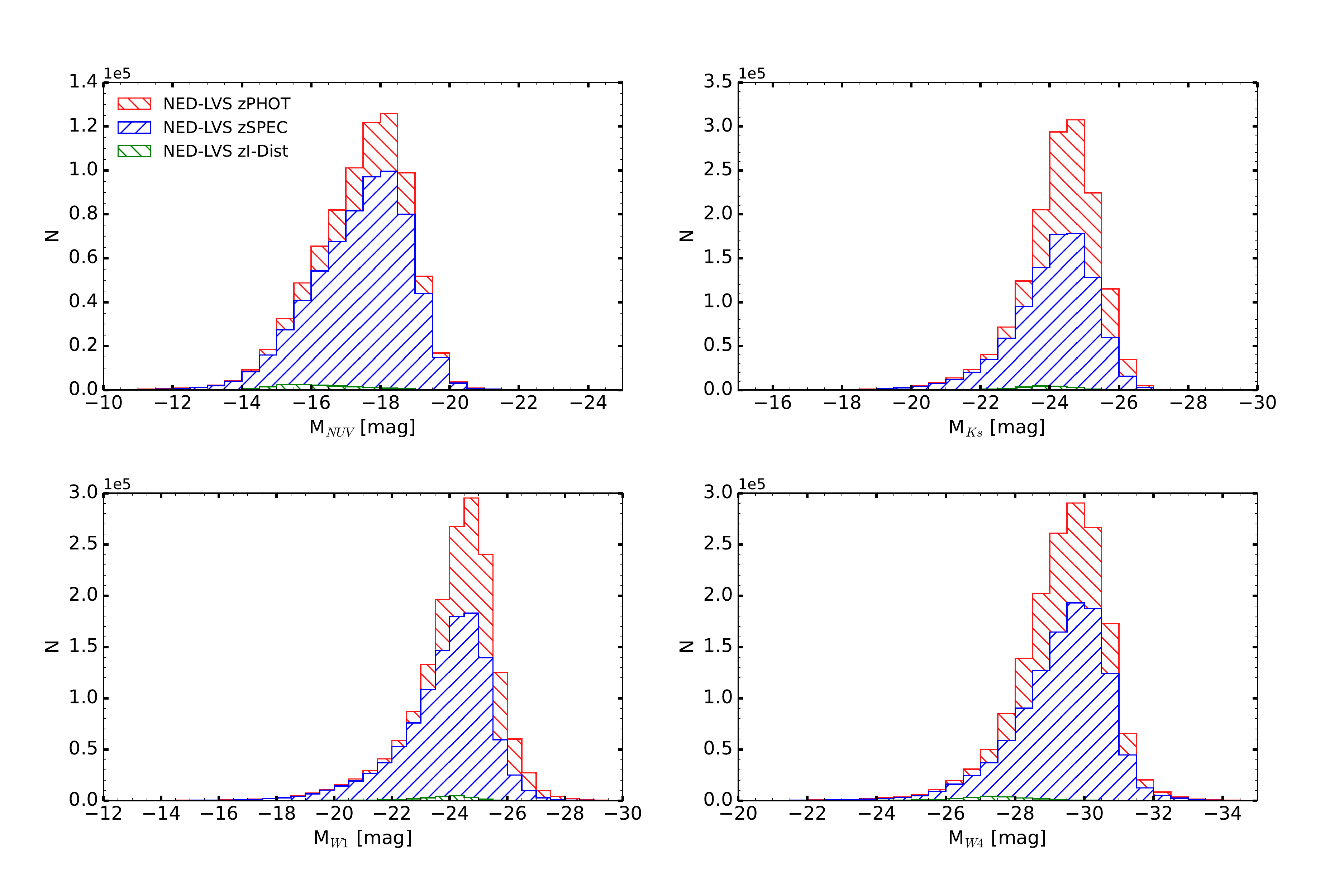}
  \caption{The absolute magnitude histograms of the NED-LVS sample. \textbf{Upper-Left:} GALEX $NUV$ histogram. \textbf{Upper-Right:} 2MASS-$Ks$ histogram. \textbf{Lower-Left:} WISE $W1$ histogram. \textbf{Lower-Right:} WISE $W4$ histogram.}
   \label{fig:absmag}
   \end{center}
\end{figure*}

The fluxes in each band have been corrected for MW extinction using the \cite{schlafly11} dust maps tabulated from IRSA\footnote{https://irsa.ipac.caltech.edu/applications/DUST/} and the \cite{fitzpatrick99} reddening law.  The absolute magnitude distributions of the NED-LVS photometry is presented in Figure~\ref{fig:absmag}. We find that the peak and range of these fluxes show broad agreement with other published local Universe samples \citep[e.g.,][]{jones06,wyder07,dale09,lee11,Dale17,jarrett17}.

\subsubsection{GALEX - UV}
There are two GALEX source catalogs that have been integrated into NED: All-Sky Survey Catalog (GASC) and Medium Imaging Survey Catalog (GMSC). The GMSC is deeper ($NUV$=23.5~mag) than GASC ($NUV$=20.5~mag), but covers a smaller area of the sky with areas of 26,300 and 5,000 deg$^2$ for GASC and GMSC, respectively. Consequently, the GMSC catalog will provide more accurate fluxes for fainter objects. Thus, we prioritize GMSC fluxes over those in GASC. The fluxes for both catalogs are those measured inside a KRON aperture. The total number of objects in NED-LVS with GALEX photometry is 789,080, where 298,456 (38\%) and 479,029 (61\%) come from the GMSC and GASC catalogs, respectively.

\subsubsection{2MASS - NIR}

There are three 2MASS catalogs that have been integrated into NED: the Large-Galaxy Atlas \citep[2MASS-LGA;][]{2mass-lga}, Extended Source Catalog \citep[2MASS-XSC;][]{2mass-xsc}, and Point Source Catalog \citep[2MASS-PSC;][]{2mass-psc}. Corresponding DOIs are: \citet{2MASSLGA_doi}, \citet{2MASSXSC_doi}, and \citet{2MASSPSC_doi}. For our purposes, it is preferable to capture as much of an object's flux as possible, thus we prioritize the fluxes for over 600 of the largest galaxies from the 2MASS-LGA catalog measured inside large, custom apertures that span a range of 2--30 arcminutes. We assigned photometry in the following priority order if they exist: 2MASS-LGA total fluxes, then 2MASS-XSC total fluxes, then 2MASS-PSC fluxes inside a 4\arcsec~aperture. The total number of objects in the final sample with 2MASS photometry is 548 (0.04\%), 1,086,368 (73\%), and 394,198 (27\%) from the LGA, XSC, and PSC catalogs, respectively.



\subsubsection{AllWISE - MIR}
There are several flux measurements available for the four WISE bands ($W1$ at 3.4$\mu m$, $W2$ at 4.6$\mu m$, $W3$ at 12$\mu m$, $W4$ at 22$\mu m$) integrated into NED from the AllWISE Source Catalog \citep{allwise_doi}. These were derived from relatively large apertures (8.25, 16.5, and 22\arcsec) due to the larger PSF of WISE imaging (6.1 and 12\arcsec~for $W1$ and $W4$ filters, respectively), as well as profile-fitted measurements. Since it is preferable to capture as much of the object's light as possible, we prioritized the 22\arcsec~aperture fluxes for extended sources while utilizing the profile-fitted fluxes for point sources. The total number of objects in the NED-LVS catalog with AllWISE photometry is 1,649,491 where 640,897 (39\%) and 992,917 (60\%) come from the larger 22\arcsec and profile-fit apertures, respectively.

\subsubsection{Highly Extended Galaxies} \label{sec:z0mgs}

While the published GALEX and AllWISE source catalogs provide fluxes for a great many galaxies in our sample, the photometry for the closest, most extended galaxies ($d\gtrsim$1 arcmin) is not available in these catalogs. However, there has recently been a significant effort by the ``z=0 Multiwavelength Galaxy Synthesis'' \citep[z0MGS;][]{z0mgs} survey to provide uniformly measured GALEX and WISE fluxes for these galaxies in apertures appropriate for their large sizes. We utilize these fluxes for all galaxies with available measurements.

The z0MGS survey convolved archival GALEX and WISE imaging for $\sim$15,000 galaxies within 50~Mpc onto maps with common Gaussian PSFs of 7\farcs5 for GALEX $FUV$/$NUV$ and WISE $W1$/$W2$/$W3$ and 15\arcsec~for WISE $W4$. The photometry procedures used by the z0MGS survey are detailed in \S7 of \cite{z0mgs}, but we provide a brief overview here. The images were first masked of bright contaminating stars and background galaxies out to several times the radius of the fiducial aperture taken from the RC3 D25 ellipses \citep{rc3} as cataloged by LEDA \citep{hyperleda} or 60\arcsec (whichever was larger), then a fitted background noise was subtracted from the images \citep[see also][]{lang14}. Next, the fluxes were summed inside radially binned annuli of half PSF sizes out to 2 times the fiducial aperture radius. The sky subtraction and integration procedure facilitated better correction for the galaxy flux present underneath the areas of bright star contamination. Finally, they applied an extended source correction to the WISE fluxes since the WISE calibration was optimized for point sources (typical corrections are a few percent). In a comparison of z0MGS fluxes to published values from studies using by-hand aperture definition and contamination removal \citep{Dale17,Clark18,Munoz15}, \cite{z0mgs} report good agreement in both GALEX and WISE bands with scatter that ranges between 0.03--0.1~dex in the logarithmic flux densities.

As a result of the uniformly measured photometry in custom large apertures performed by the z0MGS survey, these data are ideal for providing total fluxes for highly extended galaxies to be used in estimating their completeness. To incorporate these fluxes into our catalog, we perform a positional cross-match with a radius of 10\arcsec\ where the closest object results in 15,548 (98.7\%) matches; leaving 200 NED objects unmatched. We then perform a name match using the NED name resolver and match an additional 167 objects resulting a total match success rate of 99.78\%.  

\begin{figure*}
  \begin{center}
  \includegraphics[scale=0.52]{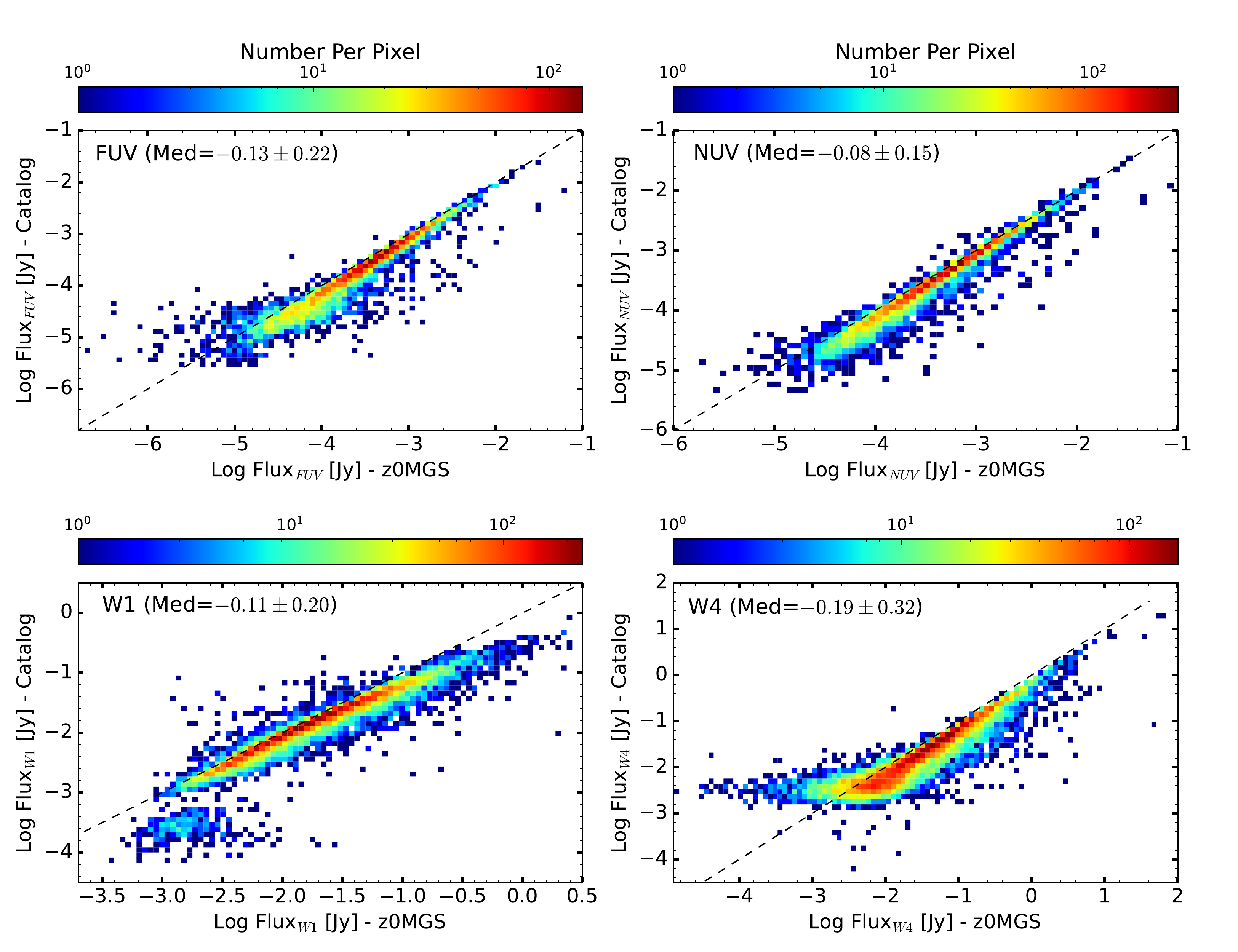}
  \caption{A comparison of GALEX and WISE fluxes for highly extended galaxies between those taken from source catalogs available in NED versus those derived from the z0MGS survey. The panels show 2D histograms of objects, where color coding represents the number of objects per pixel. We find that the z0MGS survey fluxes capture more of a galaxy's light compared to standard catalog fluxes in the GALEX and AllWISE source catalogs. The z0MGS fluxes on average capture $\sim$2 times the flux for GALEX $FUV$ and $NUV$, while capturing 1.2 and 1.6 times more flux for $W1$ and $W4$, respectively. The horizontal envelopes (more prominently seen in the $W4$ panel) are likely the effect of cleaner sky estimates computed in large annuli for z0MGS which provides more reliable fluxes for fainter objects. }
   \label{fig:z0mgscomp}
   \end{center}
\end{figure*}  


We further illustrate the need for these larger apertures in Figure~\ref{fig:z0mgscomp} where we present flux comparisons between the values in z0MGS and those from the source catalogs available in NED. We find that on average z0MGS captures 20-35\% more flux for GALEX, while capturing 25-60\% more flux for WISE. Consequently, we note that the source catalog fluxes for the faintest galaxies which show upper limit measurements (i.e, the horizontal envelopes more prominently seen in the $W4$ panel of Figure~\ref{fig:z0mgscomp}) have detection measurements in z0MGS. This is likely the effect of cleaner sky estimates computed in large annuli for z0MGS which provides more reliable fluxes for fainter objects.



    

\subsection{Derived Physical Properties} \label{sec:physprop}

Here we present the physical properties of galaxies in NED-LVS derived from the MIR fluxes, which will provide stellar mass ($M_{\star}$) and SFR estimates. We note here that we have not applied a $K-$correction to our fluxes when deriving physical properties, but that future versions of NED-LVS will incorporate such a correction. These corrections are on the order of a few tenths of a magnitude in the NIR for the most distance objects in our sample \citep{blanton07}.


The stellar masses are derived from mass-to-light ratios using $W1$ fluxes ($\Upsilon_{\star}^{\rm W1}$). This bandpass provides a robust tracer of a galaxy's stellar mass as this light is dominated by an older stellar population that makes up the majority of a galaxy's stellar mass \citep{Pahre04,meidt12a}, and is less affected by attenuation from dust than shorter wavelengths. Many studies have found that a constant mass-to-light ratio using both the Spitzer 3.6$\mu m$ and WISE $W1$ (3.4$\mu m$) bandpasses provide an accurate estimate of a galaxy's stellar mass with relatively low scatter \citep[$\sigma\sim$0.1~dex;][]{oh08,eskew12,barnes14,mcgaugh14,meidt14,norris14,mcgaugh15,Kettlety18}. 

However, the values of $\Upsilon_{\star}$ found in these studies range from 0.3--0.7, which is likely due to the composition of the galaxy samples used since NIR mass-to-light ratios are expected to increase with the stellar population age \citep{Bell01,bell03,Courteau14}. z0MGS \citep{z0mgs} used 16,000 nearby galaxies with large numbers of both early- and late-type galaxies to study the effects of galaxy properties on mass-to-light ratios by combining uniformly measured W1 luminosities with stellar masses derived from the more sophisticated SED fitting methods of the GALEX-SDSS-WISE Legacy Catalog \citep[GSWLC-2;][]{GSWLC1,GSWLC2}. This mass-to-light comparison showed the expected range of variations in $\Upsilon_{\star}^{\rm W1}$ and that these variations correlated with specific star formation rate (sSFR), where early-type galaxies (older stellar populations) showed higher mass-to-light ratios compared to later-type galaxies \citep[see also][]{meidt14,querejeta15,Hunt19}. The z0MGS study of \cite{z0mgs} also quantified a relationship between $\Upsilon_{\star}^{\rm W1}$ and an observable proxy for sSFR (i.e., $W4$-to-$W1$ luminosity ratio) which can be used to derive a mass-to-light ratio that broadly accounts for the stellar population age:


\begin{equation}
    \Upsilon_{\star}^{\rm W1} = 
    \begin{cases}
      0.5,                  &\text{if}~Q < 0 \\
      0.5 - 0.4 (Q - 0.1), &\text{if}~0.1 < Q < 0.75 \\
      0.2,                  &\text{if}~Q > 0.75\\
    \end{cases}       
\end{equation}

\noindent where $Q$ is the logarithmic $W4$-to-$W1$ luminosity ratio and $\Upsilon_{\star}^{\rm W1}$ is in units of solar masses per the solar luminosity in the $W1$ (3.4$\mu m$) filter bandpass \citep[m$_{\odot,\rm W1}$=3.24~mag;  L$_{\odot,\rm W1}=1.58\times10^{32}$~erg~s$^{-1}$;][]{jarret13}. Since we expect a mix of early- and late-type galaxies in our sample, we utilize the mass-to-light ratio prescription of \cite{z0mgs}.



The recent star formation rates (SFRs) of galaxies can be estimated from many different luminosity tracers (e.g. \ha, UV, IR, etc.), where measured luminosities are transformed into SFRs via scaling prescriptions \citep[e.g., ][]{kennicutt98,kennicutt12}. Hybrid tracers of SFR have been powerful tools which combine the unobscured (UV) and obscured (by dust and reprocessed into the IR) components of young ($<$100~Myr) stars to derive total SFRs \citep{calzetti07,kennicutt07,kennicutt12}. While NED-LVS does contain both UV and IR fluxes, the photometry in the sample is limited by the relatively shallow GALEX survey limits where only $\sim$40\% of our galaxies have UV detections (compared to 85\% with WISE detection). 

Empirical prescriptions for IR-only luminosities (WISE $W3$ or $W4$) have also been derived that can recover the SFRs of nearby galaxies with properties that span 4--5 orders of magnitude and exhibit only moderate scatter (0.2~dex) \citep{calzetti07,jarret13,Cluver17}. However, the z0MGS survey also investigated the dependence of SFR calibration coefficients on galaxy properties and found that a W3-based SFR shows a strong dependence on sSFR and stellar mass due to the presence of strong PAH features suggesting that W3 is not an ideal primary indicator for a sample of galaxies that contain both early- and late-type galaxies \citep[see also][]{cook14c}. The $W4$-based SFRs show some degree of dependence on stellar mass, but this variation is much smaller than that seen in W3 and is more similar to the variation seen in hybrid SFRs based on $FUV$+$W4$. Additionally, we performed self-consistency comparisons between IR-only and hybrid ($FUV$+$W4$) SFRs and found that W3-based SFRs shows a significant offset of 0.35 dex compared to the more reliable hybrid, and that $W4$ shows good agreement with hybrid SFRs (median offest of 0.03 and a scatter of 0.1~dex).

Given these limitations and the need for a consistent estimation of physical properties, we provide the SFRs estimated from both IR-only ($W4$) and the hybrid ($FUV$+$W4$) tracers. For the hybrid tracer, we utilize the prescription of \cite{murphy11} with dust corrections based on \cite{hao11}:

\begin{equation}
  \rm{SFR}_{hybrid} (\rm{M}_{\odot}\rm{yr}^{-1}) = (C \times \nu L_{FUV}) + (3.89 \times \nu L_{W4}),
\end{equation}

\noindent where C=4.42$\times10^{-44}$ and $\nu$L is the observed monochromatic luminosities of the $FUV$ and $W4$ bands in ergs per second. This calibration assumes solar metallicity and a \cite{kroupa03} initial mass function (IMF). The hybrid tracer is only calculated for objects with detections in both $FUV$ and $W4$ ($N=$675,901). For the IR-only SFR, we utilize the $W4$-based SFR prescriptions of \cite{jarret13}:

\begin{equation}
  \rm{SFR}_{W4} (\rm{M}_{\odot}\rm{yr}^{-1}) = C \times \nu L_{W4},
\end{equation}

\noindent where C=1.95$\times10^{-43}$ and $\nu$L is the observed monochromatic luminosity of the $W4$ band in ergs per second. This calibration assumes solar metallicity and a \cite{kroupa03} IMF.




\begin{figure}
  \begin{center}
  \includegraphics[scale=0.52]{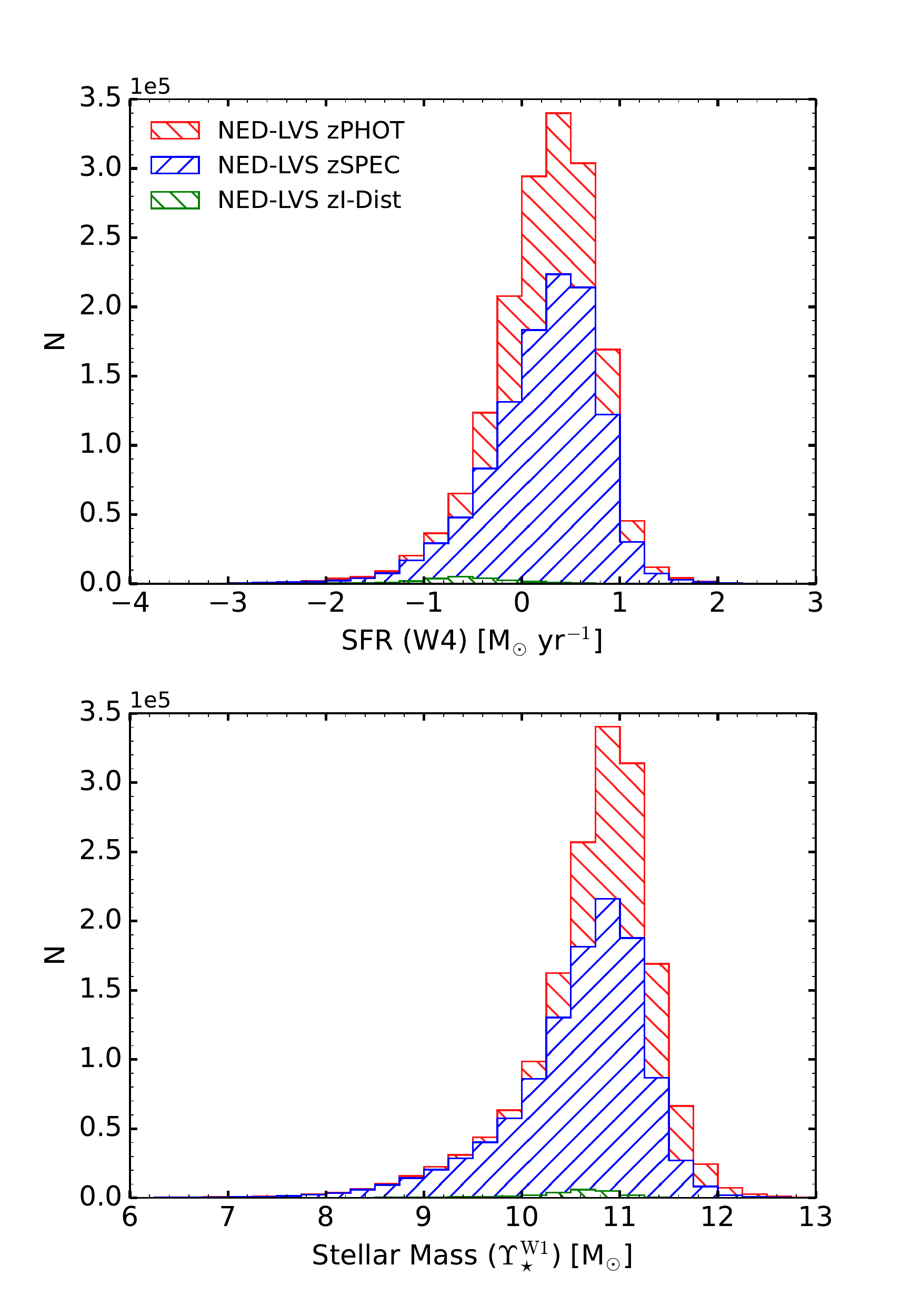}
  \caption{The derived physical properties of the NED-LVS sample. \textbf{Top:} The recent SFRs as computed via the $W4$ band luminosities as per the prescriptions of \protect\cite{jarret13}. \textbf{Bottom:} The stellar masses as computed by mass-to-light ratios using the $W1$ luminosity with a dependence on the $W4$-to-$W1$ luminosity ratio, an observable proxy for sSFR \citep{z0mgs}. We find that the distributions of both mass and SFR agree with other large samples of local Universe studies.}
  \label{fig:physprop}
  \end{center}
\end{figure}  

The SFR and stellar mass distributions of the NED-LVS sample are presented in Figure~\ref{fig:physprop}. We find that the peak and range of the NED-LVS properties show broad agreement with those from other published local Universe samples \citep[e.g.,][]{salim07,lee09b,cook14c,GSWLC1,jarrett17,cook19b}

\begin{figure}
  \begin{center}
  \includegraphics[scale=0.51]{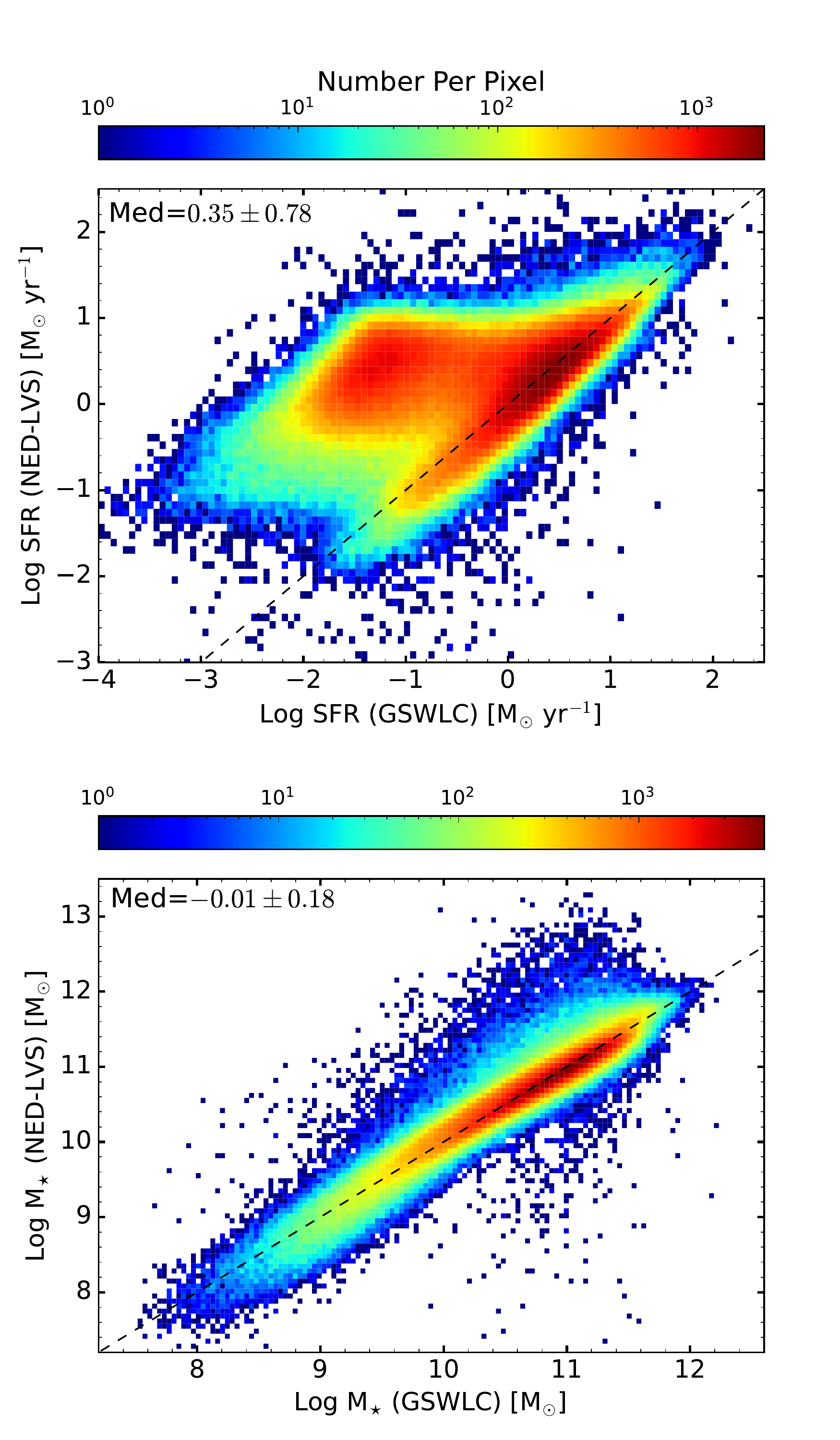}
  \caption{A comparison of galaxy physical properties between NED-LVS and those from GSWLC-2 estimated with more sophisticated SED fitting. The panels show 2D histograms of objects, where color coding represents the number of objects per pixel. \textbf{Top:} SFR comparison which shows large scatter due to the cloud of objects to the upper-left. These objects exhibit colors consistent with early-type galaxies, where the SFRs are overestimated in NED-LVS due to the heating of interstellar grains from older stars (see \S\ref{sec:disclimit}). A SFR flag has been set in the catalog to identify which NED-LVS objects are affected. \textbf{Bottom:} Stellar mass comparisons where we find that simple mass-to-light ratio methods provide good agreement with the masses derived from SED fitting with moderate scatter of 0.2~dex.}
   \label{fig:gswlc2comp}
   \end{center}
\end{figure}

To assess the accuracy of the physical properties derived here we compare our stellar masses and SFRs to those from catalogs using more advanced SED fitting methods: GSWLC-2. This catalog contains $\sim$700,000 galaxies at $z<0.3$ inside the SDSS footprint \citep{york00}, where the physical properties are derived from UV/optical/IR Baysean SED fitting using the Code Investigating GALaxy Emission \citep[CIGALE;][]{noll09,CIGALEpy}. This sample contains a wide range of galaxy types and activity from which to compare with our sample. Figure~\ref{fig:gswlc2comp} shows the comparison of galaxy physical properties between NED-LVS and GSWLC-2. The stellar masses derived in this paper via a simple mass-to-light ratio based on the NIR luminosities show good agreement (median offset of -0.01 dex) with moderate scatter (0.2 dex) to the more sophisticated SED fitting of GSWLC-2. 

The SFR comparison shows overall good agreement, but with larger scatter, much of which is due to the cloud of points located to the upper-left of the plot (i.e., 0.5--2~dex above the 1-to-1 line). The majority of these galaxies have UV-IR or IR colors (i.e., $NUV$-$W1$ $>$ 7, $W1$-$W3$ $<$ 2.5, or $W1$-$W4$ $<$ 5.5; see \S\ref{sec:disclimit}) indicating they are early-type galaxies with low sSFRs. A similar mismatch for low-sSFR galaxies was also reported in \cite{GSWLC1} when comparing their SED fitted values to those derived from scaling relationships using MIR luminosities \citep[see also][]{kovlakas21}, where low-sSFR (SFR/M$_{\star}$ $<-11$) galaxies had higher MIR SFRs with offsets as high as 2~dex. They attribute this discrepancy to the assumption made by scaling relationships that the MIR emission is dominated by dust heated by young stars. In quiescent galaxies, however, a larger fraction of the MIR emission will come from interstellar dust grains heated by older stars and thus overestimate the recent SFR \citep{Davis14,Simonian17}. We note here that the SFR investigations of z0MGS also found that even hybrid ($FUV$+$W4$) SFRs are affected by this issue, where low sSFR galaxies show offsets from SED fitted values up to a factor of five. See \S\ref{sec:disclimit} for a more detailed discussion. 

Since the NED-LVS SFRs are derived from scaling relationships (both hybrid and IR-only), our SFRs will be overestimated for quiescent, early-type galaxies. Using UV-IR and IR colors as a proxy for sSFR activity \citep{z0mgs} we estimate that 47\% of our galaxies will be affected to varying degrees. These objects have been flagged (`ET\_flag' set to True in the catalog), and we advise caution when using the SFRs derived here for quiescent galaxies. Improvements to our SFR estimates are left for future work and updates to NED-LVS \cite[see appendix of][]{z0mgs}.

\subsection{Catalog Availability} \label{sec:catalogavail}

The first version of NED-LVS corresponding to the analysis in this article will be maintained at NED \citep{nedlvs_doi}. In addition, we note that NED regularly updates its holdings with new objects and redshifts from survey catalogs and the literature, and future versions of NED-LVS with updated content will also be available on the NED website\footnote{http://ned.ipac.caltech.edu/NED::LVS/}.

\section{Results}  \label{sec:results}

In this section, we present the completeness estimates of the galaxies in NED-LVS. Completeness is computed here as the ratio of the total luminosities of galaxies in NED-LVS to the total luminosity expected from integrating under published galaxy luminosity functions \citep[LFs;][]{kopparapu08,white11,gehrels16,glade,kovlakas21}. We provide estimates in two wavelength bands (GALEX $NUV$ and 2MASS $Ks$). We note here that there does not currently exist a published LF using WISE fluxes from which to compute completeness, and such an estimate is left for future work when one is published.

\subsection{Methods} \label{sec:LFmethods}

We compute the NED-LVS completeness by luminosity in spherical shells with a thickness of 20~Mpc out to 1~Gpc, where we divide the sum of galaxy luminosities in each shell by the expected luminosity density of the local Universe scaled to the volume in that shell:

\begin{equation}
    Completeness = \frac{\displaystyle \sum_{i=1}^{N} L_{i}} {j_L \times V_{\rm shell}},
\end{equation}

\noindent where $L_i$ is the luminosity of each galaxy in the shell, $V_{s\rm hell}$ is the volume of the spherical shell, and $j_L$ is the luminosity density. The local luminosity density can be derived from published LFs which are commonly fit with a Schechter function \citep{schechter76}, the functional form of which is described by three parameters: \phistar\ is the normalization factor, \Lstar\ is the characteristic luminosity (or ``knee''), and $\alpha$ is the faint-end power-law slope.  Integration of the Schechter function over all luminosities yields the total luminosity density (in solar units of L$_{\odot}$Mpc$^{-3}$) and can be calculated from the Schechter parameters analytically via: 

\begin{equation}
    j_{L} = \int_{0}^{\infty} L \Phi(L) dL = \phi^{*}~L^{*}~\Gamma(\alpha+2),  \label{eqn:jden}
\end{equation}

\noindent where $\Gamma$ is the gamma function. We note here that the characteristic luminosity is often quoted as an absolute magnitude ($M^{*}$) which can be converted to a luminosity via:

\begin{equation}
    L^{*}/L_{\odot} = 10 ^{-0.4 (M_{\lambda}^{*} - M_{\odot},_{\lambda})}, 
\end{equation}

\noindent where $M_{\odot},_{\lambda}$ is the absolute magnitude of the Sun in a given filter. In this study we adopt the following solar absolute magnitudes: $NUV$=10.16 (AB) and K$_{s}$=3.27 (Vega)~mag \citep{Willmer18}. We use the \texttt{PYTHON} package \texttt{COSMOLOPY}\footnote{https://pypi.org/project/cosmolopy/} to calculate the luminosity densities from published Schechter functions in different filters. 

The uncertainties in our completeness estimates are derived by combining in quadrature the uncertainties from both galaxy fluxes and distance, where we also add an extra 10\% uncertainty on each galaxy's luminosity to account for systematic uncertainties \citep[e.g., calibration;][]{GildePaz07,dale09,Jarrett11}. In addition, we also take into account the uncertainties related to the Schechter LF parameters and those associated with the number of galaxies and the distribution of luminosities in each bin. 

To account for the uncertainties associated with the distribution of luminosities and number of objects in each distance bin, we perform a re-sampling of the entire NED-LVS via the `bootstrap' method \citep{Efron86}. We sample the full set of NED-LVS objects in each of the 10,000 iterations allowing for multiple draws of objects, where for each iteration we bin the resulting objects by distance and measure the total luminosity in each bin. The resulting interval holding the middle 68.3\% of total luminosities across the iterations (i.e., the interval between the 16th and 84th percentiles) is taken as the 1$\sigma$ confidence interval. The uncertainties derived from this bootstrap method account for a few percent in the closest bins but decreases for bins with higher counts.

To quantify the total luminosity error from the LF parameters, we randomly sample all three Schechter function parameters in 10,000 iterations assuming the published values and errors are the mean and sigma of a normal distribution, and then calculate the total luminosity density via Equation~\ref{eqn:jden} for each iteration. The resulting middle 68.3\% of total luminosities across the iterations (i.e., the 16th and 84th percentiles) are again taken as the 1$\sigma$ confidence interval. We find that the errors associated with the LF parameters contribute the majority of the error budget (i.e., $>$90-95\%). In other words, the uncertainty in our knowledge of the galaxy luminosity density dominates over the measurement uncertainties of the galaxies in our sample. 

While the general methodology used here is the same as that used by previous studies of galaxy completeness estimates, there is a difference in how the luminosity density is determined. In this study, we integrate the LF over all luminosities to obtain a total luminosity density, while some previous studies \citep{gehrels16,glade} have computed completeness down to $\sim$60\% of the ``knee'' of the LF \citep[i.e., 0.626$\times$\Lstar;][]{gehrels16}. Estimating completeness in this fashion will act to shift completeness to higher values as only the higher luminosity galaxies (which are more likely to be cataloged and to have a redshift measurement) are used in the numerator, and only \textit{half} of the total luminosity is used as the denominator. In addition, \cite{gehrels16} reported that this luminosity cut captures only half of the sGRB host galaxies known at the time. Given these caveats and limitations, we conclude that a completeness calculation relative to the total luminosity density more accurately reflects the true completeness of galaxies in the local Universe and is a more suitable metric to guide follow-up efforts for multi-messenger and transient events. We note here that our completeness level comparisons with previous galaxy catalogs (see \S\ref{sec:catcomp}) have been re-computed from their raw catalogs in a consistent manner using the total luminosity density. Thus, our completeness estimates for other catalogs will have lower values compared to those reported in their respective publications.



\subsection{Comparisons of Published Luminosity Functions} \label{sec:LFcomp}

\begin{table*}

{Comparison of Published Luminosity Functions}\\
\begin{tabular}{lllllrrrr}

\hline
\hline
Publication  & Filter  &  $\alpha$ & $\phi^{\star}$         & $M^{\star} - 5 \rm{log}(h)$ & N        & Area       & Redshift & Abs Mag   \\
             &         &            & ($h^3 \rm{Mpc}^{-3}$) & (mag)                  & Galaxies & (Deg$^2$)  &          & Range   \\ 
\hline

Wyder et al. 2005 & NUV & -1.16 $\pm$ 0.07 & 0.0050 $\pm$ 0.0115\dag & -18.23 $\pm$ 0.11\dag & 1124 & 56.73 & <0.1 & [-12.00,-20.00] \\
Cole et al. 2001 & Ks & -0.96 $\pm$ 0.04 & 0.0108 $\pm$ 0.0016 & -23.44 $\pm$ 0.02 & 5683 & 619.0 & <0.2 & [-18.00,-26.00] \\
Kochanek et al. 2001 & Ks & -1.09 $\pm$ 0.06 & 0.0116 $\pm$ 0.0010 & -23.39 $\pm$ 0.05 & 3878 & 6969.0 & <0.1 & [-20.25,-26.00] \\
Bell et al. 2003 & Ks & -0.77 $\pm$ 0.04 & 0.0143 $\pm$ 0.0007 & -23.29 $\pm$ 0.05 & 6282 & 414.0 & <0.078> & [-20.25,-26.00] \\
Jones et al. 2006 & Ks & -1.16 $\pm$ 0.03 & 0.0074 $\pm$ 0.0010 & -23.83 $\pm$ 0.03 & 60869 & 9075.0 & <0.2 & [-17.60,-27.40] \\
Smith et al. 2009 & Ks & -0.81 $\pm$ 0.04 & 0.0176 $\pm$ 0.0008 & -23.17 $\pm$ 0.04 & 36663 & 619.0 & <0.3 & [-20.00,-25.60] \\
Hill et al. 2010 & Ks & -0.96 $\pm$ 0.06 & 0.0156 $\pm$ 0.0015 & -23.36 $\pm$ 0.09 & 1785 & 27.99 & <0.1 & [-18.25,-25.30] \\

\hline

\end{tabular}
\caption{A list of published galaxy luminosity functions under consideration to calculate the local Universe luminosity density, which is used as the normalization in our completeness estimates. To the best of our knowledge, we list all studies of the local Universe (z$\la$0.3) with large galaxy samples (N$\ga$1000) covering large areas ($\ga$10s square degrees). For each study, this table presents the photometry filter used, the fitted Schechter LF parameters, the number of galaxies in the sample, the survey area, the redshift range probed, and the range in absolute mag of the data points used to fit the LF. There is only one suitable study in the NUV, while there are six studies available in the Ks-band. We select the \protect\cite{jones06} LF study as it contains the largest number of galaxies, covers the largest area, and has the largest range of data points used to fit their data. We note that most studies of LFs scale their parameters with ``$h$'' ($H_0$=100~$h~\rm{km s}^{-1} \rm{Mpc}^{-1}$). The $\dag$ indicates luminosity function parameters that have not been scaled with ``$h$'' and the parameters reflect an assumed $H_0$ value of 70 $\rm{km s}^{-1} \rm{Mpc}^{-1}$ (similar to the values assumed in this study).}
\label{tab:LFcomp}

\end{table*}

In order to estimate completeness for NED-LVS relative to the total luminosity density, we have identified LFs from the literature that are appropriate for the local Universe and selected a fiducial LF for each filter to estimate our final completeness. Identification of LFs is based upon the properties of the galaxy sample used to derive their parameters with the following criteria: have a large enough number of galaxies to provide adequate sampling of properties found in the local Universe and robust constraints on the LF parameters, have secure (preferably spectroscopic) redshifts for the galaxies that probe the local Universe (i.e., out to a few tenths in redshift), and probe a large enough volume to overcome cosmic variance \citep{Driver10}. The past few decades have given rise to several galaxy LF studies \citep[][ and references therein]{Johnston11}, many of which meet our criteria. The Schechter LF parameters and galaxy sample properties identified as appropriate to derive the luminosity densities in each band are presented in Table~\ref{tab:LFcomp}.

In the $NUV$ filter, there is only one published LF study that meets our local Universe criteria. We adopt the LF derived in \cite{Wyder05} which combined fluxes from the GALEX all-sky Imaging Survey (AIS) with the Two-Degree Field Galaxy Redshift Survey \cite[2dFGRS;][]{2dFGRS}. This study used $\sim$1000 galaxies with $z<$0.1 in 57~deg$^{2}$ on the sky. They removed objects outside the $NUV$ range of 17-20~mag, where the brighter galaxies were removed to avoid photometry issues with highly extended galaxies and fainter galaxies suffered from redshift incompleteness. The final list of galaxies had $>$80\% redshift completeness. The total luminosity density calculated from the LF parameters is $j_{\rm NUV}=1.40\times 10^{9}$ $L_{\rm NUV,\odot}$ Mpc$^{-3}$ \citep[assuming $M_{\rm NUV,\odot}=10.14$ and $L_{\rm NUV,\odot}$=3.749 $\times 10^{16} {\rm erg~s}^{-1} {\rm Hz}^{-1}$;][]{Willmer18}. We note that these values are derived assuming $H_0$ value of 70 $\rm{km s}^{-1} \rm{Mpc}^{-1}$, and we have scaled them to the Hubble constant assumed in this study.

In the $Ks$ filter, there are six LFs derived from galaxy samples that meet our local Universe criteria: \cite{Cole01,Kochanek01,bell03,jones06,Smith09,Hill10}. To choose our fiducial $Ks$-band LF, a secondary selection criterion is applied: the range of absolute magnitudes used to fit their Schechter functions. The most extreme luminosity ends of the fitted data can have significant effects on the derived LF parameters, and consequently on the total luminosity density measured which vary by 20-30\% for the LFs under consideration (see Figure~5 in \cite{Hill10} for a visual comparison).  We adopt the \cite{jones06} LF parameters which were derived with the largest range of luminosities ($-17.6< M_{\rm Ks}<-27.4$) among the other $Ks$-band LFs, and also contains the largest number of galaxies and covers the largest area. The galaxy sample was constructed from a combination of the 2MASS XSC source catalog and the 6dF Galaxy Survey \citep[6dFGS;][]{jones06,jones09} with spectroscopic redshifts. The total luminosity density calculated from the LF parameters is $j_{\rm Ks}=5.81\times 10^{8}$ $h L_{\rm Ks,\odot}$ Mpc$^{-3}$ \citep[assuming $M_{\rm Ks,\odot}=3.27$ and $L_{\rm Ks,\odot}$=3.926 $\times 10^{18} {\rm erg~s}^{-1} {\rm Hz}^{-1}$;][]{Willmer18}.



We note here that there does not currently exist a $W1$ LF published in the literature from which to compute completeness. It is reasonable to estimate that our completeness relative to the $W1$ fluxes would be increased based on the number of objects with $W1$ detections (88\%) compared to either GALEX (40\%) or 2MASS (70\%). However, the increase in completeness may only be modest as many of the galaxies with WISE detections over the other survey are likely to be fainter objects without viable distance estimates. An update to the NED-LVS completeness estimates relative to $W1$ is left for future work when $W1$ galaxy LF is published.

\subsection{Completeness Estimates} \label{sec:compest}

\begin{figure*}
  \begin{center}
  \includegraphics[scale=0.59]{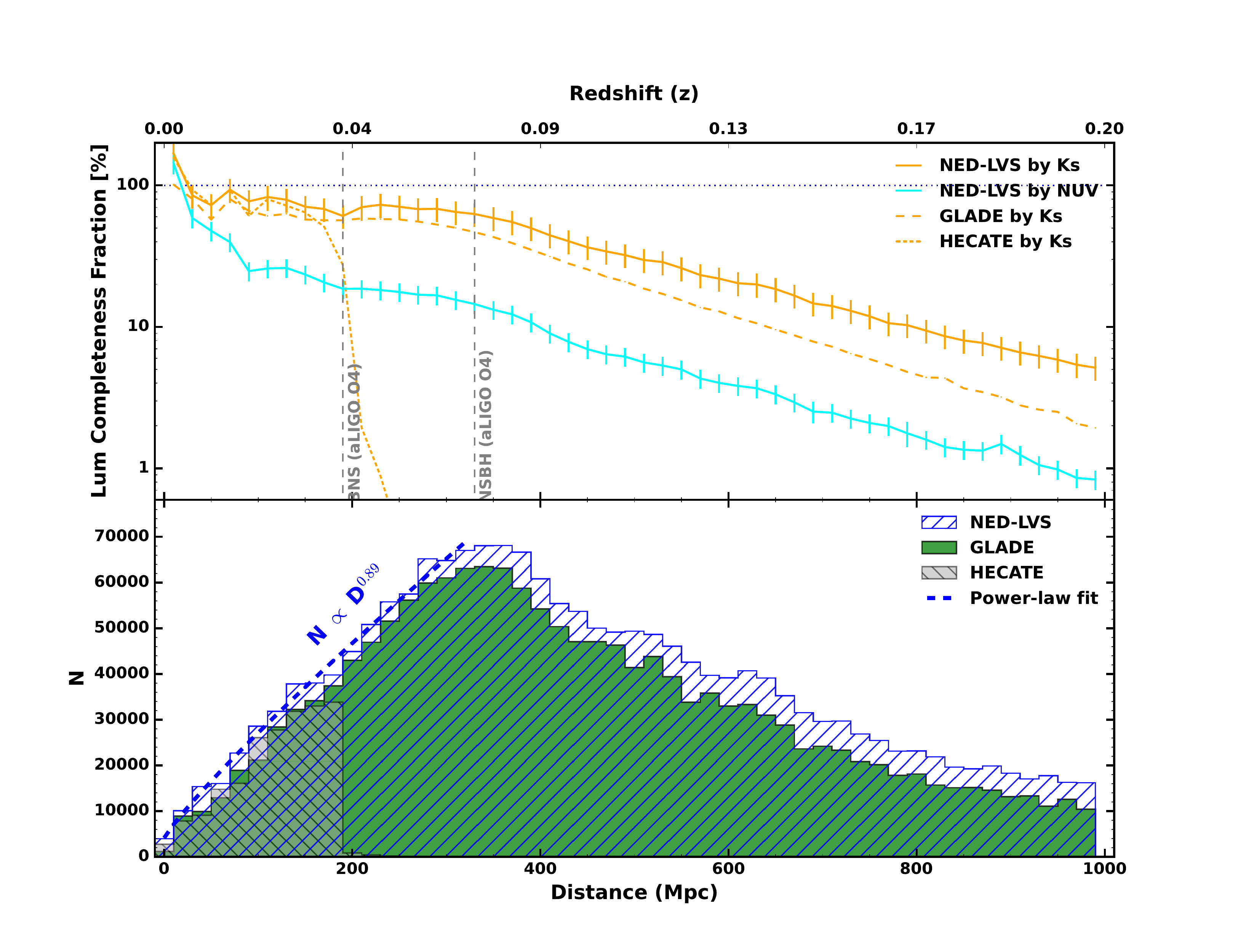}
  \caption{\textbf{Top Panel:} The completeness of the NED-LVS catalog as defined by the fraction of luminosity in available bands from 2MASS-$Ks$ (orange) and GALEX-$NUV$ (cyan) to the total luminosity derived from published luminosity functions: $NUV$ from \citep{Wyder05} and $Ks$ from \citep{jones06}. We find that NED-LVS in the NIR is highly complete at distances less than 30~Mpc and remains moderately (70\%) complete out to 300~Mpc. We also find that NED-LVS is less complete in the UV filters. The completeness of NED-LVS drops slowly past 300~Mpc to $\sim$1\% at 1000~Mpc. \textbf{Bottom Panel:} The distance histogram of both NED-LVS, GLADE, and HECATE. The number growth for NED-LVS rises linearly as quantified by an power-law fit to the data out to the peak of the histogram at D$\sim$350~Mpc ($N \propto D^{1}$). }
   \label{fig:NEDcomp}
   \end{center}
\end{figure*}

Figure~\ref{fig:NEDcomp} shows the completeness of NED-LVS in 2 bands: GALEX $NUV$ and 2MASS $Ks$. The galaxies in NED-LVS show nearly 100\% completeness (within the uncertainties) below 30~Mpc in both bands, which is to be expected as many of the galaxies in this distance range likely have a distance measurement.


Past 30~Mpc, we find that the NIR band continues to show moderate completeness levels at 70\% out to 300~Mpc, and then gradually decreases to a few percent at 1~Gpc. In the $NUV$ band, we find completeness levels that drop from 65\% at 20~Mpc to 25\% at 100~Mpc, then gradually declines to 15\% at 300~Mpc and under 1\% at 1~Gpc. The significantly lower completeness in the $NUV$ band can largely be attributed to the lower sensitivity of the GALEX survey compared to 2MASS and WISE relative to typical galaxy SEDs, where the fraction of sources with GALEX $NUV$ fluxes is roughly half compared to the sources with NIR fluxes ($N_{\rm NUV}/N_{\rm NIR}=$ 55\% and 47\% relative to $Ks$ and $W1$, respectively). 





We also find a plateau of completeness at 70\% in the NIR band between 100-300 Mpc. A flat completeness level over such a large distance range implies that the number of objects in our sample should increase with redshifts at a rate to match the growth expected by the increased volume in a shell ($N \propto D^2$). However, we measure a linear number growth ($N \propto D$) out to 300~Mpc (bottom panel of Figure~\ref{fig:NEDcomp}), which implies that there are fewer galaxies in our sample than expected given the simple assumption of constant number density. A plausible explanation for this behavior is that our sample contains the brightest galaxies which dominate the total luminosity, but is missing the fainter galaxies which dominate by number. Thus, our total luminosity ratio completeness remains constant, but the number of galaxies in each distance shell shows a deficit due to incompleteness in available redshifts.

\begin{figure*}
  \begin{center}
  \includegraphics[scale=0.6]{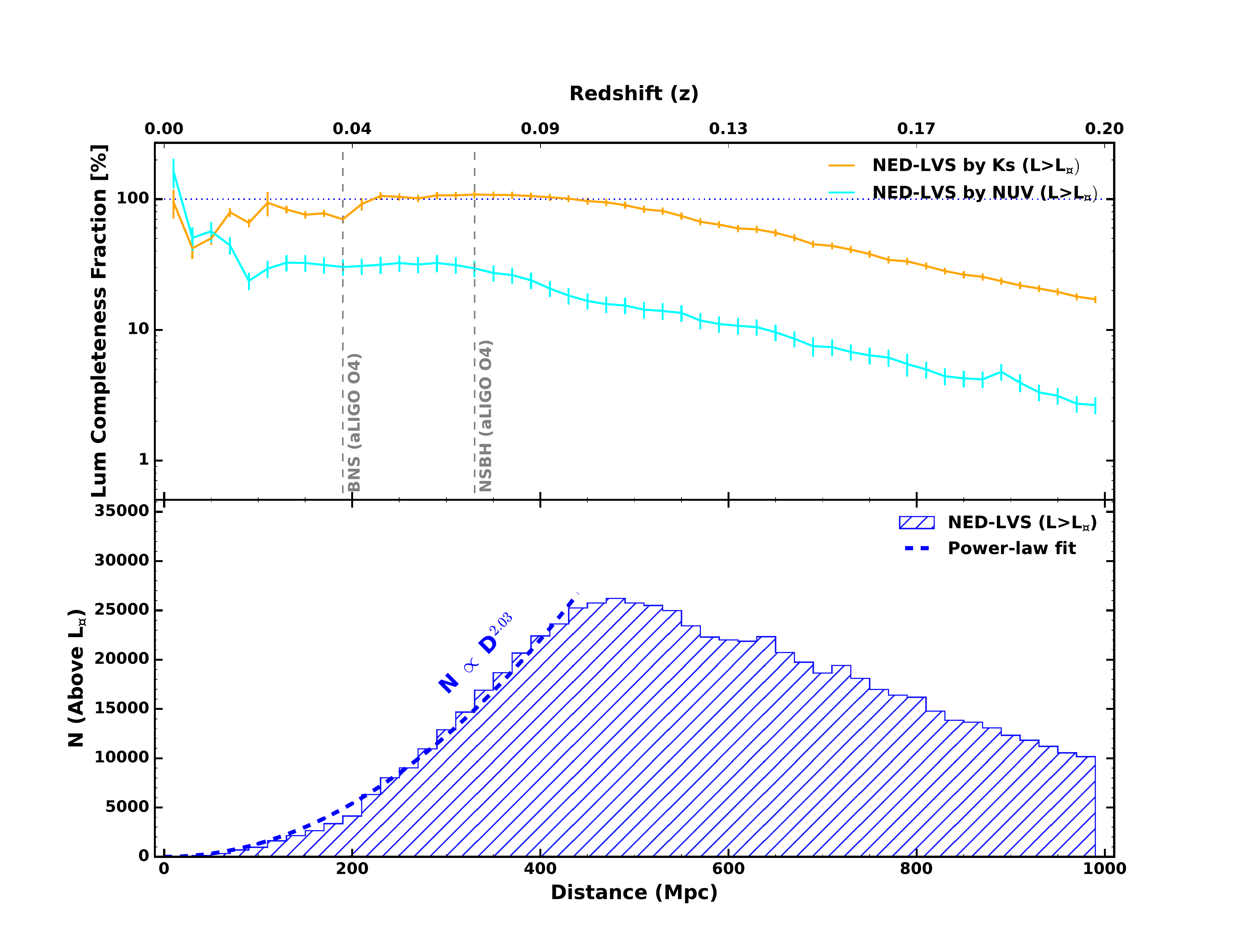}
  \caption{\textbf{Top Panel:} The completeness of the NED-LVS catalog as computed above \Lstar\ in the 2MASS-$Ks$ (orange) and GALEX-$NUV$ (cyan) bands. \textbf{Bottom Panel:} The distance histogram of NED-LVS. We find that NED-LVS in the NIR band is roughly 100\% complete for \Lstar\ and brighter galaxies out to $\sim$400~Mpc, and that the number growth of galaxies matches the rate expected by the increased volume in a spherical shell ($N \propto D^2$). }
   \label{fig:NEDcompLstar}
   \end{center}
\end{figure*}

We can test this hypothesis by re-computing completeness down to a range of luminosity limits and examine how the completeness levels and number growth changes. In this experiment, we use the same methodology described in \S\ref{sec:LFmethods}, but only consider galaxies with luminosities greater than a given limit and integrate the Schechter LF down to this same limit. The luminosity density constrained above a given luminosity can be calculated analytically by the following equation: 

\begin{equation}
    j_{L} = \int_{L_{\rm lim}}^{\infty} L \Phi(L) dL  = \phi^{*}~L^{*}~\Gamma(\alpha+2,L_{\rm lim}/L_{*}),  \label{eqn:jden_lim}
\end{equation} 

\noindent where $L_{\rm lim}$ is the limit down to which the LF is integrated and $\Gamma$ is the incomplete gamma function. We calculated completeness levels and fit power-law functions to the distance histogram for sub-samples of NED-LVS with limits between 0.1 and 2 $\times$ \Lstar. Figure~\ref{fig:NEDcompLstar} shows the results for $L_{\rm lim} = 1\times$\Lstar, where we find that NED-LVS is $\sim$100\% complete at distances out to $\sim$400~Mpc (a few bins left of the peak of the distance histogram) and the number of galaxies grows as $D^2$ for galaxies above the ``knee'' of the LF. These results suggest that our hypothesis of missing faint galaxies by number is correct, but still recover the majority of the total luminosity. The total luminosity densities calculated from the LF parameters down to \Lstar\ is $j_{\rm Ks}=1.75\times 10^{8}$ $h L_{\rm Ks,\odot}$ Mpc$^{-3}$ and $j_{\rm NUV}=4.18\times 10^{8}$ $L_{\rm NUV,\odot}$ Mpc$^{-3}$ for $Ks$ and $NUV$, respectively.

The result that NED-LVS is nearly 100\% complete for \Lstar\ galaxies and brighter out to $\sim$400~Mpc can be verified by examining completeness estimates by number. To make this estimate, we compare the NED-LVS luminosity histograms to the expected number of objects from the published Schechter LFs in the distance bins studied here. The details of this comparison are presented in Appendix~\ref{sec:LFnum}, where Figure~\ref{fig:LFnum_1} shows the luminosity histograms in distance bins between 0--500~Mpc. We find that the number of NED-LVS objects with luminosities at \Lstar\ and above meets the expected number under the LF at distances in and below the 380--400~Mpc bin, but begins to significantly diverge at distances past this bin. Thus, we can confirm that NED-LVS is nearly 100\% complete by number for galaxies at \Lstar\ and brighter out to $\sim$400~Mpc.

\section{DISCUSSION}  \label{sec:disc}
In this section, we compare the properties and completeness estimates of NED-LVS to previously published galaxy catalogs with distance limits of 200~Mpc or greater which are suitable for use in future LVK observing runs (i.e., O4 and O5). We also highlight the science use cases for NED-LVS and discuss the limitations of our sample and completeness estimates.

\subsection{Comparison to Other Local Galaxy Samples} \label{sec:catcomp}

There are two publicly available catalogs that we utilize for comparison: GLADE and HECATE. The basic properties of each catalog in the appropriate distance ranges (due to the different limits of each catalog) are given in Table~\ref{tab:catcomp}.

\begin{table*}
\centering

{Comparison of Galaxy Catalogs}\\
\begin{tabular}{lrrrr}

\hline
\hline
  & HECATE      & NED-LVS      & GLADE        & NED-LVS        \\ 
  & (D<200 Mpc) & (D<200 Mpc)  & (D<1000 Mpc) & (D<1000 Mpc)   \\ 
\hline

N (Total) & 204,733 & 268,514 & 1,703,812 & 1,872,544 \\
f (GALEX) & 0.0\% & 49.7\% & 0.0\% & 42.1\% \\
f (B-band) & 90.6\% & 0.0\% & 89.4\% & 0.0\% \\
f (SDSS) & 60.4\% & 0.0\% & 0.0\% & 0.0\% \\
f (2MASS) & 70.1\% & 66.6\% & 57.4\% & 79.1\% \\
f (WISE) & 60.1\% & 76.9\% & 0.0\% & 88.1\% \\
f (IRAS) & 9.6\% & 0.0\% & 0.0\% & 0.0\% \\
f (M$_{\star}$) & 46.0\% & 76.8\% & 0.0\% & 88.0\% \\
f (SFR) & 65.0\% & 76.9\% & 0.0\% & 88.1\% \\
\hline
\end{tabular}

\caption{A comparison of the total number of objects in NED-LVS, HECATE, and GLADE along with the fraction (`f') of objects with basic properties: UV fluxes (GALEX), optical fluxes (B-band and SDSS bands), NIR fluxes (2MASS), MIR fluxes (WISE), FIR fluxes (IRAS), and physical properties (stellar masses and SFRs). We present these properties out to two distances corresponding to the limits imposed by HECATE (D$<$200~Mpc) and NED-LVS (D$<$1000~Mpc). We find that NED-LVS has 32\% and 10\% more objects that HECATE and GLADE in the 200~Mpc and 1000~Mpc distances, respectively. In addition, NED-LVS has 17\% more objects with WISE fluxes (and similarly physical properties) than HECATE, and 15\% more objects with 2MASS fluxes than GLADE.}
\label{tab:catcomp}

\end{table*}

The GLADE catalog was constructed by cross-matching HyperLEDA, 2MASS-XSC, GWGC, 2MPZ \citep{bilicki14}, and SDSS-DR12 quasar catalog \citep{Paris17}, which results in 3.26 million objects with no explicit limits on distance. The basic properties provided in GLADE include: names, positions, distances, and fluxes in the optical $B$ and 2MASS NIR bands. GLADE did not derive physical properties for the galaxies in their catalog. 

A direct comparison between GLADE and NED-LVS in the same distance range ($D<1$~Gpc) reveals that NED-LVS contains $\sim$10\% more galaxies by number: N(GLADE)=1.7 million and N(NED-LVS)=1.9 million. We also estimate the completeness by total luminosity for both NED-LVS and GLADE with identical methods and in the same $Ks$-band filter (see \S\ref{sec:LFmethods}). The dashed line in Figure~\ref{fig:NEDcomp} shows the GLADE completeness re-computed from their publicly available catalog relative to the total luminosity density; rather than down to $0.626 \times L^{*}$ as computed in \cite{glade}. Below 30~Mpc, both samples are roughly 100\% complete, but NED-LVS is 10-15\% more complete between 40-600~Mpc.

Both GLADE and NED-LVS contain 2MASS fluxes in $J$, $H$, and $Ks$ bands, while NED-LVS contains GALEX $FUV$ and $NUV$ fluxes but is missing optical fluxes. While the NIR fluxes from both samples cover the whole sky, there is likely a significant fraction of GLADE sources missing NIR photometry since the GLADE catalog only contains 2MASS-XSC fluxes. NED-LVS, on the other hand, uses an aperture-prioritized selection of fluxes from three 2MASS source catalogs, where $\sim25$\% of our objects with NIR fluxes are derived from 2MASS catalogs not used by GLADE (2MASS-PSC or the larger-aperture 2MASS-LGA). Using the LGA and PSC fluxes in NED for GLADE objects that are missing 2MASS fluxes, we can compute the fraction of missing luminosity in our 20~Mpc distance bins relative to the total luminosity under the Schechter function (i.e., the percent luminosity completeness that is missing). We find that the absence of the fluxes from the LGA/PSC catalogs contributes to roughly half of the higher completeness levels exhibited by NED relative to GLADE in the $Ks$-band. This suggests that GLADE's lower completeness estimates must be due in part to objects with missing redshifts, not just missing fluxes.

The HECATE catalog uses galaxies from HyperLEDA as the basis of their sample, which contains 204,733 objects at distances less than 200~Mpc. These galaxies are subsequently cross-matched to: SDSS DR12 photometric catalog, various IRAS catalogs \citep{Sanders03,Wang14}, a catalog of forced photometry on unWISE images inside SDSS apertures \citep{Lang16}, 2MASS (LGA, XSC, and PSC catalogs), and GSWLC-2 physical properties (stellar mass, SFR, and metallicity) derived from SED fitting \citep{GSWLC2}.  The basic properties provided in HECATE include: names, positions, distances, fluxes in the optical (HyperLEDA collected UBVI and SDSS ugriz), NIR (2MASS $J$, $H$, and $Ks$-bands), MIR (WISE $W1$-$W4$), and FIR (IRAS 12, 25, 60, and 100 $\mu m$), and physical galaxy properties (stellar mass, SFR, and metallicity). HECATE also derives additional stellar masses and SFRs based on scaling relations similar to NED-LVS.

A direct comparison between HECATE and NED-LVS reveals that NED-LVS contains $\sim$30\% more galaxies by number inside 200~Mpc: N(HECATE)$\sim$200k and N(NED-LVS)$\sim$270k (Table~\ref{tab:catcomp}). We also estimate the completeness by total luminosity for both NED-LVS and HECATE with identical methods and in the $Ks$-band filter (for comparison across all three catalogs). The dotted line in Figure~\ref{fig:NEDcomp} shows the HECATE completeness re-computed from their publicly available catalog relative to the total luminosity density. HECATE shows similarly high completeness levels to NED-LVS below 80~Mpc by luminosity; however, NED-LVS is 10-20\% more complete between 80-200~Mpc.



Both HECATE and NED-LVS contain 2MASS ($J$, $H$, and $Ks$) and WISE ($W1$-$W4$) fluxes, while NED-LVS contains GALEX $FUV$ and $NUV$ fluxes but is missing optical and IRAS fluxes. The 2MASS fluxes in both catalogs are chosen with similar prioritization of the same source catalogs, thus both catalogs should have similar NIR fluxes. However, there is likely to be a significant difference in sky coverage for the WISE fluxes. The HECATE survey incorporated WISE fluxes only for sources with apertures defined from SDSS (i.e., only within the SDSS footprint). Conversely, NED-LVS incorporates fluxes from the AllWISE catalog with sources from the entire sky. A consequence of the SDSS footprint constraint is that the HECATE WISE fluxes will be absent for $\sim$3/4 of the sky. A caveat to the AllWISE fluxes used in NED-LVS is the lack of an extended, elliptical aperture. However, we mitigate this issue by prioritizing WISE fluxes from the z0MGS catalog which measured fluxes inside large, custom apertures for the largest galaxies within $\sim$50~Mpc, and utilized the largest circular apertures (22\arcsec) when available.  

HECATE has also compiled and computed physical properties for their galaxy sample. These properties include stellar masses, SFRs, and metallicities derived from SED fitting and SDSS spectra for sources within the SDSS footprint. A major limitation of these properties is the restriction to the SDSS footprint which only covers $\sim$1/4 of the sky. In addition, stellar masses and SFRs are also computed from luminosity scaling relationships, which should help to alleviate the HECATE sky restrictions on physical properties. However, these properties also have caveats associated with them. HECATE stellar mass estimates are derived from the 2MASS $Ks$-band \citep{bell03}, where the fainter survey limits of 2MASS (compared to WISE) will result in a reduced number of galaxies with stellar masses (46\%) compared to the NED-LVS masses derived from WISE fluxes (77\% out to 200~Mpc). The HECATE SFRs are computed from a variety of luminosities: WISE ($W3$ and $W4$) and IRAS (60$\mu m$, FIR, and TIR). While the IRAS-based SFRs span the entire sky, these catalogs were limited to brighter objects and only account for $\sim$10\% of the HECATE sample. The WISE-based SFRs of HECATE add values for fainter objects, but these are confined to the SDSS footprint resulting in only $\sim$40\% of HECATE objects with a WISE-based SFR. HECATE also provides a best-available SFR labeled as ``homogenized" which totals 65\% of the catalog. In comparison, $\sim$77\% of the galaxies in the NED-LVS have a SFR (within the distance limit of HECATE of 200~Mpc).

In summary, while these other catalogs have unique aspects, NED-LVS is more complete by number than both GLADE and HECATE by 10\% and 30\%, respectively, in the appropriate limiting distance. All three catalogs contain fluxes in several filters where only 2MASS measurements are common to all three. Compared to the other catalogs, NED-LVS is missing optical B-band due the lack of homogeneous, all-sky optical surveys, but neither GLADE nor HECATE contain GALEX UV fluxes. When comparing the completeness of the 3 catalogs relative to the total $Ks$-band luminosity density computed with the same methods, all three approach 100\% completeness below 30~Mpc. At distances beyond $\sim$80~Mpc, NED-LVS is more complete than both GLADE and HECATE (limited to 200~Mpc) by $\sim$10-20\%. Both NED-LVS and HECATE provide derived physical properties (e.g., stellar mass and SFR) which are useful for prioritising GW follow-up targets. However, only $\sim$46\% and $\sim$65\% of the HECATE galaxies have stellar mass and SFR estimates (largely limited to the SDSS footprint), respectively, whereas NED-LVS derives physical properties from all-sky GALEX and WISE fluxes resulting in $\sim$88\% of galaxies having both a SFR and stellar mass out to 1000~Mpc.

\subsection{Science Use Cases}

We anticipate that NED-LVS will be a valuable resource for the community to use in many areas of astrophysical research (star formation, galaxy evolution, etc.). In this section, we focus on facilitating the rapid follow-up to gravitational wave events and on studies of large scale structure where a more thorough understanding of the variation of redshift completeness across the sky is required.

\subsubsection{Host Galaxies to MMA Events}

Finding the electromagnetic counterpart, or kilonova, to GW events requires a rapid response from a well-organized astronomical community (ENGRAVE; \cite{Ackley20}, GOTO; \cite{GOTO}, GRANDMA; \cite{grandma}, GROWTH; \cite{Coughlin19a}, MASTER-NET; \cite{MASTER}, and others), where coordinated follow-up observations probing multiple wavelength regimes are needed to correctly identify the kilonova signature from BNS events \citep{EMGWcapstone} and possibly NSBH events \citep[][]{kruger20,zhu20}. Since the discovery of the kilonova for GW170817, no further kilonovae have been found. 

A recent analysis of the events during the O3 run by \cite{Petrov21} shows that improvements to the data analysis by LVK gave rise to the detections of events with weaker signals, which resulted in larger than expected sky localization areas (median areas of $\sim$2000~\degsq\ for BNS and NSBH events) and greater distances (median distances of 175 and 340~Mpc for BNS and NSBH events, respectively). Utilizing these new detection sensitivities, \cite{Petrov21} also made predictions for future GW events, where the median sky areas for both O4 and O5 will continue to be greater than 1000~\degsq, and that the median distances are likely to be 350 and 620~Mpc during O4 for BNS and NSBH events, respectively. These large areas and distances illustrate the need for continued optimization of follow-up strategies.

Incorporating galaxy catalogs with up-to-date data into follow-up strategies can aid in kilonova searches. In addition to reducing the search effort (by either number of targets or time of observations), a galaxy-targeted approach can increase efficiency via: weeding out background and foreground transient contamination \citep{kasliwal14,Andreoni20}, optimizing tiling strategies \citep{Antolini17,Arcavi17,Rana19,coughlin19b,ducoin20}, and prioritizing the galaxies by observed or derived properties \citep{Arcavi17,ducoin20,kovlakas21}. In this discussion, we focus on galaxy-targeted, prioritization strategies.  

\cite{coulter17} used a galaxy-targeted approach to find the kilonova of GW170817 in the nearby galaxy NGC~4993. Using information from NED, they found a total of 49 galaxies in the 90\% probability volume (including the distance probability), and used optical $B-$band luminosity \citep[a combination of stellar mass and SFR;][]{Phinney91} to prioritize their list where NGC~4993 was ranked \#12. The kilonova was later found within 11 hours of the event's alert. However, with a sample size of just one kilonova host galaxy, it is not yet clear what galaxy property, or properties, best track the merger rate of BNS events. 

To first order, the merger rate of BNS events is a combination of the host-galaxy's star formation history and the delay time to the merger event, where short merger delay times suggest that BNS events will happen in high-SFR galaxies and long delay times would occur in high-mass galaxies \citep{Adhikari20}. Using the only host galaxy known to date, several studies have found that NGC~4993 (with a high stellar mass and low SFR) is consistent with a merger delay time of $>$1~Gyr \cite[e.g.,][]{Levan17,Pan17,Belczynski18,Adhikari20}. Hence, many previous prioritization strategies use mass and SFR. However, another galaxy property that inversely correlates with the age of the stellar population is sSFR \citep{salim07,Rowlands12,Kannappan13,Guo19}, which is effectively a joint probability metric on stellar mass and SFR that may better track the BNS merger rate. Here, we use NED-LVS to test which properties ($M_{\star}$, SFR, or sSFR) provide higher rankings of the known host galaxy inside the GW170817 sky localization.

We cross match, in 3D, the NED-LVS galaxies with the BAYESTAR \citep{singer16} sky localization of GW170817, which has a 90\% probability area of 31~\degsq\ and a distance of 40$\pm$8~Mpc. We utilize the \texttt{CROSSMATCH} function in the \texttt{PYTHON} package \texttt{ligo.skymap}\footnote{https://lscsoft.docs.ligo.org/ligo.skymap/\\postprocess/crossmatch.html} to compute the probability density (per volume; $P_{\rm V}$) for each galaxy resulting in 46 galaxies in the 90\% probability event volume \citep{singer16}. We then use the following formalism to ``grade'' galaxies by different physical properties \citep{Arcavi17,ducoin20}, for example the grade using stellar mass ($P_{\rm M_{\star}}$) is: 

\begin{equation}
    P_{\rm M_{\star}} = \frac{\rm M_{\star,\rm galaxy}}{\sum \rm M_{\star,\rm galaxy}}
\end{equation}
    
\noindent where the denominator is the sum of stellar mass of galaxies in the event's 2D probability and whose distances fall within 3$\sigma$ of the event's distance error \citep{ducoin20}. We note here that we require lower sSFR values to have higher marginalized probabilities since it is inversely proportional to stellar population age, and consequently we use the inverse (sSFR$^{-1}$), which is a timescale quantity with units of years, to calculate this probability. We can then combine the galaxy property grades with the 3D probabilities of the galaxy's locations to derive a total combined probability:

\begin{equation}
    P_{\rm tot} = P_{\rm property} \times P_{\rm V},
\end{equation}

\noindent where $P_{\rm property}$ is the grade of the galaxy based on one of three physical properties examined in this study: $P_{\rm M_{\star}}$, $P_{\rm SFR}$, $P_{\rm sSFR^{-1}}$. We then rank the galaxies by sorting their combined probabilities. 

We present the results in Table~\ref{tab:gw170817} where we show the top 10 galaxies when sorting by the total probabilities for each of the three properties. We find that NGC~4993 ranks high on the list when sorted by stellar mass (at \#6) and sSFR (at \#3), but SFR puts the correct host in the middle of the list (at \#24). 

\begin{table*}
\centering

{Galaxy Lists in GW170817 Sky Localizations} \\ 
\begin{tabular}{lcllcl}

\hline
\multicolumn{3}{l}{Final Sky Map (31~deg$^2$)} & \multicolumn{3}{l}{Sky Map w/o Virgo (187~deg$^2$)} \\ 
\hline

Galaxy & Rank & $P_{M_{\star}} \times P_{V}$ & Galaxy & Rank & $P_{M_{\star}} \times P_{V}$ \\ 
\hline
NGC 4763 & 1 & 4.34e-02 & NGC 5061 & 1 & 5.03e-02 \\
NGC 4970 & 2 & 3.08e-02 & NGC 4763 & 2 & 2.13e-02 \\
NGC 4830 & 3 & 2.83e-02 & NGC 5114 & 3 & 1.87e-02 \\
IC 4197 & 4 & 2.26e-02 & NGC 3976 & 4 & 1.65e-02 \\
IC 4180 & 5 & 2.07e-02 & NGC 4970 & 5 & 1.64e-02 \\
NGC 4993 & 6 & 1.81e-02 & NGC 4830 & 6 & 1.58e-02 \\
MCG -02-33-036 & 7 & 1.56e-02 & NGC 5967 & 7 & 1.43e-02 \\
NGC 4968 & 8 & 1.30e-02 & IC 4197 & 8 & 1.30e-02 \\
ESO 508- G 033 & 9 & 8.67e-03 & NGC 4993 & 9 & 1.18e-02 \\
ESO 575- G 029 & 10 & 8.37e-03 & IC 4180 & 10 & 1.10e-02 \\

\hline
Galaxy & Rank & $P_{\rm SFR} \times P_{V}$ & Galaxy & Rank & $P_{\rm SFR} \times P_{V}$ \\ 
\hline
NGC 4968 & 1 & 1.12e-01 & NGC 4968 & 1 & 8.03e-02 \\
ESO 508- G 033 & 2 & 6.59e-02 & ESO 508- G 033 & 2 & 3.86e-02 \\
NGC 4763 & 3 & 4.05e-02 & NGC 4763 & 3 & 2.58e-02 \\
MCG -02-33-036 & 4 & 6.01e-03 & NGC 2782 & 4 & 2.55e-02 \\
ESO 575- G 029 & 5 & 5.96e-03 & NGC 5967 & 5 & 2.55e-02 \\
IC 4180 & 6 & 5.91e-03 & NGC 3810 & 6 & 2.21e-02 \\
UGCA 331 & 7 & 5.53e-03 & NGC 3976 & 7 & 1.90e-02 \\
ESO 508- G 019 & 8 & 5.36e-03 & NGC 4123 & 8 & 1.88e-02 \\
ESO 508- G 003 & 9 & 3.28e-03 & NGC 4658 & 9 & 1.02e-02 \\
ESO 508- G 010 & 10 & 3.14e-03 & NGC 3686 & 10 & 8.84e-03 \\
\ldots & \ldots &\ldots &\ldots &\ldots &\ldots \\
NGC 4993 & 24 & 8.16e-04 & NGC 4993 & 53 & 6.86e-04 \\

\hline
Galaxy & Rank & $P_{\rm sSFR^{-1}} \times P_{V}$ & Galaxy & Rank & $P_{\rm sSFR^{-1}} \times P_{V}$ \\ 
\hline
NGC 4970 & 1 & 5.12e-02 & NGC 4970 & 1 & 4.34e-02 \\
NGC 4830 & 2 & 3.97e-02 & NGC 4830 & 2 & 3.54e-02 \\
NGC 4993 & 3 & 8.69e-03 & IC 0874 & 3 & 2.89e-02 \\
IC 4197 & 4 & 5.69e-03 & UGCA 289 & 4 & 1.57e-02 \\
ESO 575- G 061 & 5 & 4.16e-03 & NGC 5048 & 5 & 1.37e-02 \\
WISEA J125732.70-194200.8 & 6 & 4.04e-03 & NGC 5114 & 6 & 1.33e-02 \\
WISEA J125251.08-152929.7 & 7 & 3.07e-03 & MCG -02-32-026 & 7 & 1.26e-02 \\
MCG -02-33-036 & 8 & 2.16e-03 & NGC 5061 & 8 & 1.09e-02 \\
IC 4180 & 9 & 2.02e-03 & NGC 4993 & 9 & 9.02e-03 \\
2MFGC 10461 & 10 & 1.85e-03 & IC 4197 & 10 & 5.24e-03 \\
\hline
\end{tabular}

\caption{The top 10 galaxies prioritized by physical properties (stellar mass, SFR, and inverse sSFR) inside two sky localizations for GW170817: the final BAYESTAR map with a 90\% probability area of 31~deg$^2$ and a 90\% volume that contains 46 NED-LVS galaxies (\textbf{left columns}), and the initial sky map without the Virgo detector constraints with an area of 187~deg$^2$ and a 90\% volume that contains 207 NED-LVS galaxies (\textbf{right columns}). The host galaxy, NGC~4993, for GW170817 is in the top 10 when sorting by stellar mass and sSFR, but half way down the list when using SFR alone.}
\label{tab:gw170817}

\end{table*}

While a validation of NED-LVS galaxies and rankings in the final sky localization of GW170817 is a useful exercise, the localizations in the upcoming O4 run will be larger and a validation to such events would be more revealing. Following the test performed by \cite{ducoin20}, we cross match NED-LVS with the larger BAYESTAR sky localization of GW170817 where the Virgo detector constraints were not applied. This localization has a 90\% probability area of 187~\degsq\ with a distance of 34$\pm$9~Mpc. We find 207 galaxies in the 90\% probability event volume, where NGC~4993 continues to rank in the top 10 by stellar mass (at \#9) and sSFR (at \#9) but ranks low by SFR (at \#53). Thus, prioritizing by stellar mass and sSFR continues to perform well even when larger numbers of galaxies are present in the event's volume.

Our rankings by stellar mass are in broad agreement with the rankings of HECATE \citep{kovlakas21} and MANGROVE \citep{ducoin20}. In addition, the low ranking of NGC 4993 by SFR in combination of the high ranking by mass provides supporting evidence that longer delay times trace BNS merger event rates. It is interesting to note that the ranking of NGC 4993 by sSFR performs just as well as stellar mass since sSFR correlates with the age of the stellar population, a property closely tied to theoretical predictions of BNS merger rates. A caveat with the rankings provided here by sSFR is the SFRs used (scaling relations from WISE luminosities), which can be overestimated in early-type galaxies (see \S\ref{sec:physprop}). However, by simply sorting the galaxies by observable sSFR proxies ($NUV$-$W1$ and $W1$-$W4$ colors) to mitigate this issue, we find that NGC 4993 ranks in the top 3 and top 6 by color for the updated and initial sky maps, respectively. A study on the prioritization of galaxy lists by more direct stellar population age measurements (e.g., depth of the 4000~\AA~break ($D_{N}4000$), the EW of Balmer lines, and sSFR color proxies) is encouraged.





Another aspect of kilonova searches that is of significant importance is the need for rapid follow-up observations. While the fast-fading optical transient \citep[1~mag per day;][]{Arcavi17optical} to the GW170817 kilonova was found within half a day, the even faster fading UV component that can shed light on the merger pathways and dynamics \citep{Metzger15} was not observed in time. To help mitigate rapid response barriers, NED has developed a gravitational wave follow-up service (NED-GWF).\footnote{https://ned.ipac.caltech.edu/NED::GWFoverview/} This service continuously listens for GW event alerts on the Gamma-ray Coordination Network (GCN)\footnote{https://gcn.gsfc.nasa.gov/}, automatically downloads each GW event's HEALpix \citep[Hierarchical Equal Area isoLatitude Pixelization;][]{Gorski05} sky map, and cross matches to the latest version of NED-LVS to produce a web page for each event with the following: basic information on the GW event, an all-sky image of the probability contours and locations of the galaxies in the event's volume, a link to download the entire galaxy list in the volume, and a list of the top 20 galaxies sorted by galaxy properties. There is also an API function for automatic downloads of the galaxy lists.\footnote{https://ned.ipac.caltech.edu/Documents/Guides/Interface/GWF} All of these tasks are automated and take on average a few minutes from the time of the alert to publishing the website. This service is open to the public and can be used by anyone in the world to search for the kilonova counterparts to GW events.

While NED-LVS (and subsequently NED-GWF) can be used to increase the efficiency of GW follow-up efforts, there is a potential for galaxy-targeted strategies to be applied to other multi-messenger events, such as high-energy neutrino events \citep[e.g.,][]{Stein21} or poorly localized transient events likely associated with GW events \citep[i.e., short duration gamma-ray bursts; sGRBs;][]{Rastinejad21}. However, since neither sGRBs nor neutrino events have independent constraints on the distance to the event, the potential galaxy lists for these events will be relatively large due to sky area-only constraints \citep[by a factor of $\sim$50;][]{ducoin20,kovlakas21}. These events would require additional study of prioritization metrics for host-galaxy properties and this work is left for a future study.


\subsubsection{Large-Scale Structure and Galaxy Environments}

\begin{figure*}
  \begin{center}
  \includegraphics[scale=0.65]{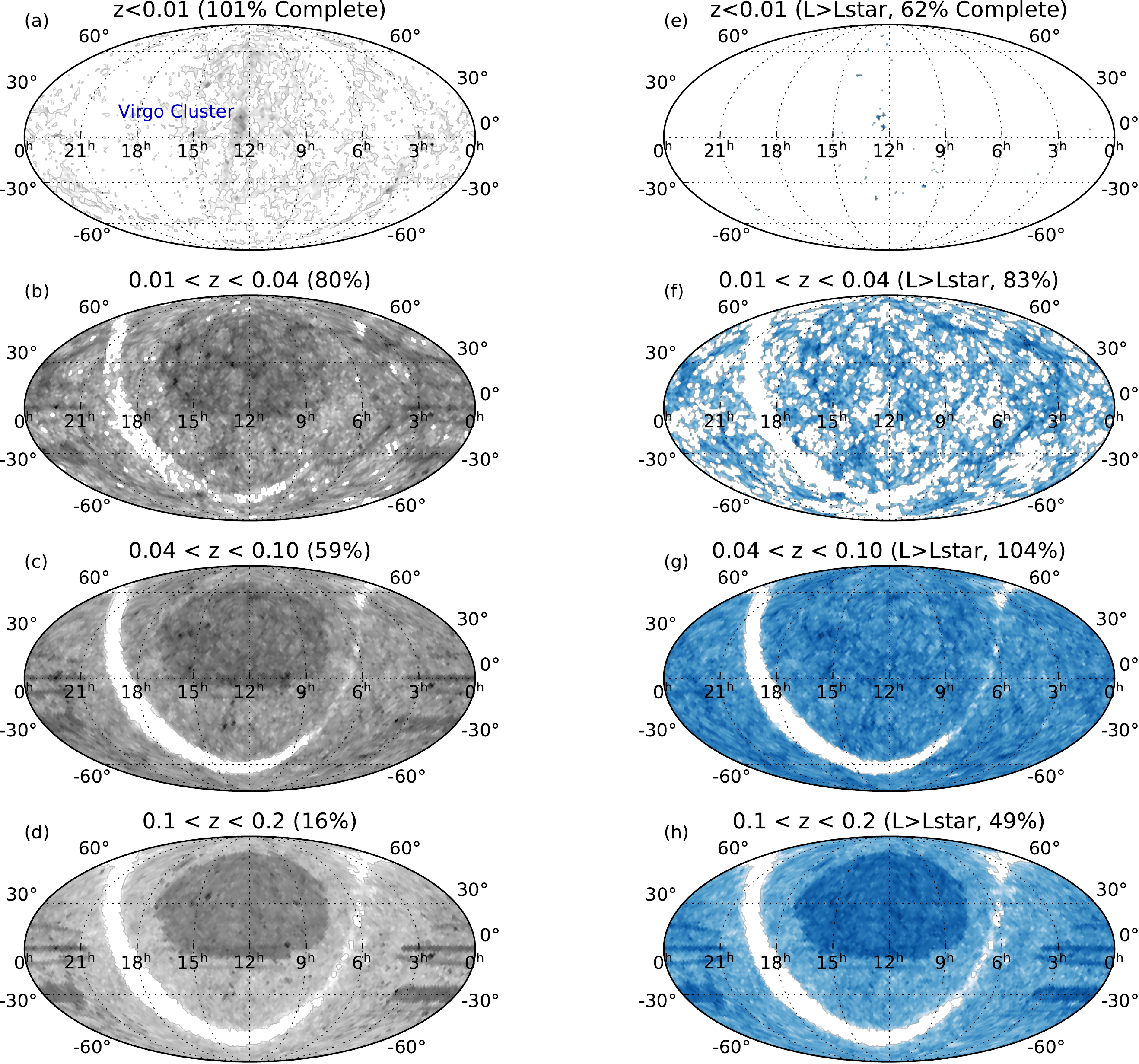}
  \caption{ All-sky density plots of NED objects in equatorial (J2000) coordinates, where the \textbf{left column} presents objects with no luminosity cut and the \textbf{right column} presents objects with $Ks$-band luminosity greater than \Lstar. Each row shows the objects in redshift shells: $z<$0.01 ($D<40$~Mpc), 0.01$<z<$0.04 ($40<D<180$~Mpc), 0.04$<z<$0.10 ($180<D<460$~Mpc), and  0.1$<z<$0.2 ($460<D<1000$~Mpc), where the percentages in parentheses provide the completeness of objects in each shell. The HEALPix maps have a $\rm 2~deg^2$ resolution for each pixel. The artificial patterns (survey footprints) seen in panels with low completeness (c, d, and h) suggest that studies of galaxy environment or large scale structure should exercise caution when interpreting results. }
   \label{fig:lss}
   \end{center}
\end{figure*}

Primary science motivations for the great efforts required to perform large galaxy redshift surveys include: mapping and quantifying the large scale structure (LSS) of the Universe in 3D, also known as cosmic cartography \citep[e.g.,][]{2009MNRAS.400..183K}, studying the influence of local and large scale environments on the physical properties of galaxies, and providing a better understanding of galaxy populations from an unbiased sampling of objects. Extensive research in this field over the past few decades has resulted in the acquisition of millions of spectroscopic redshifts from galaxy surveys such as the CfA redshift survey \citep{CFA}, 2dFGRS \citep{2dFGRS}, 6dFGRS, GAMA \citep{Driver09,Driver11}, Las Campanas Redshift Survey \citep{LCRS}, SDSS \citep{sdss7,sdss12}, and others \citep[e.g.,][]{2020IAUS..341....1O}, all of which are represented in the NED-LVS. Published studies of these surveys have revealed, in successively greater detail, a complex network of galaxy groups, clusters, and superclusters configured in a cosmic web of walls and filaments separated by vast voids of underdense regions tens of Mpc in diameter \citep[e.g.,][and references therein]{2013AJ....146...69C,2020IAUS..341....1O}. Furthermore, these surveys have yielded significant discoveries on the relationships between galaxy evolution and environment \citep[e.g., morphology-density relationship;][]{1980ApJ...236..351D,1998ApJ...499..589H,2017A&A...602A..15C,2018ApJ...852..142C}.



Despite the impressive sky coverage and increasing depth of redshift surveys, and the scientific progress made recently in understanding galaxy environments and the cosmic web, future attempts to more accurately quantify variations in galaxy properties and evolutionary effects as a function of environment over larger volumes of space are limited by significant gaps in available redshift measurements. It is therefore crucial for researchers to understand when apparent variations in the density of galaxies in any particular 3D volume are due to actual physical cosmic variance, or variations in the completeness of redshift measurements, or some complex mixture.

To provide a global view of this issue, Figure~\ref{fig:lss} shows all-sky density maps of galaxies in the NED-LVS in different redshift shells. The most recognizable gap that is visible in most panels of Figure~\ref{fig:lss} is the plane of the Galaxy, which most certainly lowers our completeness estimates averaged over the whole sky (see Section~\ref{sec:disclimit}). The panels on the left contain all galaxies with available redshifts, where we have shown that completeness drops from essentially 100\% in the nearest redshift shell to only 16\% in the outer shell. Panels (c) and (d) show clearly that this drop is due to incomplete sky coverage of the redshift surveys combined in this work. The redshift measurements are most complete in the north Galactic cap in the footprint of the SDSS, covering about one forth of the sky. In contrast, the panels in the right column contain only the galaxies with $L > L^{*}$, for which we have shown the completeness (Fig.~\ref{fig:NEDcompLstar}) is near 100\% out to z $\sim 0.1$ (D = $\sim$400~Mpc), dropping to 20\% at the edge of the sample at z = 0.2 (D = 1000 Mpc). Zooming into Figure~\ref{fig:lss} panels (e), (f) and (g) shows that local galaxies with $L > L^{*}$ trace very well the ``tip of the iceberg'' of LSS revealed in many studies \citep[e.g.,][and references therein]{2013AJ....146...69C,2020IAUS..341....1O}.

Given the limited sky coverage of current redshift surveys, a robust approach used by most investigators of LSS is to confine studies to galaxies within the boundaries of a specific sky survey within which there is uniform sky coverage of photometric observations and follow-up spectroscopic observations.  The footprint of the SDSS spectroscopic redshift sample, for example, is most clearly visible in the outer redshift shells of the NED-LVS in Figure~\ref{fig:lss}(d) and (h).  However, LSS studies over the whole sky are possible using NED-LVS to the extent that investigations should remain confined to galaxies with $L > L^{*}$ and $z<0.1$ where there exist high levels of completeness and roughly uniform coverage.





\subsubsection{Other Science Cases}
The merger of panchromatic fluxes and distance information across thousands of journal articles and survey catalogs is an invaluable resource for the astronomical community. In addition, given that most objects in NED-LVS have data with independent measurements from multiple sources which have been vetted during the merging process by the NED team over many decades (including the additional efforts performed here; see \S\ref{sec:app_visclass}), NED-LVS is a highly reliable compilation suitable for addressing a number of statistical questions in astrophysics. For instance, the 3D locations of objects can be used to compute correlation functions to test theories of dark matter, the panchromatic colors can identify different galaxy types and serve as proxies for star formation activity to better understand both star formation and galaxy evolution, and these data can be used to inform sample selection and exposure times for future observational studies.



\subsection{Limitations of NED-LVS} \label{sec:disclimit}

In this section we discuss the limitations of using NED-LVS during multi-messenger and poorly-localized transient events, and those related to our completeness estimates. 

A general limitation of any galaxy sample in the local Universe stems from distances based on redshifts, where the peculiar velocities of the nearest galaxies ($D\lesssim$30-40~Mpc) can be a significant fraction of the observed velocity. We mitigate these issues by prioritizing redshift-independent distances over redshifts, where 25\% of our sample with D$<$20~Mpc have redshift-independent measurements. However, the largest peculiar velocities occur in the members of galaxy clusters, where deviations from the Hubble flow will be more pronounced for the nearest clusters \citep[e.g., Virgo at D$\sim$16.5~Mpc;][]{Mei07}. For instance, \cite{kovlakas21} used a Kernel Regression technique to correct 650 Virgo cluster members and found that the local average distance deviated from the Hubble flow by as much as a factor of two. These corrections are left for a future version, but we note that only a small number of galaxies would be affected \citep[as many as 1500 in Virgo;][]{Kim14}. The velocities for galaxies in more distant clusters (e.g., Coma, Shapely, etc.) would be affected to a lesser degree. 

A caveat in our luminosity completeness estimates stems from the use of a Schechter function to define the expected total luminosity density, where this functional form may not fully reflect the true luminosity distribution of galaxies \citep{salim12}, especially at the high-luminosity end where poor statistics can lead to poor fits and a potential over/under estimation of the total expected luminosity. An inspection of the numbers of of rare, bright ($L>10\times$\Lstar) NED-LVS galaxies (after contamination removal) compared to those expected from our fiducial $Ks$-band LF  (Figures~\ref{fig:LFnum_1} and \ref{fig:LFnum_2}) shows an excess of these objects past 200~Mpc that increases with distance: a few at 200~Mpc and a total of $N\sim$1100 out to 1000~Mpc. However, the luminosity of NED-LVS galaxies with $L>10$\Lstar\ makes up $<$1\% of the expected total luminosity ($j_{Ks}$) in each of the distance shells, suggesting that the potential overestimation of our completeness is small. It is unclear if there is a true or spurious excess of luminous galaxies in our sample or if the LF used was poorly constrained at the bright end.

Further investigation of $L>10\times$\Lstar\ objects reveals a mixture of early-type galaxies, AGN or QSOs, and even some larger spiral galaxies \citep{Ogle16}. The luminosities of QSOs or galaxies that host an AGN will be boosted \citep{Montero09}, in which case our completeness estimates may be overestimated. However, we find that only 90 of these luminous objects are potentially AGN (including QSOs) whose light typically accounts for only 0.01\% of the total luminosity in their respective distance shells. It is interesting to note that $\sim$200 of the luminous NED-LVS objects have UV detections to indicate active star formation or have readily identifiable spiral arms during our visual inspection. These so called ``super spirals'' generally have large sizes ($>$50~kpc), stellar masses ($>10^{11}~{\rm M}_{\odot}$), and SFRs ($>$5-65$~{{\rm M}_{\odot}{\rm yr}^{-1}}$), where these objects are thought to be a remnant population of unquenched, massive disk galaxies and may be the precursors of some giant elliptical galaxies found in low-density environments \citep{Ogle16,Ogle19}. Other explanations for excess luminosities may include close galaxy pairs that are not resolved and potential unreliable distances that are not easily flagged.


There is also a limitation of our completeness estimates that stems from the non-uniform coverage of galaxies across the entire sky (e.g.,  heterogeneous flux limits for spectroscopic surveys, unobservable galaxies behind the MW, etc.). Consequently, our completeness estimates may be an underestimate, especially due to the likely large numbers of galaxies located in the Zone of Avoidance \citep{Staveley16}. To quantify the effects of missing galaxies behind the MW, we have recomputed completeness by excluding all NED-LVS objects with $|b|<10^{\circ}$ (i.e., in the Galactic Plane) and adjusted the volumes accordingly. We find that the completeness by $Ks$-band increases an additional $\sim$5\% between 20--500~Mpc. The heterogeneous collection of galaxies from a wide range of surveys and individual measurements thus results in completeness that is a complicated function of sky location. This is further evidenced by Figure~\ref{fig:lss}, where (excluding the MW) we see a relatively uniform distribution at distances less the 300 Mpc, but begin to see a dearth of objects at greater distances especially in the southern sky. In a future work, we plan to derive completeness maps across the sky that would allow users to account for the varying levels of completeness. 




One of the intended uses for NED-LVS is to increase the search efficiency for kilonovae associated with GW events. The median sky localizations for these events will be relatively large for the next few years \citep{Petrov21}, and will require that large lists of potential hosts be prioritized by either observable or derived physical properties. NED-LVS contains flux measurements from UV and IR taken from uniform, all-sky surveys, and provides physical properties primarily based on MIR WISE fluxes which are available for 88\% of the sample. 

A potential caveat with regard to WISE fluxes (and subsequently the derived properties) is that there are no WISE extended source catalogs with apertures that depend on the extent of a galaxy's light (D25, half-light radius, etc.) to ensure robust sorting by properties. Our construction of MIR fluxes instead relies on a strategy to recover as much of the total flux for an object as possible using the largest apertures available. Deviations from the total flux will depend on the angular size of the galaxy, where more extended galaxies will likely be affected to a greater degree. To mitigate this effect, we have opted to prioritize the MIR fluxes derived from large, custom apertures in the z0MGS survey \citep[see \S\ref{sec:z0mgs};][]{z0mgs}, where the NED-LVS `GALEXphot' or `WISEphot' flags are set to `z0MGS'. These fluxes are measured inside isophotal apertures where the distribution of semi-major axes extends to values as low as 15\arcsec ($\sim$16, 33 and 72~kpc at z=0.05, 0.1, and 0.2, respectively). Our next highest priority aperture is 22\arcsec\ from the AllWISE catalog which should cover the majority of light for extended objects without fluxes in z0MGS. Thus, sorting NED-LVS galaxy lists by MIR fluxes or physical properties derived from MIR fluxes should provide reasonably robust results.  


Estimating physical properties of galaxies using the photometry compiled in NED-LVS comes with its own limitations. One caveat related to our estimate of stellar mass stems from using the same scaling relationships for the wide range of galaxy types seen in our sample. The mass-to-light ratios used to scale NIR fluxes to stellar mass can be significantly affected by the age of the stellar population \citep{Bell01,bell03,Courteau14}, where the ratios can differ by as much as 0.5~dex between late- and early-type galaxies \citep{z0mgs}. We have adopted the prescription of \cite{z0mgs} that employs a mass-to-light ratio dependent upon an observable proxy for sSFR which traces stellar population age. Our resulting stellar masses show good agreement with those derived using more sophisticated Bayesian SED fitting methods \citep{GSWLC2} with a median offset and scatter of -0.01 and 0.2~dex. We conclude that the stellar masses computed for NED-LVS provide reasonably robust results.


The scaling relationships for SFRs have also shown biases amongst some galaxy types, where the MIR SFRs based on bands that are contaminated by PAH and silicate absorption features (i.e., $W3$) can show a strong metallicity dependence \citep{Relano07,calzetti07}.  These lower-metallicity galaxies can have their SFRs significantly underestimated due to the dearth of PAH emission \citep{engelbracht05,engelbracht08,draine07,smith07b}. SFRs estimated from $W4$ fluxes have shown good agreement with low scatter compared to Balmer-decrement corrected \ha\ and SED fitted values even for dwarfs \citep{Brown17,z0mgs}. In our sample we utilize the $W4$-based SFRs and choose not to provide the W3-based estimates. 

However, there are caveats associated with using $W4$-based scaling relationships resulting in additional MIR luminosity contributions from active galactic nuclei (AGN) and/or older stellar populations in early-type galaxies, where the SFRs for these galaxies will be overestimated \citep{Bendo10,GSWLC2,z0mgs}. Using the conservative WISE-only color selection of $W1$-$W2$ > 0.8~mag from \cite{Stern12}, we estimate that only 1\% (20k) of our sample will be affected by AGN contamination. In addition, early-type galaxies represent a much wider challenge where SFRs can be overestimated by as much as two orders of magnitude \citep[see also HECATE;][]{kovlakas21}. Using a combination of $NUV$-$W1$ > 7 mag, $W1$-$W3$ <2.5 mag, and $W1$-$W4$ < 5.5 mag colors (which do not suffer these same biases) to select low sSFR galaxies, we estimate that 47\% of our sample will be affected to varying degrees. Recent studies of early-type galaxies have found that some show $W4$ luminosities that are 2 orders of magnitude above what is expected from stellar population models given their the $Ks$ or $W1$ luminosities \citep[][]{Davis14,Simonian17}. This MIR excess is not well understood given the low star formation activity of these systems, but has been attributed to dust from AGB stars \citep{Athey02} or very small grains created during the destruction of larger dust grains \citep{Xilouris04}. To date, no published SFR prescriptions have successfully accounted for the excess dust present in early-type galaxies. Such an effort merits its own study and is beyond the scope of this work. We advise caution when using the $W4$-based SFRs for early-type galaxies, and have populated a flag (`ET\_flag'$=$True) in our sample which uses a combination of UV-IR and IR colors. 

Given the limitations and caveats discussed above related to derived properties, we emphasize here that users can utilize the measured fluxes compiled in NED-LVS to estimate their own physical properties in a manner appropriate for their own science requirements. In addition, NED-LVS could also be augmented with new, better behaved derived properties emerging in the literature.



Challenges encountered while constructing the NED-LVS, as well as future improvements as new data become available, point to the  importance of following best practices when publishing redshifts and other measurements \citep{Chen22}. This includes listing accurate and meaningful uncertainty estimates for redshifts, with a realistic number of significant figures, and clearly flagging unreliable redshifts due to low signal-to-noise, contamination in the aperture, limitations of photometric redshift techniques or model fitting assumptions, etc. This will lead to improvements to automated star/galaxy/AGN classifiers in NED, and result in more reliable derived quantities such as selection of fiducial redshift (from multiple published measurements) and estimates of distances and physical properties that depend on robust redshift measurements.

\section{SUMMARY \& Future Plans} \label{sec:summary}

In this study we present the NASA/IPAC Extragalactic Database-Local Volume Sample (NED-LVS) which is composed of nearly 1.9 million extragalactic objects with distances less than 1 Gpc. The distances selected for NED-LVS are based on both redshifts (98.6\%) and redshift-independent distances (1.4\%), where redshift-independent measurements are prioritized at distance less than 200~Mpc. We have also performed automated assessment and visual inspection to clean over 200,000 contaminating objects and unreliable redshifts. The re-classified objects and redshift quality flags assigned during the vetting process will be folded back into the NED database.

We extract photometry from all-sky surveys (GALEX, 2MASS, and AllWISE) that have been joined into NED, where we prioritize the largest available aperture fluxes to account for resolved galaxies. In addition, we utilize the large, custom aperture fluxes in the GALEX and WISE filters for highly extended galaxies (out to 50~Mpc) as measured by z0MGS \citep{z0mgs} when available. The resulting fluxes in NED-LVS are used to derive physical properties ($M_{\star}$ and SFR) via integrated-light scaling relationships from WISE luminosities, where 88\% of objects have estimated values. The physical properties derived here broadly agree with those derived from more sophisticated SED fitting methods. However, the SFRs for early-type galaxies show varying degrees of overestimation which is a known issue with integrated-light scaling relationships in the MIR. We provide a flag (`ET\_flag') to indicate which objects may have overestimated SFRs and these values should be treated with caution. 

The main goal of this study is to estimate the completeness of the galaxy census per unit volume in the NED local Universe compilation, where completeness is defined as the ratio of the total luminosities of galaxies to the total luminosity expected from integrating under published galaxy luminosity functions. The completeness of NED-LVS relative to the GALEX $NUV$ and 2MASS $Ks$-bands are $\sim$100\% out to $\sim$30~Mpc. Past this distance, the completeness relative to NIR luminosities (which trace a galaxy's stellar mass) remains moderate (70\%) out to 300~Mpc. In addition, when considering only galaxies with luminosities brighter than \Lstar, NED-LVS is $\sim$100\% complete out to $\sim$400~Mpc. We confirm this result via completeness estimates by number, where we find that the number of NED-LVS galaxies with $L>$\Lstar\ roughly matches the number expected under the LF at these greater luminosities. These estimates do not account for the missing galaxies obscured by the MW disk, which accounts for 5\% of the sky for a Galactic latitude restriction of $|b|<10^{\circ}$.

A comparison to other local Universe samples shows that NED-LVS is more complete by number than both GLADE and HECATE by 10\% and 30\%, respectively, in the appropriate limiting distances. When comparing the luminosity completeness of the 3 catalogs it is important to provide estimates based on the same expected luminosity scale (i.e., the luminosity in the denominator). Previous studies have provided estimates that are relative to only 0.6$\times$\Lstar\ whereas the estimates computed in this study are relative to the total luminosity under the LF. To make a proper comparison, we recompute the completeness of previous samples relative to the total luminosity and find that all three approach 100\% completeness below 30~Mpc. At distances beyond $\sim$80~Mpc, NED-LVS is more complete than both GLADE and HECATE by $\sim$10-20\%.

The galaxies in NED-LVS can be used for studies in many areas of astrophysics that include: star formation, galaxy evolution, large scale structure, galaxy environments, and searches for the electromagnetic counterparts to gravitational wave events, to name a few. When applying NED-LVS toward these or other science goals, users should note that the completeness of NED-LVS (and likely of any local Universe sample) can vary significantly with distance and location on the sky. Other limitations and caveats pertaining to the use and completeness estimates of NED-LVS include: an overestimation of SFRs based on MIR luminosities for early-type galaxies, and that investigations of large scale structure or galaxy environment should limit their focus to areas with high levels of completeness or to galaxies with luminosities greater than \Lstar.

We also explore the prioritization of galaxy properties when searching for the counterparts to GW events, where we include previously studied properties (stellar mass and SFR) and include for the first time sSFR since it is inversely correlated with the stellar population age of a galaxy. We find that prioritization by both stellar mass and sSFR put the host galaxy of GW170817 (NGC 4993) consistently in the top ten of the list of potential hosts, but SFR puts the host galaxy significantly lower on the list. These results persist even when using the larger, initial sky localization which contains four times the number of galaxies in the event's volume. The high prioritization of stellar mass and sSFR suggests long delay times for BNS merger events. Further study is encouraged on the prioritization of galaxy lists by more direct stellar population age measurements (e.g., depth of the 4000~\AA~break ($D_{N}4000$), the EW of Balmer lines, etc.).

The version of NED-LVS corresponding to the analysis presented here (see \S\ref{sec:catalogavail} and Appendix\ref{sec:coldesc}) can be found on the NED website \citep{nedlvs_doi}. As data from the literature and next generation of large redshift surveys (DESI; \cite{DESI}, Euclid; \cite{Euclid}, Nancy Grace Roman Space Telescope; \cite{wfirst}, SPHEREx; \cite{spherex}, etc.) become available in the coming years, we plan to fold them into NED and facilitate the extraction of updates to NED-LVS by users directly from the database, with updates to the increasing completeness estimates also made available on the website. NED-LVS is currently more complete than the galaxy samples used during LVK O3, but will grow in completeness due to the ingestion of new data, and will be a valuable resource for many other areas of astrophysical science.




\section*{Acknowledgements}

We thank Brad Cenko for helpful comments on the manuscript. This work was funded by the National Aeronautics and Space Administration through a cooperative agreement with the California Institute of Technology. This research has made use of the NASA/IPAC Extragalactic Database (NED), which is funded by the National Aeronautics and Space Administration and operated by the California Institute of Technology.

The Legacy Surveys consist of three individual and complementary projects: the Dark Energy Camera Legacy Survey (DECaLS; Proposal ID \#2014B-0404; PIs: David Schlegel and Arjun Dey), the Beijing-Arizona Sky Survey (BASS; NOAO Prop. ID \#2015A-0801; PIs: Zhou Xu and Xiaohui Fan), and the Mayall z-band Legacy Survey (MzLS; Prop. ID \#2016A-0453; PI: Arjun Dey). DECaLS, BASS and MzLS together include data obtained, respectively, at the Blanco telescope, Cerro Tololo Inter-American Observatory, NSF’s NOIRLab; the Bok telescope, Steward Observatory, University of Arizona; and the Mayall telescope, Kitt Peak National Observatory, NOIRLab. The Legacy Surveys project is honored to be permitted to conduct astronomical research on Iolkam Du’ag (Kitt Peak), a mountain with particular significance to the Tohono O’odham Nation.

The Pan-STARRS1 Surveys (PS1) and the PS1 public science archive have been made possible through contributions by the Institute for Astronomy, the University of Hawaii, the Pan-STARRS Project Office, the Max-Planck Society and its participating institutes, the Max Planck Institute for Astronomy, Heidelberg and the Max Planck Institute for Extraterrestrial Physics, Garching, The Johns Hopkins University, Durham University, the University of Edinburgh, the Queen's University Belfast, the Harvard-Smithsonian Center for Astrophysics, the Las Cumbres Observatory Global Telescope Network Incorporated, the National Central University of Taiwan, the Space Telescope Science Institute, the National Aeronautics and Space Administration under Grant No. NNX08AR22G issued through the Planetary Science Division of the NASA Science Mission Directorate, the National Science Foundation Grant No. AST-1238877, the University of Maryland, Eotvos Lorand University (ELTE), the Los Alamos National Laboratory, and the Gordon and Betty Moore Foundation.

Funding for SDSS-III has been provided by the Alfred P. Sloan Foundation, the Participating Institutions, the National Science Foundation, and the U.S. Department of Energy Office of Science. The SDSS-III web site is http://www.sdss3.org/. SDSS-III is managed by the Astrophysical Research Consortium for the Participating Institutions of the SDSS-III Collaboration including the University of Arizona, the Brazilian Participation Group, Brookhaven National Laboratory, Carnegie Mellon University, University of Florida, the French Participation Group, the German Participation Group, Harvard University, the Instituto de Astrofisica de Canarias, the Michigan State/Notre Dame/JINA Participation Group, Johns Hopkins University, Lawrence Berkeley National Laboratory, Max Planck Institute for Astrophysics, Max Planck Institute for Extraterrestrial Physics, New Mexico State University, New York University, Ohio State University, Pennsylvania State University, University of Portsmouth, Princeton University, the Spanish Participation Group, University of Tokyo, University of Utah, Vanderbilt University, University of Virginia, University of Washington, and Yale University.


\facilities{ADS,IRSA,NED}

\bibliographystyle{texstuff/aasjournal}   
\bibliography{texstuff/all,texstuff/joe}

\clearpage
\appendix

\section{Quality Checks of NED-LVS} \label{sec:app_visclass}

In this section, we provide more detailed information on our quality checks of NED-LVS. We have updated the object types and/or the redshift quality flags of over 200,000 objects via automated assessments and visual inspection. NED-LVS entries whose object types are not consistent with a galaxy (i.e., stars or nebulae in the Milky Way or parts of nearby galaxies) were removed from the sample, and are listed in Table~\ref{tab:objtypeup}. NED-LVS entries with redshifts that have been determined to be unreliable (i.e., based on noisy spectra, unphysical recessional velocities, or unphysical patterns in their redshift distributions) have been given a quality flag (z\_qual) of `True', and are listed in Table~\ref{tab:zqualup}. Incorporating these updates into the NED database is underway. We note here that our efforts to clean NED-LVS focuses on obvious cases, but should be considered as `likely' classifications and quality flags. Further assessments and cleaning will be performed in future versions.


\begin{figure}
  \begin{center}
  \includegraphics[scale=0.52]{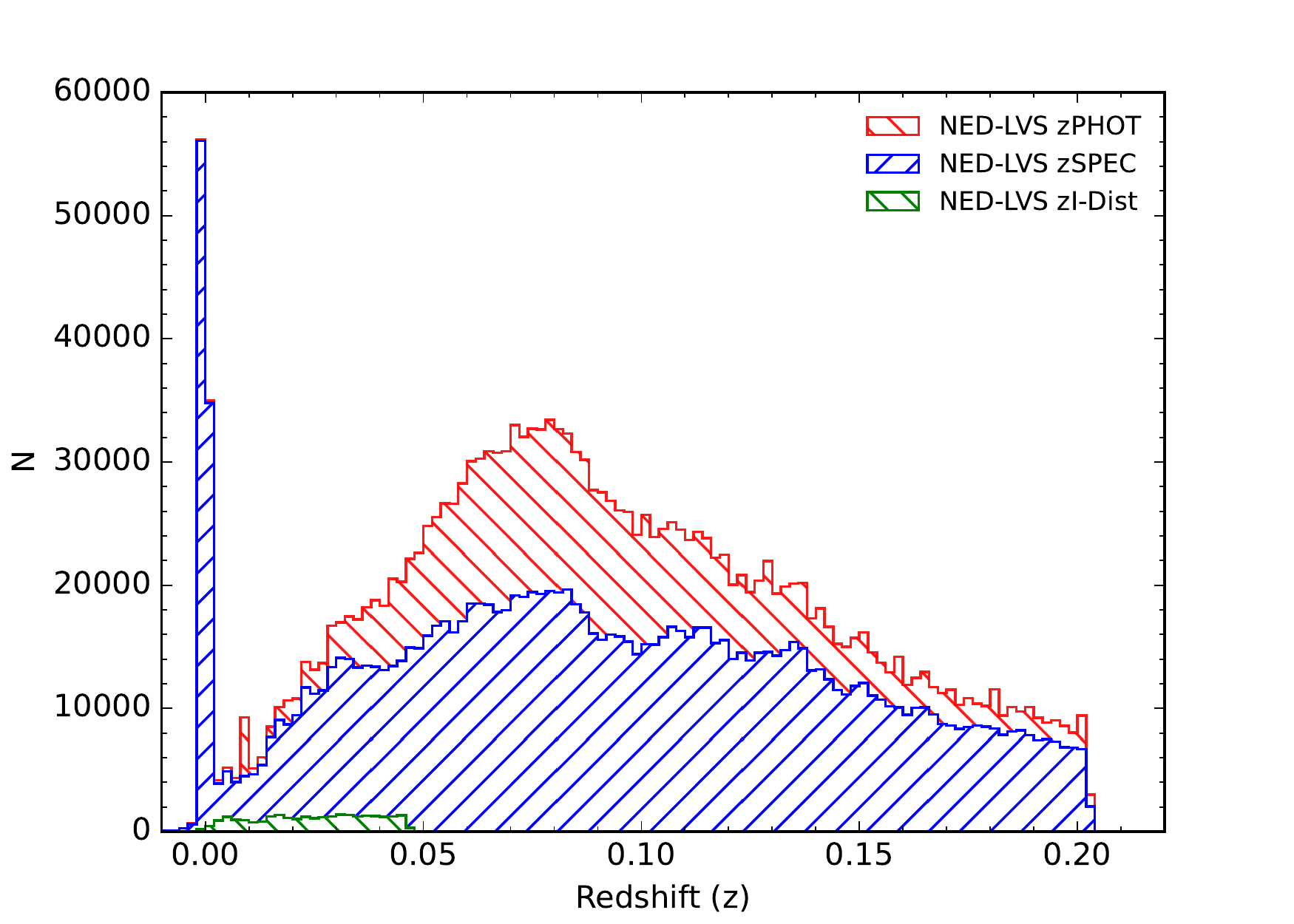}
  \caption{The redshift histogram for objects in NED-LVS prior to vetting. The large number of objects at z$\sim$0 is unphysical and prompted our efforts to clean the NED-LVS sample of stars misclassified as galaxies, unreliable redshifts, and other issues. These data are presented as a stacked histogram chart with zSPEC on top of zI-Dist and zPHOT on top of zSPEC in each bin. }
   \label{fig:zhist_precut}
   \end{center}
\end{figure} 

\subsection{Automated Assessments}
Prior to the vetting of NED-LVS objects, the redshift histogram showed a large, unphysical spike in the number of objects (N$\approx$100,000) near a redshift of zero indicating that these objects are Galactic in nature or that they have unreliable redshifts (Figure~\ref{fig:zhist_precut}). Inspection of a subset of objects in this spike showed that a significant fraction originated from SDSS DR13 ingested by NED in 2017 as generic visual sources ('VisS') with no distinction between `GALAXY' or `STAR'. Consequently, these objects were not removed from NED-LVS during construction based on their object types. We also found objects whose updated SDSS DR17 redshifts were greater than the distance limit of NED-LVS ($z>0.2$). 

\begin{figure}
  \includegraphics[scale=0.245]{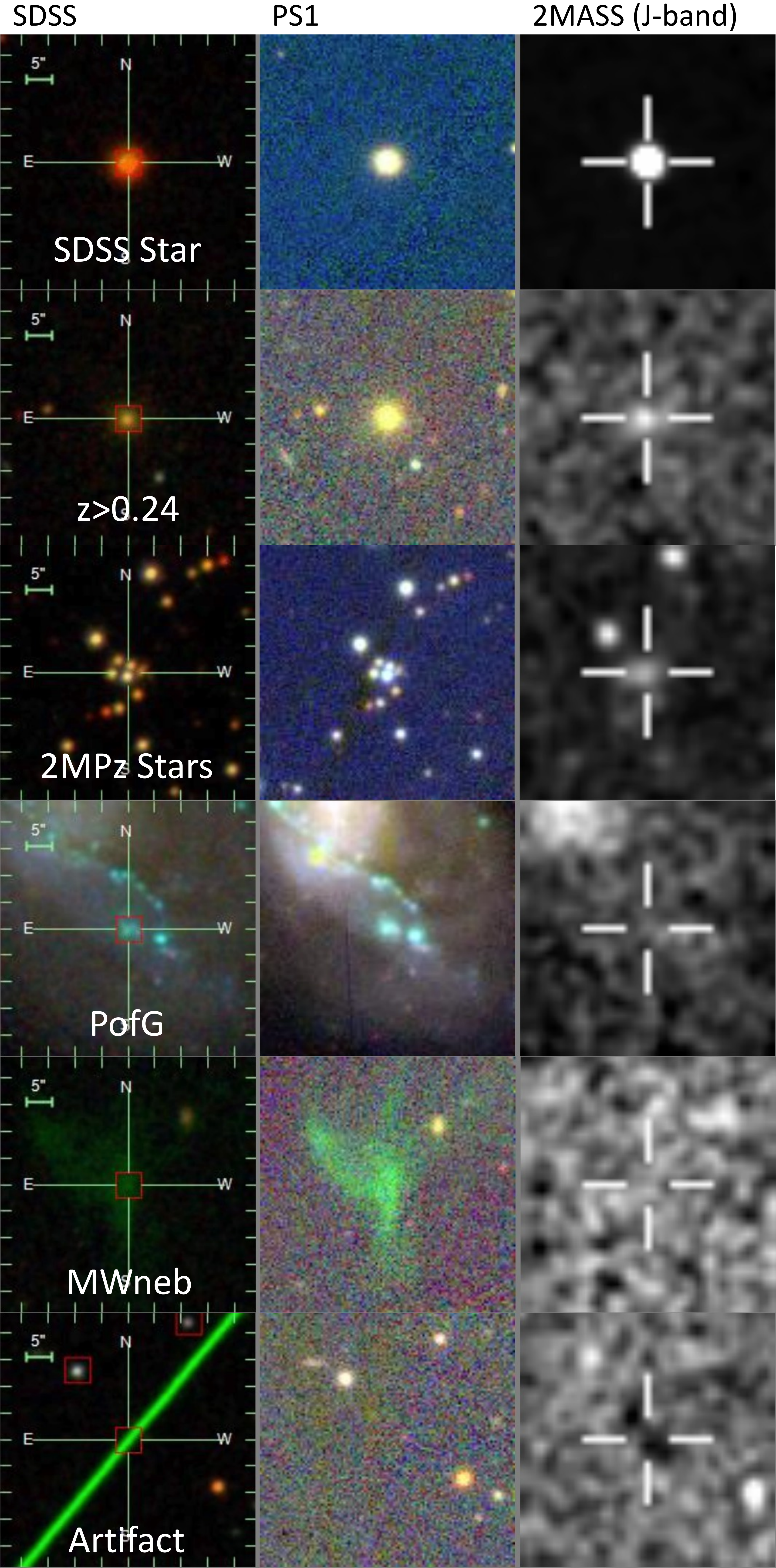}
  \caption{Examples of contamination found in NED-LVS. \textbf{Row 1} shows a generic `VisS' object in NED whose SDSS DR17 spectra confirm a classification of `STAR', \textbf{row 2} is a an object whose updated SDSS redshift is beyond the limits of our sample, \textbf{row 3} shows a grouping of stars given photo-z's since they are listed as extended objects in the 2MASS-XSC catalog, \textbf{row 4} shows a part of a galaxy (e.g., an \hii~reg), \textbf{row 5} shows a nebular region inside the MW, and \textbf{row 6} shows an artifact in SDSS where a spectroscopic fiber was placed. All of these types of objects have been removed from NED-LVS, and updates to the NED database are underway.}
   \label{fig:badobj}
\end{figure}

Our first attempt to clean these contaminants and outdated redshifts, is to perform an automated assessment given the SDSS DR17 values. We performed a cross match between all NED-LVS objects with $z<0.01$ to find 87k NED-LVS objects with updated spectroscopic classifications of `STAR', and $\approx$1,500 objects whose updated redshift in SDSS DR17 was above our distance limit. All of these objects have been removed from NED-LVS, and examples of these contaminants are shown in the top two rows of Figure~\ref{fig:badobj}.


While the automated updates from SDSS DR17 greatly reduced the number of objects at z$\sim$0, there remained a small overdensity (N$\approx$20k) with the majority at z$<$0.002. Visual inspection of these objects revealed continued contamination from obvious stars but with SDSS spectroscopic classifications of `GALAXY', galaxies with updated SDSS DR17 redshifts that fall inside our distance limits, faint objects with photometric redshifts, and parts of nearby galaxies (\hii~regions, star clusters, etc.). Given the heterogeneous sources of the remaining contamination, we have performed visual inspection of all NED-LVS objects with redshifts near zero, and in a few cases have expanded our visual inspection to include significant fractions of objects from individual publications whose data warranted additional investigation.

\subsection{Visual Inspection} \label{sec:app_visinspection}

We inspected all objects with z<0.002 to remove contaminants and to ensure a robust sample of nearby galaxies. Visual classification was performed via inspection of an object's  meta data (object types, redshifts, flags, redshift origin, etc.) from NED and optical color images from large-area surveys covering both the northern and southern sky: Legacy Survey \citep{Legacy}, SDSS \citep{york00,sdss7,sdss12}, and Pan-STARRS \citep{ps1}. In addition to these data, we also utilized updated SDSS DR17 redshifts, redshift quality flags, and object classifications based on spectra. These data were displayed on a web-based tool for quick inspection which facilitated updated classifications and redshift quality flags for tens of thousands of objects. The results of our visual inspection can be categorized by either updates to object types or to redshift quality flags, and are discussed separately below. 

\subsubsection{Object Types}

There was a wide variety of objects whose object types required updating. The contaminating objects most commonly found were: obvious stars or closely grouped stars with redshifts, nebular regions in the MW, and parts of nearby galaxies. We discuss identification of each type below. The full list of objects whose object types were updated is shown in Table~\ref{tab:objtypeup}, and examples are shown in Figure~\ref{fig:badobj}.

\begin{table*}

{Updated Object Types in NED-LVS}\\

\begin{tabular}{lrrrcc}
\hline
\hline
Object         & RA      &  Dec      & b    & Obj Type   & Obj Type  \\ 
Name           & (Deg)    &  (Deg)    & (Deg)  & (Old)      & (new)     \\ 
               & (J2000) &  (J2000)  &        &            &           \\ 
\hline

WISEA J201025.79-485254.3 & 302.6071 & -48.8817 & -32.77 & G & STAR \\
WISEA J005002.89+124301.5 & 12.5120 & 12.7171 & -50.15 & IrS & STAR \\
AGC 122835 & 31.3875 & 29.2328 & -30.91 & G & MWneb \\
AGC 227898 & 182.0042 & 5.9744 & 66.46 & RadioS & MWneb \\
SDSS J165713.92+251009.2 & 254.3080 & 25.1693 & 35.37 & G & Artifact \\
SDSS J034931.46-010314.5 & 57.3811 & -1.0540 & -40.00 & G & Artifact \\
2MASS J00412577+4110191 & 10.3577 & 41.1719 & -21.66 & VisS & *Cl \\
2MASS J00422968+4119533 & 10.6238 & 41.3315 & -21.51 & VisS & *Cl \\
GALEXMSC J004847.83+425136.7 & 12.1993 & 42.8602 & -20.01 & UvS & PofG \\
GALEXASC J004648.73+420005.1 & 11.7031 & 42.0014 & -20.86 & UvS & PofG \\
\hline
\end{tabular}

\caption{Ten Examples of updated object types where we list the first two instances of each object type. The updated object types include stellar contaminants in our Galaxy (`STAR'), nebulae associated with our Galaxy ('MWneb'), SDSS image artifact where spectra were taken (`Artifact'), objects specifically identified in a publication as star clusters associated with nearby galaxies (`*Cl'), and generic objects considered as parts of nearby galaxies (`PofG'). We provide the names, coordinates, Galactic latitudes, object types given in NED, and updated object types (i.e., `\textbf{STAR}'). The full list of objects is available online.}
\label{tab:objtypeup}

\end{table*}

\begin{table*}

{Updated Redshift Quality Flags in NED-LVS}\\
\begin{tabular}{lrrcll}

\hline
\hline
Object         & RA      &  Dec     & z       & reference code & z\_qual    \\ 
Name           & (Deg)   &  (Deg)   &         &                &          \\ 
               & (J2000) &  (J2000) & (helio) &                &          \\ 
\hline

SDSS J153958.64+233534.8 & 234.9943 & 23.5930 & $-2.6000e-04 \pm 1.78000e-04$ & 2007SDSS6.C...0000: & ? \\
CANDELS J033228.96-274656.1 & 53.1207 & -27.7822 & $+0.0000e+00 \pm \ldots$~~~~~~~~~~~~~~ & 2008ApJ...682..985W & ? \\
WISEA J195959.36-534630.0 & 299.9973 & -53.7751 & $-3.0700e-04 \pm 1.50000e-04$ & 20096dF...C...0000J & ? \\
WISEA J011743.33-545542.7 & 19.4300 & -54.9286 & $-1.1070e-03 \pm 3.94000e-04$ & 2009A\&A...499..357G & ? \\
SDSS J155603.29+111504.6 & 239.0137 & 11.2513 & $+1.5000e-04 \pm \ldots$~~~~~~~~~~~~~~ & 2011ApJS..193...28P & ? \\
FAIRALL 0232 & 47.2534 & -73.9611 & $-4.3030e-03 \pm 7.00000e-04$ & 1980MNRAS.192..389F & Contaminated \\
UGCA 390 & 216.9255 & 12.7602 & $+0.0000e+00 \pm \ldots$~~~~~~~~~~~~~~ & 1992CORV..C...0000F & Contaminated \\
GALEXASC J030255.39-603300.4 & 45.7313 & -60.5503 & $-6.1440e-03 \pm 1.77000e-04$ & 1996ApJS..107..201L & Contaminated \\
2dFGRS N325Z254 & 193.1543 & 0.3344 & $-2.0000e-04 \pm 2.97000e-04$ & 20032dF...C...0000C & Contaminated \\
WISEA J122224.16+431122.3 & 185.6009 & 43.1897 & $-3.4000e-03 \pm 2.80000e-03$ & 2004AJ....127.1943G & Contaminated \\
\hline
\end{tabular}

\caption{Ten Examples of updated redshift quality flags (`zQual') where we list the first five instances of each categorization. The updated flags include redshifts considered to be unreliable (`?') and redshifts derived from fluxes that have been contaminated by bright, nearby stars (`Contaminated'). We provide the names, coordinates, redshifts, reference code for that redshifts, and the updated quality flag. The full list of objects is available online.}
\label{tab:zqualup}

\end{table*}

One source of stellar contamination after automated updates are objects with SDSS spectroscopic classifications of `GALAXY', but whose spectral features (i.e., Hydrogen absorption lines) and radial optical profiles clearly indicated a stellar object (in addition to a redshift consistent with the MW). We conservatively classified $\sim$1000 objects as `STAR' only in obvious cases where strong spectral features in high signal-to-noise SDSS spectra and obvious point-like radial profiles indicated an object type of `STAR'.    

We also found closely grouped stars that were given photometric redshifts from the 2MASS Photometric Redshift catalog \citep[2MPZ;][]{bilicki14} based on their extended light profiles in 2MASS images \citep[see;][]{2mass-xsc}. However, visual inspection of higher-resolution images (SDSS, PS1, and Legacy survey) showed that these objects were clearly groups of stars (see Figure~\ref{fig:badobj}).  We visually inspected all NED-LVS objects whose final distances come from 2MPZ out to $z<0.05$ to find $\sim$6000 contaminants and labeled them as `STAR' in Table~\ref{tab:objtypeup}. More than 90\% of these $\sim$6000 objects have Galactic latitudes of |b| < 10~degrees which is further evidence of their stellar nature. To be conservative, we did not classify any 2MPZ object as `STAR' that spatially overlapped with an extended object. We note here that there exist spatial gaps in the optical surveys used for identification where no images were available, and consequently a visual classification could not be robustly determined. 

We also encountered $\sim$2500 objects in the vicinity of large, nearby galaxies with locations within the optical extent of the galaxy and/or those explicitly labeled as parts of galaxies in their publication description. The objects found in this category are: planetary nebulae, star clusters, globular clusters, \hii\ regions, and individual stars inside Local Group galaxies. All of these objects were labeled as `PofG' in Table~\ref{tab:objtypeup}. 

A similar type of contamination to the `PofGs' in nearby galaxies are objects inside our own Galaxy. Some of these are bright nebular knots within MW \hii\ regions where an SDSS spectroscopic fiber was placed, but the majority ($\sim$800) are HI radio detections from the Arecibo Legacy Fast ALFA Survey \citep[ALFALFA;][]{alfalfa}. These objects are known as high velocity clouds (HVCs) and are likely associated with the MW \citep{Wakker01,Lehner11,Lehner22}. We combined the ALFALFA lists of HVCs from  \cite{Martin09,Saintonge08,haynes11}, and classified these objects as MW nebula (`MWneb') and have subsequently removed them from NED-LVS. We also verified that none of these objects appear in the ALFALFA HI source catalog of \cite{Haynes18} which are considered to be extragalactic due to their velocities and associations with an optical counterpart. Only a handful of ALFALFA objects with low redshifts remain in NED-LVS that are associated with optical counterparts.

We also found N=33 instances where an SDSS spectroscopic fiber was placed on an artifact in the SDSS images (satellite streak, diffraction spike, etc.) or was clearly a sky fiber, and whose spectra contained high levels of noise and no obvious spectral features. These contaminants were labeled `Artifact' in Table~\ref{tab:objtypeup}.

Finally, we inspected objects with extreme luminosities as they could have a large effect on our completeness estimates. We inspected all objects with $Ks$-band luminosities greater than ten times $L^{*}$, and found $\sim$150 contaminants dominated by stars or nebular objects in the MW that were given distances outside of the Galaxy. These contaminants were labeled in Table~\ref{tab:objtypeup} with an appropriate classification as discussed earlier in this section.

\subsubsection{Redshift Quality Flags}

In addition to updated object types, we also flagged individual redshift measurements that were deemed unreliable. However, unlike the updated object types, setting a redshift quality flag does not require removal from NED-LVS since a more reliable redshift measurement may still be available in NED. Consequently, we have simply flagged these measurements as `True' and should be used with caution. We note that these quality flags are being incorporated back into NED and will inform future iterations of our fiducial redshift algorithm, where these redshifts will be given lower rank. Future versions of NED-LVS will likely have many of these redshifts replaced with more reliable measurements.

Examples of unreliable redshifts found in NED-LVS are those with unphysical recessional velocities: those with values less than $-$500 km/s. All redshifts with these values were treated as suspect as only a few objects in the literature have such extreme peculiar velocities \citep{Kourkchi17}. Roughly $\sim$800 objects with velocities less than -500 km/s were given z\_qual=`True'. 

Another common occurrence of an unreliable redshift is a redshift measurement of exactly zero and no uncertainty given. Data presented in this form often do not represent real measurements except in cases where a value of zero is used as a placeholder for a Galactic object or star \citep[e.g.,][]{Croom04,Conselice11,Bacon15,Sluse17,sanchez19,Boutsia20}. Of the objects with z=0 and no uncertainty given, we found $\sim$13k objects where the publication described these values as placeholders for stars and were given an object type of `STAR', and found $\sim$800 objects where no description of the object nor redshift quality were given and subsequently were given `z\_qual' of `True'.

We also examined objects with SDSS DR17 spectra in our visual inspection redshift range ($z<0.002$) to find $\sim$2000 objects whose spectra are either dominated by noise or whose redshift was based on a few lines with significant noise, where the redshift determination was not robust. In many cases the SDSS `zWarning' flag for these spectra was set to a value indicating a non-unique line identification. The `z\_qual' flag for these objects were set to `True'.


In addition, during our visual inspection of NED-LVS, we also encountered faint objects whose photometric redshifts exhibited discrete values at 0.001 intervals. We determined that these redshifts came from two main sources: 1) an early HST-COSMOS photometry catalog \citep{Capak07} and 2) the Red-Sequence Cluster Survey \citep[RCS;][]{RCS}. The photometric redshifts from \cite{Capak07} targeted higher redshift ranges than NED-LVS and were later updated by the COSMOS team \citep{Ilbert09,ilbert14} where several issues were addressed. We note that the \cite{Capak07} version of the redshifts remain in NED as they have been used by other authors before the updated values were published. The photometric redshifts from the RCS survey targeted and tuned their measurements to a range of 0.2--0.5, where the scatter around known spectroscopic redshifts is large and shows systematic offsets at z$<$0.2. Given that the intended useable redshift ranges of both data sets fall beyond the NED-LVS distance limits, the unphysical quantized values, and the large scatter when compared to spectroscopic measurements, we have determined that these objects are not suitable for our purposes and have removed them from the sample.

While we have put significant effort into vetting NED-LVS objects (especially at z$<$0.002), resource constraints prohibit checking of all 1.9 million objects. We anticipate further quality checks in future iterations of the sample.

\section{Completeness Estimates by Number} \label{sec:LFnum}

Here we provide an estimate of completeness by number in order to confirm the result from \S\ref{sec:compest} that NED-LVS is nearly 100\% complete for \Lstar\ galaxies and brighter. We compare the number of objects in NED-LVS to those expected from our fiducial $Ks$-band LFs \citep{jones06} via luminosity histograms in distance bins \citep[see also][]{gehrels16,glade,ducoin20,kovlakas21}. The expected number of objects under a Schechter LF in a given luminosity bin can be computed with the differential form of the Schechter function integral (see Figure 2 of \cite{gehrels16}) via:

\begin{equation}
N_{bin} = \Delta N = \phi^{*} x^{\alpha} e^{-x} \Delta V dx,
\end{equation} 

\noindent where $x=L/L^{*}$, $\Delta V$ is the volume element for the spherical shell, and $dx$ is the width of the luminosity bin.





\begin{figure*}
  \includegraphics[scale=0.52, trim=2.5cm 2.5cm 1cm 1cm]{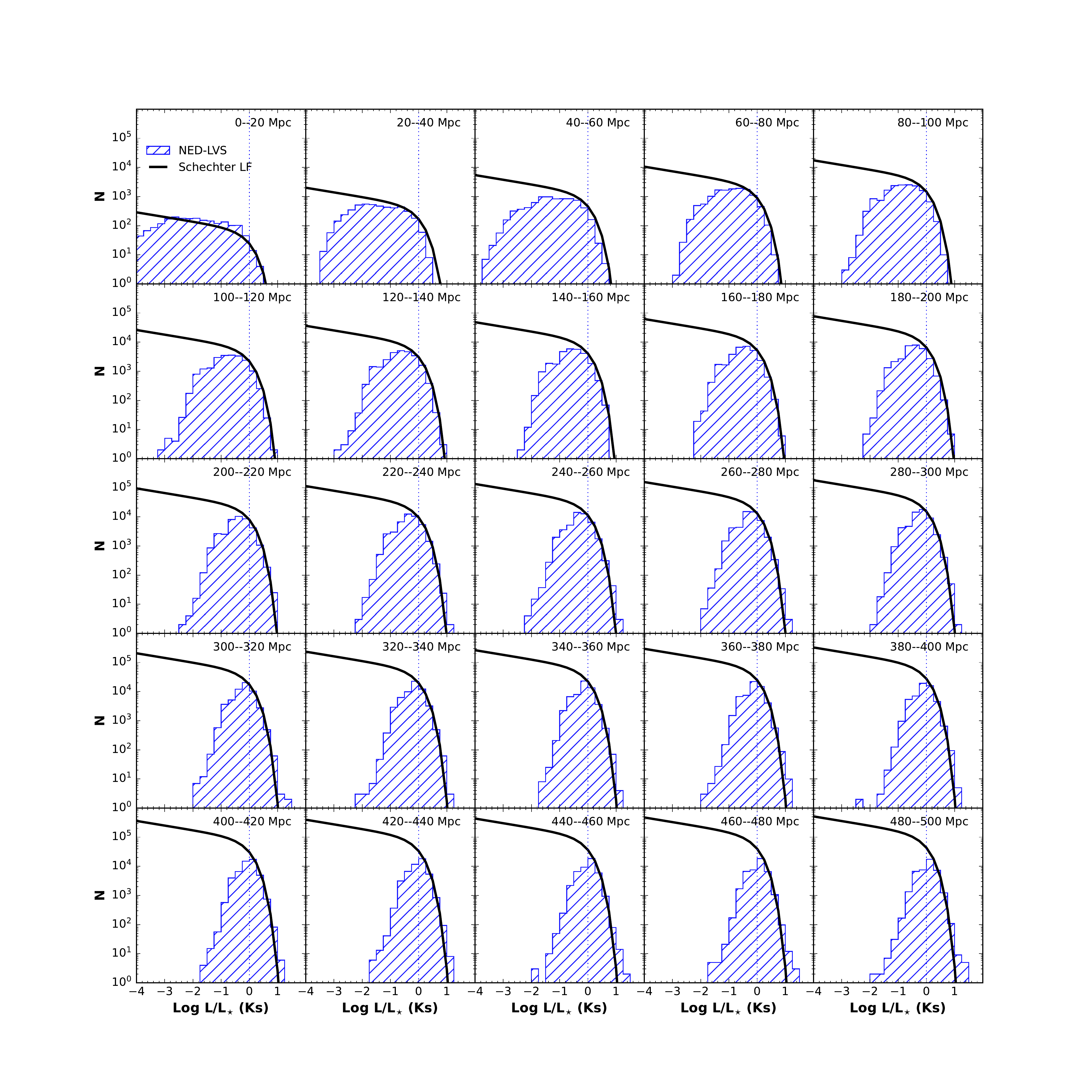}
  \caption{The distribution of $Ks$-band luminosities for objects in NED-LVS (blue-hashed filled histogram) in distance bins from 0--500~Mpc. The solid-black lines are the volume-scaled prediction of objects derived from the published $Ks$-band LF of \citep[][]{jones06}. }
   \label{fig:LFnum_1}
\end{figure*}
\begin{figure*}
  \includegraphics[scale=0.52, trim=2.5cm 2.5cm 1cm 1cm]{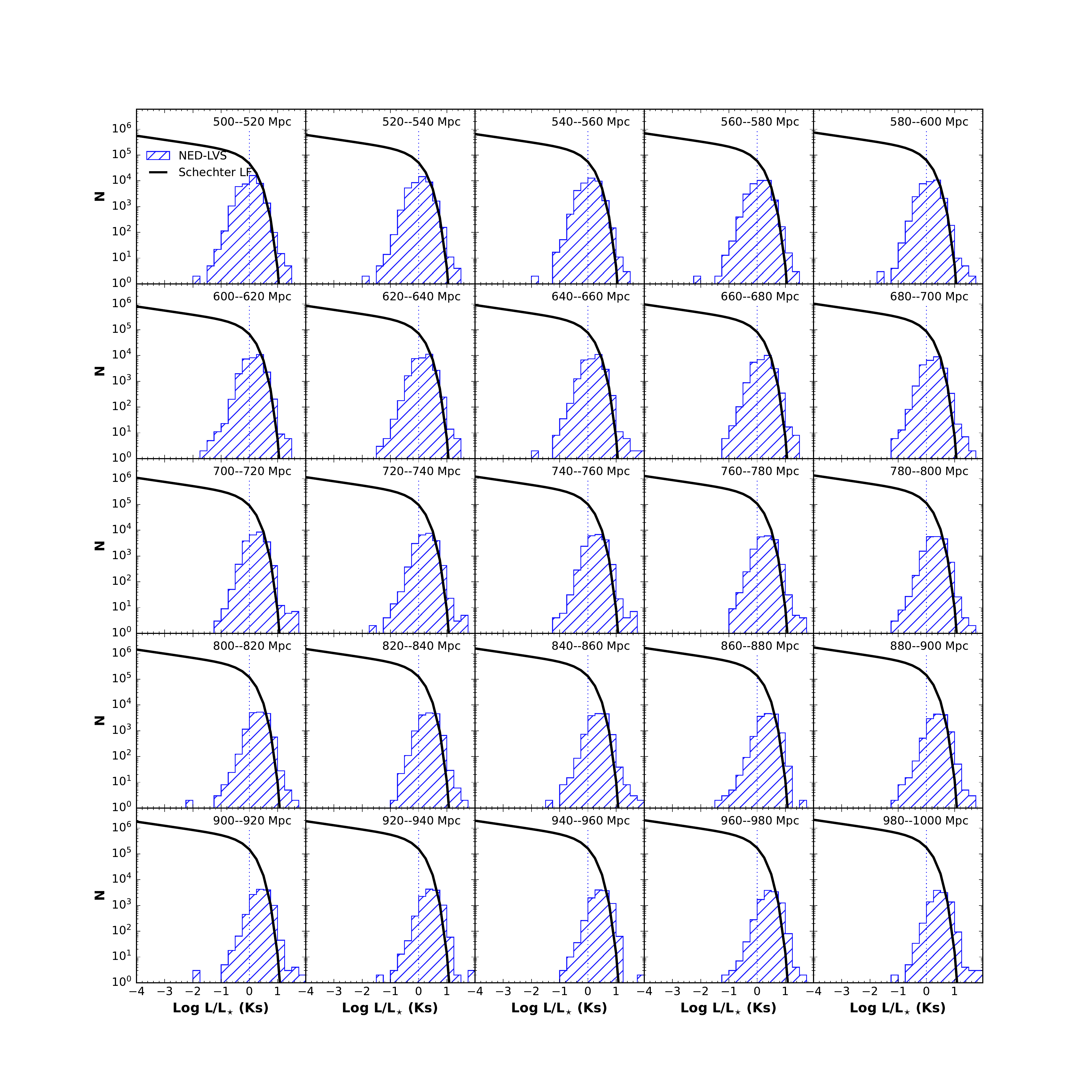}
  \caption{The distribution of $Ks$-band luminosities from NED-LVS (blue-hahsed filled histogram) in distance bins from 500--1000~Mpc. The solid-black lines are the volume-scaled prediction of objects derived from the published $Ks$-band LF of \citep[][]{jones06}.}
   \label{fig:LFnum_2}
\end{figure*}

Figure~\ref{fig:LFnum_1} (0--500~Mpc) and Figure~\ref{fig:LFnum_2}  (500--1000~Mpc) show the luminosity histograms of objects in NED-LVS and those expected from the Schechter function. We find that the number of \Lstar\ and brighter objects in NED-LVS meet the number expected from the LF in all distance bins below $\sim$400~Mpc. However, significant deviations from the number of objects expected from the LF are seen in the distance bins above 400~Mpc. Thus, we conclude that NED-LVS is nearly 100\% complete by number for \Lstar\ galaxies and brighter out to $\sim$400~Mpc.

We also note that there is a non-insignificant number of high-luminosity objects ($L>10\times$\Lstar); especially at distances beyond 300~Mpc. We have visually inspected (see \S\ref{sec:app_visinspection}) all objects with $Ks$-band luminosities greater than 10 times \Lstar. The contaminants found when inspecting high-luminosity objects were dominated by stars and nebular regions in the MW that were given extragalactic distances. However, we note that the vast majority of objects with $Ks$-band luminosities greater than 10 times \Lstar\ were large early-type and spiral galaxies \citep{Ogle16} or bright QSOs whose high luminosities are reasonable given the object's properties. Since we have visually inspected all objects with $L>10\times$\Lstar, this could indicate limitations in computing completeness estimates by using Schechter LFs. See \S\ref{sec:disclimit} for more discussion.

\section{Column Description of NED-LVS} \label{sec:coldesc}

In this section we present the column descriptions of NED-LVS. Table~\ref{tab:ColDesc} presents a description of the  columns, and the following paragraphs provide additional details for columns not described in the main text or in other appendix sections. The NED-LVS table is available at \citep{nedlvs_doi} in FITS format ($\sim$1~GB).

The values populated in the redshift technique column `z\_tech' are defined as follows: `SPEC' is a redshift estimated by spectroscopy, `PHOT' is estimated by photometry, `MOD' is a modelled value, `MULT' is determined using multiple techniques, `INFD' is value inferred from other redshifts (e.g., nearest neighbor), `UNKN' indicates that a publication did not clearly state the technique used to measure the redshift. The last technique in NED-LVS is labeled as `None' has a NULL value, which for historical reasons is treated as spectroscopic. Measurements ingested with a `None' technique are the result of early ingestion (prior to approximately 10 years ago) of redshifts into NED when the vast majority of measurements were derived from spectroscopy, and subsequently were given an empty or blank redshift technique. However, given the orders of magnitude rise in redshift measurements and the increase in diversity of techniques, NED now explicitly indicates the measurement technique from the list above.

The redshift quality flag `z\_qual' in NED is generally set when the publication describing a measurement (see `z\_refcode' column) explicitly states that the redshift value may be or is unreliable. However, in this study we have used automated assessment and visual inspection to find an additional 110,000 redshifts that are unreliable for various reasons (see \S\ref{sec:app_visclass}). These redshift values have had their `z\_qual' flag set, and we have set an additional flag (`z\_qual\_flag') to indicate that the quality is unreliable as determined in our study and the original publication of the data. 

The apparent magnitudes in NED-LVS are left as those reported in the original catalogs and do not have MW extinction applied. However, the luminosities reported in this study do have their MW extinctions applied before conversion. We have provided the extinction values used for each filter in the columns with names of the form: `A\_FILTER\_MWext'. We also note that upper limit apparent magnitudes have their uncertainties set to 99. 

\newpage
\LTcapwidth=\textwidth
\begin{longtable}{l@{\hskip -1.1cm}r@{\hskip 0.1cm}rp{11cm}}
\caption{Description of NED-LVS columns. Columns 2 and 3 of this table present the data types (B=Boolean, D=double precision floating point, F=single precision floating point, and S=string) and the units for the NED-LVS columns. }
\label{tab:ColDesc}\\

\hline
\hline
 Column  & Data Type & Unit & Description      \\
\hline
\endfirsthead

\hline
\hline
 Column  & Data Type & Unit & Description      \\
\hline
\endhead

\hline
\endfoot

\hline
\endlastfoot

objname & S & -- & Preferred object name in NED \\
ra, dec & D & deg & Equatorial coordinates of right ascension and declination [J2000, fk5] \\
objtype & S & -- & Preferred object type in NED (see Table 1 in section 2.1) \\
z, z\_unc & F & -- & Fiducial redshift [heliocentric] \\
z\_tech & S & -- & Technique used to measure redshift [SPEC, PHOT, UNKN, INFD, MOD, MULT, None or NULL] \\
z\_qual & B & -- & Qualifier flag indicating reliability of redshift [True=unreliable] \\
z\_qual\_flag & B & -- & Boolean flag indicating that the zqual flag has been updated in NED-LVS [True=updated] \\
z\_refcode & S & -- & Reference code for the publication that provided the redshift \\
ziDist, ziDist\_unc & F & Mpc & Redshift-independent luminosity distance computed as the weighted average of measurements in either the primary or secondary indicators \\
ziDist\_method & S & -- & Method used to measure the redshift-independent distance when a single method is used to compute the weighted average, otherwise "Wavg" when a mixture of methods are used [Cepheid, SNIa, TRGB, Wavg, etc.] \\
ziDist\_indicator & S & -- & Measurement type indicator for redshift-independent distance [Primary or Secondary] \\
ziDist\_refcode & S & -- & Reference code for the publication that provided the distance when a single measurement was chosen, otherwise "Mix" indicates an average of multiple measurements was used [refcode or Mix] \\
DistMpc, DistMpc\_unc & F & Mpc & Selected luminosity distance used in NED-LVS \\
DistMpc\_method & S & -- & Method used in the selected distance [Redshift or zIndependent] \\
ebv & F & mag & Foreground MW reddening E(B-V) from Schlafly \& Finkbeiner (2011) \\
A\_FUV\_MWext & F & mag & MW extinction in GALEX FUV-band assuming Fitzpatrick et al. (1999) \\
A\_NUV\_MWext & F & mag & MW extinction in GALEX NUV-band assuming Fitzpatrick et al. (1999) \\
A\_J\_MWext & F & mag & MW extinction in 2MASS J-band assuming Fitzpatrick et al. (1999) \\
A\_H\_MWext & F & mag & MW extinction in 2MASS H-band assuming Fitzpatrick et al. (1999) \\
A\_Ks\_MWext & F & mag & MW extinction in 2MASS Ks-band assuming Fitzpatrick et al. (1999) \\
A\_W1\_MWext & F & mag & MW extinction in WISE W1-band assuming Fitzpatrick et al. (1999) \\
A\_W2\_MWext & F & mag & MW extinction in WISE W2-band assuming Fitzpatrick et al. (1999) \\
A\_W3\_MWext & F & mag & MW extinction in WISE W3-band assuming Fitzpatrick et al. (1999) \\
A\_W4\_MWext & F & mag & MW extinction in WISE W4-band assuming Fitzpatrick et al. (1999) \\
m\_FUV, m\_FUV\_unc & F & mag & Apparent magnitude in GALEX FUV-band  in AB system \\
m\_NUV, m\_NUV\_unc & F & mag & Apparent magnitude in GALEX NUV-band  in AB system \\
Lum\_FUV, Lum\_FUV\_unc & D & erg/s & Monochromatic luminosity ($\nu F_{\nu}$) in GALEX FUV-band \\
Lum\_NUV, Lum\_NUV\_unc & D & erg/s & Monochromatic luminosity ($\nu F_{\nu}$) in GALEX NUV-band \\
GALEXphot & S & -- & Flag indicating which catalog was used for GALEX photometry [ASC, MSC, z0MGS] \\
m\_J, m\_J\_unc & F & mag & Apparent magnitude in 2MASS J-band  in Vega system \\
m\_H, m\_H\_unc & F & mag & Apparent magnitude in 2MASS H-band  in Vega system \\
m\_Ks, m\_Ks\_unc & F & mag & Apparent magnitude in 2MASS Ks-band  in Vega system \\
Lum\_J, Lum\_J\_unc & D & erg/s & Monochromatic luminosity ($\nu F_{\nu}$) in 2MASS J-band \\
Lum\_H, Lum\_H\_unc & D & erg/s & Monochromatic luminosity ($\nu F_{\nu}$) in 2MASS H-band \\
Lum\_Ks, Lum\_Ks\_unc & D & erg/s & Monochromatic luminosity ($\nu F_{\nu}$) in 2MASS Ks-band \\
tMASSphot & S & -- & Flag indicating which catalog was used for 2MASS photometry [PSC, XSC, LGA] \\
m\_W1, m\_W1\_unc & F & mag & Apparent magnitude in WISE W1-band  in Vega system \\
m\_W2, m\_W2\_unc & F & mag & Apparent magnitude in WISE W2-band  in Vega system \\
m\_W3, m\_W3\_unc & F & mag & Apparent magnitude in WISE W3-band  in Vega system \\
m\_W4, m\_W4\_unc & F & mag & Apparent magnitude in WISE W4-band  in Vega system \\
Lum\_W1, Lum\_W1\_unc & D & erg/s & Monochromatic luminosity ($\nu F_{\nu}$) in WISE W1-band \\
Lum\_W2, Lum\_W2\_unc & D & erg/s & Monochromatic luminosity ($\nu F_{\nu}$) in WISE W2-band \\
Lum\_W3, Lum\_W3\_unc & D & erg/s & Monochromatic luminosity ($\nu F_{\nu}$) in WISE W3-band \\
Lum\_W4, Lum\_W4\_unc & D & erg/s & Monochromatic luminosity ($\nu F_{\nu}$) in WISE W4-band \\
WISEphot & S & -- & Flag indicating which catalog was used for WISE photometry [Pfit, APER, z0MGS] \\
SFR\_W4, SFR\_W4\_unc & F & Msun/yr & Star formation rate using W4 luminosity scaling relationship (Section 2.4) \\
SFR\_hybrid, SFR\_hybrid\_unc & F & Msun/yr & Star formation rate using a FUV+W4 luminosity scaling relationship (Section 2.4) \\
ET\_flag & B & -- & Boolean flag indicating an object is an early-type galaxy determined via IR or UV-IR colors (Section 4.3) [True=colors indicate an early-type galaxy], and that the SFRs may be overestimated to varying degrees.  \\
Mstar, Mstar\_unc & F & Msun & Stellar mass using a W1 luminosity scaling relationship (Section 2.4) \\
MLratio & F & Msun/erg/s & The mass-to-light ratio used to calculate Mstar (Section 2.4) \\
\hline

\end{longtable}



\end{document}